\documentclass[journal]{IEEEtran}
\ifCLASSINFOpdf
\else
\fi
\usepackage{cite}
\usepackage{amsmath,amssymb,amsfonts}
\usepackage{algorithmic}
\usepackage{graphicx}
\usepackage{textcomp}
\usepackage{xcolor}
\def\BibTeX{{\rm B\kern-.05em{\sc i\kern-.025em b}\kern-.08em
		T\kern-.1667em\lower.7ex\hbox{E}\kern-.125emX}}

\DeclareMathOperator*{\argmax}{arg\,max}
\DeclareMathOperator*{\argmin}{arg\,min}
\usepackage{mathtools}

\usepackage{steinmetz}
\usepackage{authblk}

\usepackage{blindtext}
\usepackage{caption}
\usepackage{subcaption}

\newtheorem{remark}{Remark}
\newtheorem{lemma}{Lemma}
\newtheorem{corollary}{Corollary}
\newtheorem{proposition}{Proposition}

\begin{document}
%\title{RSMA for Overloaded MIMO Networks: Analysis and Low-Complexity Design}
\title{RSMA for Overloaded MIMO Networks: Low-Complexity Design for Max-Min Fairness}

\author{Onur~Dizdar,~\IEEEmembership{Member,~IEEE,}
        Ata~Sattarzadeh,
        Yi~Xien~Yap,
        and~Stephen~Wang,~\IEEEmembership{Senior~Member,~IEEE}% <-this % stops a space
\thanks{The authors are with VIAVI Marconi Labs, VIAVI Solutions Inc., Stevenage, UK. e-mail: (\{onur.dizdar, ata.sattarzadeh, yap.yixien, stephen.wang\}@viavisolutions.com).}% <-this % stops a space
%%\thanks{Manuscript received April 19, 2005; revised August 26, 2015.}
}

\maketitle

\begin{abstract}
Rate-Splitting Multiple Access (RSMA) is a robust multiple access scheme for multi-antenna wireless networks. In this work, we study the performance of RSMA in downlink overloaded networks, where the number of transmit antennas is smaller than the number of users. 
RSMA has been investigated in overloaded networks in previous works by formulated optimization problems and their solutions by interior-point methods. This limits the practical use of such designs in practical systems.
Our aim is to develop low-complexity precoding, rate, and power allocation techniques for RSMA for its use in practical overloaded networks.
First, we provide analysis and closed-form expressions for optimal power and rate allocations considering max-min fairness when low-complexity precoders are employed. The derived closed-form solutions are used to propose a low-complexity RSMA system design for precoder selection and resource allocation for arbitrary number of users and antennas under perfect and imperfect Channel State Information at the Transmitter (CSIT). We compare the performance of the proposed design with benchmark designs based on Space Division Multiple Access (SDMA) with and without user scheduling. By numerical results, we show that the proposed low-complexity RSMA design achieves a significantly higher rate compared to the SDMA-based benchmark designs under perfect and imperfect CSIT.
\end{abstract}


\IEEEpeerreviewmaketitle

\section{Introduction}
\label{sec:introduction}
\IEEEPARstart{T}{h}e demand for increased connectivity and user density has become a key performance indicator for state-of-the-art communications standards, and the increasing trend of these parameters is expected to continue for the foreseeable future \cite{cisco_2020, rajatheva_2020}. %A ubiquitous coverage over a massive number of devices is necessary to realize the envisioned wireless networks of the future. 
This brings forth several challenges in the system design, such as, user scheduling, interference management, and channel state information (CSI) acquisition from each device to be served. 
A key enabling technology to support connectivity in state-of-the-art communication systems is Multi-User Multiple-Input Multiple-Output (MU-MIMO). MU-MIMO makes use of the spatial dimension to support multiple users in the same time and frequency resources. 
The most common form of MU-MIMO is Space Division Multiple Access (SDMA), which performs linear precoding to separate user streams at the transmitter and treats multi-user interference as noise at the receivers. 
The Degree-of-Freedom (DoF) analysis show that when perfect CSI at transmitter (CSIT) is available for all users, SDMA is optimal in underloaded multi-antenna broadcast channel (BC), where the number of antennas at the transmitter is larger than the number of users \cite{clerckx_2021}. This makes SDMA a popular choice in state-of-the-art communication standards, even though the optimality conditions are difficult to satisfy in practical systems. 

The trend of increasing user density imply that overloaded systems will be a common occurrence in the next generation networks. In overloaded systems, the number of devices to be served is larger than the number of transmit antennas. Overloaded systems have become common especially in emerging application scenarios, such as, massive Internet-of-Things (IoT) and satellite communications \cite{mao_2021, yin_2021, yin_2021_2, yin_2023, si_2022}. 
Such systems suffer from increased overhead and complexity due to signalling and user scheduling, which may bring a significant burden to the systems with increasing number of users \cite{3gpp_2023}. 
As the multi-user interference cannot be fully cancelled even with perfect CSIT, the DoF of overloaded SDMA systems collapse to zero, leading to a saturating rate with increasing Signal-to-Noise Ratio (SNR) \cite{clerckx_2021, yin_2021, joudeh_2017}. 
Additionally, acquiring accurate CSIT, which has already been a major challenge in practical systems, gets even more challenging with increasing user density due to the increased overhead and signalling requirements. Imperfect CSIT stems from numerous factors in multi-antenna networks, such as, channel estimation errors, finite-length CSI feedback, user mobility and network latency, pilot contamination, and so on. Using imperfect CSIT for multiple-access transmission leads to multi-user interference between the transmitted streams and performance degradation, the impact of which depends on the source and severity of CSIT inaccuracy \cite{zhang_2009, papazafeiropoulos_2015, truong_2013, hao_2015, dai_2017, mao_2018, dizdar_2021, mishra_2022_5}. These factors call for advanced schemes to address the requirements of next generation wireless networks.

In this work, we consider Rate-Splitting Multiple Access (RSMA) to perform downlink multiple-access communications in overloaded networks with perfect and imperfect CSIT. 
RSMA is a multiple access technique for multi-antenna BC that relies on linearly precoded Rate-Splitting (RS) at the transmitter and Successive Interference Cancellation (SIC) at the receivers. 
Enabled by these operations, RSMA can partially decode the multi-user interference and partially treat it as noise, and can bridge the extremes of treating multi-user interference as noise, as in SDMA, and fully decoding the multi-user interference, as in Non-Orthogonal Multiple Access (NOMA).
It has been shown that RSMA encapsulates and outperforms existing multiple access techniques, such as SDMA, NOMA, Orthogonal Multiple Access (OMA), and multicasting, and can address the problems in state-of-the-art communication systems due to its flexible nature and interference management capabilities \cite{clerckx_2021, mao_2018, clerckx_2016, mao_2019, dizdar_2020,  mao_2022_2, clerckx_2023}. 
RSMA has also been shown to have a strong interplay with other enabling technologies due to its flexible structure and strong interference management capabilities, such as satellite communications \cite{yin_2021, yin_2021_2, yin_2023, si_2022}, massive MIMO \cite{hao_2015, dai_2017, dizdar_2021, mishra_2022_5, mishra_2022_2, dai_2016, papazafeiropoulos_2018}, and semantic communications \cite{yang_2023}\footnote{For a comprehensive overview of RSMA, please see \cite{dizdar_2020, mao_2022_2, clerckx_2023, mishra_2022_4, yin_2022, li_2022} and the references therein}. It has been proven that RSMA achieves a larger DoF than SDMA and NOMA under perfect and imperfect CSIT \cite{clerckx_2021, mao_2021, yin_2021, joudeh_2017, mishra_2022_3}.  
\vspace{-0.6cm}
\subsection{Related Work}
\label{eqn:relatedwork}
\vspace{-0.2cm}
Design and investigation of max-min rate performance of RSMA in overloaded networks with perfect or imperfect CSIT has been conducted in \cite{yin_2021, yin_2021_2, yin_2023, si_2022, joudeh_2017, mao_2020, xu_2023, yu_2019, yalcin_2020, chen_2021, chen2_2020}. 
The max-min fairness DoF achieved by RSMA in overloaded systems with perfect CSIT is studied in \cite{joudeh_2017} in the context of multigroup multicasting. It is shown that RSMA achieves higher rate than SDMA using partitioned beamforming and message splitting with perfect CSIT. Further investigation of RSMA performance for multigroup multicasting is conducted in \cite{yalcin_2020, chen_2021, chen2_2020} in terms of achievable rate and error performance.
The works in \cite{yin_2021, yin_2021_2, yin_2023, si_2022} investigate the performance of RSMA in overloaded networks in the context of multigroup multicasting and multibeam satellite systems. Max-min fairness DoF analysis for imperfect CSIT is performed in \cite{yin_2021} to prove that RSMA achieves a higher DoF than SDMA in such systems. 
The authors investigate the max-min fairness performance of RSMA with cooperative relaying in \cite{mao_2020}. In \cite{xu_2023}, the max-min fairness performance with and without cooperative relaying is investigated for finite blocklength coding. Finally, max-min fairness performance of RSMA in cloud radio access network (C-RAN) systems is studied in \cite{yu_2019}.

The conclusion from all of the abovementioned works is that RSMA can outperform SDMA and NOMA in terms of max-min fairness in overloaded networks. However, the major drawback of these works is that they use optimization problem formulations which can only be solved by interior-point methods for design and performance analysis. Thus, their application in practical systems is limited.
In order to obtain a practical design, it is essential to reduce the computational complexity of the proposed design by using low-complexity precoders, such as, Zero-Forcing (ZF) and Maximum Ratio Transmission (MRT), and obtain low-complexity rate/power allocation methods to perform RS. For underloaded systems, low-complexity precoders have been employed in \cite{hao_2015, dai_2017, dizdar_2021, dai_2016, papazafeiropoulos_2018, clerckx_2020} for analysis and to obtain low-complexity RSMA designs. 
Since the abovementioned works consider underloaded scenarios, only a single power allocation coefficient needs to be optimized once the precoders are set. As will be evident in Section~\ref{sec:formulation_ZF}, our formulated problem for overloaded scenarios consists of two parameters to be optimized, namely power and rate allocation coefficients, which makes the problem more challenging.
\vspace{-0.6cm}
\subsection{Contributions}
\label{eqn:contributions}
\vspace{-0.2cm}
The contributions of the paper can be listed as follows:
\begin{enumerate}
\item We propose a system model where a multi-antenna RSMA transmitter serves multiple single-antenna users in an overloaded system for downlink transmission. We consider an arbitrary number of transmit antennas, number of users, total transmit power, varying user channel disparity, and CSIT quality in the system. The splitting of the user rates is controlled by a power allocation coefficient that distributes the total transmit power between common and private precoders. In one of the considered settings, a group of users are served only by the common stream and the rate allocated to those users are controlled by a rate allocation coefficient.

\item We formulate a max-min fairness optimization problem over ergodic rates for RSMA in overloaded networks with perfect and imperfect CSIT. We adopt low-complexity ZF and MRT precoders for the private streams to limit the design space of the formulated problem. Depending on the type of precoders employed for the private streams, we propose two different system designs, which transform the formulated problem into two novel problems of different forms. 

\item We derive expressions and bounds for the approximate ergodic rates for the common and private streams of RSMA to be used in solving the formulated problems. Using these bounds under different assumptions, we obtain closed-form solutions for rate and power allocation in terms of the number of users, transmit antennas, total transmit power, user channel disparity, and CSIT quality in the system for perfect and imperfect CSIT cases separately. Then, we make use of the obtained closed-form solutions and the two proposed system designs to design the first low-complexity RSMA system for overloaded multi-antenna networks.

\item We analyse the performance of the proposed RSMA design by numerical results. We perform performance comparison of the proposed RSMA design with that of SDMA using ZF and MRT precoding with and without scheduling for small and large scale MIMO. To the best of our knowledge, this is the first work to investigate the performance of RSMA for a large number of transmit antennas and users in overloaded MIMO systems. We verify that the proposed low-complexity design achieves significant performance gain compared to SDMA under various settings and parameters.
\end{enumerate}

To the best of our knowledge, this is the first paper analysing RSMA in an overloaded network with low-complexity precoders and proposing a low-complexity solution for precoder selection, rate and power allocation for max-min fairness with perfect and imperfect CSIT.

The rest of the paper is organized is as follows. Section~\ref{sec:system} gives the system model. We formulate the max-min fairness problem in Section~\ref{sec:formulation}. We derive expressions and bounds for the achievable rates in Section~\ref{sec:boundRate}. 
Closed-form expressions for optimal resource allocation and the proposed low-complexity design for perfect and imperfect CSIT are given in Sections~\ref{sec:solution_perfectcsit}~and~\ref{sec:solution_imperfectcsit}, respectively. Section~\ref{sec:numerical} gives numerical results. Section~\ref{sec:conclusion} concludes the paper.

\textit{Notation:} Vectors are denoted by bold lowercase letters. 
The operations $|.|$ and $||.||$ denote the cardinality of a set or absolute value of a scalar and $l_{2}$-norm of a vector, respectively. 
The operation $\mathbf{a}^{H}$ denotes the Hermitian transpose of a vector $\mathbf{a}$. 
$\mathcal{CN}(0,\sigma^{2})$ denotes the Circularly Symmetric Complex Gaussian distribution with zero mean and variance $\sigma^{2}$. $\mathbf{I}$ denotes the identity matrix. $\left \lfloor{.}\right \rceil $ denotes the round operation. Logarithms are natural logarithms, {\sl i.e.,} $\log(.)=\log_{e}(.)$, unless stated otherwise. $\mathrm{Gamma}(D,\theta)$ represents the Gamma distribution with the probability density function $f(x)=\frac{1}{\Gamma(D)\theta^{D}}x^{D-1}e^{-\frac{x}{\theta}}$ and $\chi^{2}_{k}$ represents the chi-squared distribution with the probability density function $f(x)=\frac{1}{2^{k/2}\Gamma(k/2)}x^{k/2-1}e^{-x/2}$. For any complex $x$ with a positive real part, $\Gamma(x)=\int_{0}^{\infty}t^{x-1}e^{-t}dt$ is the gamma function and $\Gamma^{\prime}(x)$ is the derivative of $\Gamma(x)$ with respect to $x$ and $\psi(x)=\frac{\Gamma^{\prime}(x)}{\Gamma(x)}$. 
\vspace{-0.3cm}
\section{System Model}
\label{sec:system}
\vspace{-0.1cm}
\begin{figure}[t!]
	\centerline{\includegraphics[width=2.8in,height=2.8in,keepaspectratio]{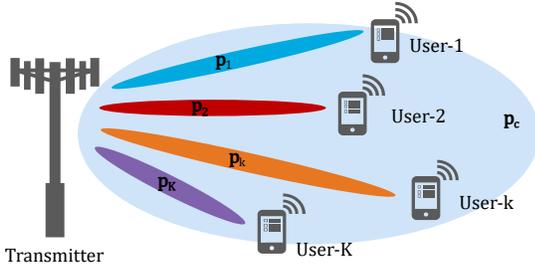}}
	\vspace{-0.1cm}
	\caption{Transmission model of $K$-user 1-layer RSMA.}
	\label{fig:system}
	\vspace{-0.5cm}
\end{figure}
We consider the system model in Fig.~\ref{fig:system} consisting of one transmitter with $N$ transmit antennas and $K$ single-antenna users indexed by $\mathcal{K}=\{1,\ldots,K\}$, with $N < K$. We employ 1-layer RSMA for downlink multi-user transmission. 1-layer RSMA performs message splitting at the transmitter to obtain a single common stream for all users. Accordingly, the message for user-$k$, denoted by $W_{k}$, is split into two independent parts, namely common part $W_{c,k}$ and private part $W_{p,k}$, $\forall k\in\mathcal{K}$. 
The common parts of all user messages are combined into a single common message, denoted by $W_{c}$. The messages $W_{c}$ and $W_{p,k}$ are independently encoded into streams $s_{c}$ and $s_{k}$, respectively. After linear precoding, the transmit signal is written as \vspace{-0.1cm}
\begin{align}	
\mathbf{x}=\sqrt{P(1-t)}\mathbf{p}_{c}s_{c}+\sqrt{Pt}\sum_{k \in \mathcal{K}}\sqrt{\mu_{k}}\mathbf{p}_{k}s_{k},
%	\label{eqn:1}
\end{align} 
where $||\mathbf{p}_{c}||^{2}=1$, $||\mathbf{p}_{k}||^{2}=1$, and \mbox{$\mathbb{E}\left\lbrace \mathbf{s}\mathbf{s}^{H}\right\rbrace =\mathbf{I}$} for  \mbox{$\mathbf{s}=\left[s_{c}, s_{1}, \ldots, s_{K}\right]$}. The power allocation coefficient $0 \leq t \leq 1$ determines the distribution of power among the common and private precoders and the coefficient $0 \leq \mu_{k} \leq 1$ determines the portion of the private streams' power allocated to user-$k$. 
Setting $t=1$ turns off the common stream and RSMA boils down to SDMA.
The signal received by user-$k$ is written as \vspace{-0.1cm}
\begin{align}
	y_{k}=\sqrt{v_{k}}\mathbf{h}_{k}^{H}\mathbf{x}+n_{k}, \quad k\in\mathcal{K},  
\end{align}
where $\mathbf{h}_{k} \in \mathbb{C}^{N}$ the channel vector representing the small scale fading  and $0 < v_{k} \leq 1$ is the channel coefficient representing the large scale fading between the transmitter and user-$k$. The term $n_{k} \sim \mathcal{CN}(0,1)$ is the Additive White Gaussian Noise (AWGN) component at user-$k$. 

Each user receives its signal by first decoding the common stream to obtain $\widehat{W}_{c}$ and extracting $\widehat{W}_{c,k}$ from the decoded signal. Then, SIC is performed by reconstructing the common stream using $\widehat{W}_{c}$ and removing the reconstructed signal from the received signal. Finally, each user decodes its intended private stream and obtains  $\widehat{W}_{p,k}$ by treating the interference from other users' private streams as noise.
The whole decoded message of user-$k$, $\widehat{W}_{k}$, is reconstructed by combining the decoded common part, $\widehat{W}_{c,k}$, and the decoded private part, $\widehat{W}_{p,k}$.
We assume perfect CSI at receiver (CSIR) and SIC throughout this work. Accordingly, the Signal-to-Interference-plus-Noise Ratio (SINR) and the ergodic rates for the common and private streams at user-$k$ are written as\vspace{-0.1cm}
\begin{align}
	&\gamma_{c,k}\hspace{-0.1cm}=\hspace{-0.1cm}\frac{P(1-t)v_{k}|\mathbf{h}_{k}^{H}\mathbf{p}_{c}|^{2}}{1\hspace{-0.1cm}+\hspace{-0.1cm}Ptv_{k}\hspace{-0.1cm}\sum_{j \in \mathcal{K}}\hspace{-0.07cm}\mu_{j}\hspace{-0.03cm}|\mathbf{h}_{k}^{H}\hspace{-0.07cm}\mathbf{p}_{j}|^{2}},  
	\gamma_{k}\hspace{-0.1cm}=\hspace{-0.1cm}\frac{\mu_{k}Ptv_{k}|\mathbf{h}_{k}^{H}\mathbf{p}_{k}|^{2}}{1\hspace{-0.1cm}+\hspace{-0.1cm}Ptv_{k}\hspace{-0.1cm}\sum_{\substack{j \in \mathcal{K} \\ j \neq k}}\hspace{-0.07cm}\mu_{j}\hspace{-0.03cm}|\mathbf{h}_{k}^{H}\hspace{-0.07cm}\mathbf{p}_{j}|^{2}},\nonumber\\
&R_{c}(t)\hspace{-0.1cm}=\hspace{-0.1cm}\mathbb{E}\left\lbrace \hspace{-0.1cm}\log_{2}\hspace{-0.1cm}\left(\hspace{-0.1cm} 1\hspace{-0.1cm}+\hspace{-0.1cm}\min_{k\in\mathcal{K}}\gamma_{c,k} \right) \hspace{-0.15cm}\right\rbrace\hspace{-0.1cm}, \ 
	R_{k}(t)\hspace{-0.1cm}=\hspace{-0.1cm}\mathbb{E}\left\lbrace \log_{2}\hspace{-0.05cm}\left( 1\hspace{-0.1cm}+\hspace{-0.1cm} \gamma_{k} \right)\right\rbrace\hspace{-0.1cm},
	\label{eqn:rates}
\end{align}
with expectations defined over $\mathbf{h}_{k}$. We consider ergodic rates to design the system instead of instantaneous rates, since we study both perfect and imperfect CSIT cases in this work, and transmitter does not have access to instantaneous achievable rates for the imperfect CSIT case \cite{yin_2021, zhang_2009, papazafeiropoulos_2015, truong_2013, hao_2015, dizdar_2021}. 

The precoders $\mathbf{p}_{c}$ and $\mathbf{p}_{k}$ are calculated based on CSIT, which is imperfect in practical scenarios. The imperfections in CSIT stem from numerous reasons, such as, channel estimation errors \cite{mao_2018, mishra_2022_3}, mobility and latency in the network \cite{zhang_2009, dizdar_2021, papazafeiropoulos_2015, truong_2013}, pilot contamination \cite{mishra_2022_5, papazafeiropoulos_2015}, and so on.  
We adopt a general model to represent the relation between the actual channel and CSIT for analytical calculations, and perform analyses with different types of CSIT errors in the Section~\ref{sec:numerical}.
Accordingly, the small scale fading component of the channel for user-$k$ is expressed as
\begin{align}
	\mathbf{h}_{k}=\sqrt{\epsilon^{2}} \widehat{\mathbf{h}}_{k} + \sqrt{1-\epsilon^{2}} \mathbf{e}_{k},
	\label{eqn:channel}
\end{align}
where $\mathbf{h}_{k}$ denotes the spatially uncorrelated Rayleigh flat fading channel with i.i.d. entries distributed according to $\mathcal{CN}(0,1)$\footnote{Note that the large scale fading effect for the overall channel of user-$k$ is captured by the coefficient $v_{k}$, thus, identical small scale fading variance is assumed for all users.}. The terms $\widehat{\mathbf{h}}_{k}$ and $\mathbf{e}_{k}$ are independent random variables (r.v.s) denoting the CSIT and CSIT error, respectively, both with i.i.d. entries also distributed according to $\mathcal{CN}(0,1)$. The coefficient $0 \leq \epsilon \leq 1$ represents the CSIT quality and can be dependent on various parameters in practical systems, such as, SNR, user speed, number of quantization bits for feedback, depending on the considered scenario. 
\vspace{-0.3cm}
\section{Problem Formulation}
\label{sec:formulation}
We write a max-min fairness problem to ensure a non-zero ergodic rate for each user as \vspace{-0.3cm}
\begin{align}
	\max_{\mathbf{P}, \mathbf{c}, \boldsymbol{\mu}, t} \min_{k \in \mathcal{K}} C_{k}(t)+R_{k}(t), 
	\label{eqn:problem}
\end{align}
where $C_{k}(t)$ is the portion of the common stream rate allocated to user-$k$, {\sl i.e}, $\sum_{k\in\mathcal{K}}C_{k}(t)=R_{c}(t)$, $\mathbf{c}=[C_{1}, C_{2}, \ldots, C_{K}]$, $\mathbf{P}=[\mathbf{p}_{c}, \mathbf{p}_{1}, \mathbf{p}_{2}, \ldots, \mathbf{p}_{K}]$, and $\boldsymbol{\mu}=[\mu_{1}, \mu_{2}, \ldots, \mu_{K}]$. 
Our target is to obtain a low-complexity design as a solution of problem \eqref{eqn:problem}. We start simplifying the solution by limiting the design space for optimal precoders. Accordingly, we consider two types of precoders, specifically ZF and MRT, to be used for transmission of private streams. MRT and ZF precoders are chosen as they provide useful statistical properties for analytical analysis \cite{zhang_2009, papazafeiropoulos_2015, truong_2013, hao_2015, dai_2017, dizdar_2021, dai_2016, papazafeiropoulos_2018, clerckx_2020}, are optimal at low and high SNR respectively \cite{bjornson_2014}, and achieve optimal DoF at high SNR in overloaded systems \cite{joudeh_2017, yin_2021, clerckx_2021}.  
As done in \cite{dizdar_2021, hao_2015}, we set the common precoder as a random precoder to obtain tractable expressions for the rate calculations.

The type of employed precoders affect the expressions for the rates $C_{k}(t)$ and $R_{k}(t)$, which in turn affect the system design and resource allocation. Limiting the design space for precoders allows us to formulate separate problem formulations for different types of precoders employed. 

Let us define $C^{\mathrm{MRT}}_{k}(t)$ and $R^{\mathrm{MRT}}_{k}(t)$ to denote the portion of the common stream rate and the private stream rate for user-$k$, respectively, achieved when MRT precoders are used for private streams, and $C^{\mathrm{ZF}}_{k}(t)$ and $R^{\mathrm{ZF}}_{k}(t)$ to denote the respective rates when ZF precoders are used for private streams. Then, we can reformulate problem \eqref{eqn:problem} as 
\begin{align}
\max \left(\max_{\mathbf{c}, \boldsymbol{\mu}, t} \min_{k \in \mathcal{K}} C^{\mathrm{MRT}}_{k}(t)+R^{\mathrm{MRT}}_{k}(t), \right. \nonumber \\[-6pt]
\ \left. \max_{\mathbf{c}, \boldsymbol{\mu}, t} \min_{k \in \mathcal{K}} C^{\mathrm{ZF}}_{k}(t)+R^{\mathrm{ZF}}_{k}(t)\right).
	\label{eqn:problem2}
\end{align}

We give more details on the system design and problem formulation for each case in the following subsections. 
\vspace{-0.3cm}
\subsection{MRT Precoding}
\label{sec:formulation_MRT}
\vspace{-0.2cm}
In this section, we consider the system design where each user is served by a private stream encoded by MRT precoders and a portion of the common stream. Different from the design in the previous section, users are not divided into two groups in the considered setting. We assume equal power allocation among the private streams, such that, $\mu_{k}=\frac{1}{N}$, $\forall k \in \mathcal{K}$ and that the common rate is distributed equally among all users. Consequently, the rate expression for each user is identical, and the second term in \eqref{eqn:problem2} becomes\vspace{-0.1cm}
\begin{align}
	\max_{t\in (0,1]} \ \frac{1}{K}R^{\mathrm{MRT}}_{c}(t)+R^{\mathrm{MRT}}_{\hat{k}}(t), \  \hat{k}=\argmin_{k \in \mathcal{K}}R^{\mathrm{MRT}}_{k}(t).
	\label{eqn:MRTproblem}
\end{align}

Defining $\mathbf{p}^{\mathrm{MRT}}_{k}=\widehat{\mathbf{h}}_{k}/|\widehat{\mathbf{h}}_{k}|$, $\forall k \in \mathcal{K}$, the rate expressions in \eqref{eqn:MRTproblem} can be written as
\begin{align}
	&R^{\mathrm{MRT}}_{c}(t)\hspace{-0.1cm}=\hspace{-0.1cm}\mathbb{E}\left\lbrace \hspace{-0.1cm}\log_{2}\hspace{-0.1cm}\left(\hspace{-0.1cm} 1\hspace{-0.1cm}+\hspace{-0.1cm}\min_{k\in\mathcal{K}}\frac{P(1-t)v_{k}|\mathbf{h}_{k}^{H}\mathbf{p}_{c}|^{2}}{1+\frac{Pt}{K}v_{k}\sum_{j \in \mathcal{K}}|\mathbf{h}_{k}^{H}\mathbf{p}^{\mathrm{MRT}}_{j}|^{2}} \right) \hspace{-0.1cm}\right\rbrace, \nonumber\\ 
	&R^{\mathrm{MRT}}_{\hat{k}}(t)\hspace{-0.1cm}=\hspace{-0.1cm}\mathbb{E}\left\lbrace \hspace{-0.1cm}\log_{2}\hspace{-0.1cm}\left(\hspace{-0.1cm}1\hspace{-0.1cm}+\hspace{-0.1cm} \frac{\frac{Pt}{K}v_{\hat{k}}|\mathbf{h}_{\hat{k}}^{H}\mathbf{p}^{\mathrm{MRT}}_{\hat{k}}|^{2}}{1+\frac{Pt}{K}v_{\hat{k}}\sum_{\substack{j \in \mathcal{K}, \\ j \neq \hat{k}}}|\mathbf{h}_{\hat{k}}^{H}\mathbf{p}^{\mathrm{MRT}}_{j}|^{2}} \hspace{-0.1cm}\right) \hspace{-0.1cm}\right\rbrace\hspace{-0.1cm}.
	\label{eqn:rates_MRT}	
\end{align}

\subsection{ZF Precoding}
\label{sec:formulation_ZF}
The DoF analysis in \cite{joudeh_2017, yin_2021, clerckx_2021} reveals that the optimal max-min DoF is achieved by serving $N$ users by private streams with ZF precoding and a portion of the common stream, while the remaining $K-N$ users are served only by a portion of the common stream. To this end, let us define user groups $\mathcal{G}_{1}$ and $\mathcal{G}_{2}$, satisfying $|\mathcal{G}_{1}|=N$, $|\mathcal{G}_{2}|=K-N$, $\mathcal{G}_{1}\cap\mathcal{G}_{2}=\emptyset$, and, $\mathcal{G}_{1}\cup\mathcal{G}_{2}=\mathcal{K}$. The set $\mathcal{G}_{1}$ is the index set for users which are served by private and common streams, while $\mathcal{G}_{2}$ is the index set for users served by common stream only. 
We assume equal power allocation among the private streams\footnote{Equal power allocation among precoders in MU-MIMO is employed both in literature and practical systems, see \cite{hao_2015, dai_2016, dai_2017, 3GPP8214} and the references therein.}, such that, $\mu_{k}=\frac{1}{N}$ if $k \in \mathcal{G}_{1}$ and $\mu_{k}=0$ if $k \in \mathcal{G}_{2}$. 
Let us define the rate allocation coefficient, $0\leq \beta < \frac{1}{N}$, to adjust the proportion of common rate allocated to each user in $\mathcal{G}_{1}$.
Accordingly, we reformulate the first term in \eqref{eqn:problem2} as \vspace{-0.2cm}
\begin{align}
	\max_{\substack{\beta \in [0,1/N),\\ t\in [0,1)}}\min_{k \in \mathcal{G}_{1}} \left\{\beta R^{\mathrm{ZF}}_{c}(t)+R^{\mathrm{ZF}}_{k}(t), \frac{1-N\beta}{K-N}R^{\mathrm{ZF}}_{c}(t)\right\}\hspace{-0.1cm}.
	\label{eqn:ZFproblem_1}
	\vspace{-0.2cm}
\end{align}
The first and second terms in parenthesis in \eqref{eqn:ZFproblem_1} represent the total rate allocated to users in $\mathcal{G}_{1}$ and $\mathcal{G}_{2}$, respectively. 
Then, \eqref{eqn:ZFproblem_1} can be expressed in a simpler form as\vspace{-0.2cm}
\begin{align}
	\max_{\substack{\beta \in [0,1/N),\\ t\in [0,1)}}\min &\left(\beta R^{\mathrm{ZF}}_{c}(t)+R^{\mathrm{ZF}}_{\tilde{k}}(t), \frac{1-N\beta}{K-N}R^{\mathrm{ZF}}_{c}(t)\right), \nonumber \\[-6pt]
	 &\tilde{k}=\argmin_{k \in \mathcal{G}_{1}}R^{\mathrm{ZF}}_{k,p}(t).
	\label{eqn:ZFproblem_2}
\end{align}

Let us denote the ZF precoder for user-$k$ as $\mathbf{p}^{\mathrm{ZF}}_{k}$, such that, $||\mathbf{p}^{\mathrm{ZF}}_{k}||^{2}=1$ and $\widehat{\mathbf{h}}^{H}_{j}\mathbf{p}^{\mathrm{ZF}}_{k}=0$ for $\forall j,k \in \mathcal{G}_{1}$ and $j \neq k$. Accordingly, the expressions in \eqref{eqn:ZFproblem_2} can be written as 
\begin{align}
	&R^{\mathrm{ZF}}_{c}(t)\hspace{-0.1cm}=\hspace{-0.1cm}\mathbb{E}\left\lbrace \hspace{-0.1cm}\log_{2}\hspace{-0.1cm}\left(\hspace{-0.1cm} 1\hspace{-0.1cm}+\hspace{-0.1cm}\min_{k\in\mathcal{K}}\frac{P(1-t)v_{k}|\mathbf{h}_{k}^{H}\mathbf{p}_{c}|^{2}}{1\hspace{-0.1cm}+\hspace{-0.1cm}\frac{Pt}{N}v_{k}\sum_{j \in \mathcal{G}_{1}}|\mathbf{h}_{k}^{H}\mathbf{p}^{\mathrm{ZF}}_{j}|^{2}} \right) \hspace{-0.1cm}\right\rbrace\hspace{-0.1cm}, \nonumber\\ 
	&R^{\mathrm{ZF}}_{\tilde{k}}(t)\hspace{-0.1cm}=\hspace{-0.1cm}\mathbb{E}\left\lbrace \hspace{-0.1cm}\log_{2}\hspace{-0.05cm}\left(\hspace{-0.1cm}1\hspace{-0.1cm}+\hspace{-0.1cm} \frac{\frac{Pt}{N}v_{\tilde{k}}|\mathbf{h}_{\tilde{k}}^{H}\mathbf{p}^{\mathrm{ZF}}_{\tilde{k}}|^{2}}{1\hspace{-0.1cm}+\hspace{-0.1cm}\frac{Pt}{N}v_{\tilde{k}}\sum_{j \in \mathcal{G}_{1}, j \neq \tilde{k}}|\mathbf{h}_{\tilde{k}}^{H}\mathbf{p}^{\mathrm{ZF}}_{j}|^{2}}\hspace{-0.1cm}\right)\hspace{-0.1cm}\right\rbrace\hspace{-0.1cm}.
	\label{eqn:rates_ZF}
\end{align}
With the simplified formulations in \eqref{eqn:MRTproblem} and \eqref{eqn:ZFproblem_2}, \eqref{eqn:problem} becomes\vspace{-0.1cm}
\begin{align}
	\max\hspace{-0.1cm}&\left(\hspace{-0.1cm}\max_{\substack{\beta \in [0,1/N),\\ t\in [0,1)}}\hspace{-0.2cm}\min \left(\hspace{-0.1cm}\beta R^{\mathrm{ZF}}_{c}(t)\hspace{-0.1cm}+\hspace{-0.1cm}R^{\mathrm{ZF}}_{\tilde{k}}(t), \frac{1\hspace{-0.1cm}-\hspace{-0.1cm}N\beta}{K\hspace{-0.1cm}-\hspace{-0.1cm}N}R^{\mathrm{ZF}}_{c}(t)\hspace{-0.1cm}\right)\hspace{-0.1cm},\right. \nonumber \\[-6pt]
	&\quad \ \left. \max_{t\in (0,1]} \frac{1}{K}R^{\mathrm{MRT}}_{c}(t)\hspace{-0.1cm}+\hspace{-0.1cm}R^{\mathrm{MRT}}_{\hat{k}}(t)\hspace{-0.05cm}\right)\hspace{-0.1cm}.
	\label{eqn:problem2}
\end{align}
\vspace{-0.5cm}
\section{Expressions and Bounds for Achievable Rates}
\label{sec:boundRate}
In this section, we provide lower bounds for the ergodic rates in \eqref{eqn:rates_MRT} and \eqref{eqn:rates_ZF} as the next step towards obtaining a low-complexity solution. The obtained bounds are used to solve problem \eqref{eqn:problem2} and find closed-form solutions for optimal power and rate allocation coefficients, $t^{*}$ and $\beta^{*}$. 

We start by providing several useful properties that will be used to derive the bounds. 
Consider the case that ZF precoding performed among the users in $G_{1}$. The precoder $\mathbf{p}^{\mathrm{ZF}}_{k}$ for user-$k$ is chosen to be orthogonal to the channel of user-$j$, $\forall k,j \in \mathcal{G}_{1}$, $k \neq j$. Owing to the isotropic nature of i.i.d Rayleigh fading, the orthogonality reduces the degrees of freedom at the transmitter to $N-|\mathcal{G}_{1}|+1=1$. Accordingly, 
\mbox{$|\sqrt{2}\widehat{\mathbf{h}}_{k}^{H}\mathbf{p}^{\mathrm{ZF}}_{k}|^{2}$} has a chi-squared distribution with a degree of freedom $2$, i.e.,
\mbox{$|\sqrt{2}\widehat{\mathbf{h}}_{k}^{H}\mathbf{p}^{\mathrm{ZF}}_{k}|^{2}\sim \chi^{2}_{2}$}. Equivalently, $|\widehat{\mathbf{h}}_{k}^{H}\mathbf{p}^{\mathrm{ZF}}_{k}|^{2}\sim \mathrm{Gamma}(1,1)$\cite{jindal_2011, jaramilloramirez_2015}. %jindal_2009, 
Furthermore, the ZF precoder $\mathbf{p}^{\mathrm{ZF}}_{k}$ is isotropically distributed and independent of the Gaussian error $\mathbf{e}^{H}_{j}$, leading to $|\mathbf{e}^{H}_{j}\mathbf{p}^{\mathrm{ZF}}_{k}|^{2}\sim \mathrm{Gamma}(1,1)$, $\forall j,k \in \mathcal{K}$ (\cite{zhang_2009, papazafeiropoulos_2015, truong_2013, dizdar_2021}).
For MRT precoding, we assume that the user channels and error vectors are independent. Accordingly, $|\widehat{\mathbf{h}}_{k}^{H}\mathbf{p}^{\mathrm{MRT}}_{k}|^{2}\sim \mathrm{Gamma}(N,1)$, $|\widehat{\mathbf{h}}_{j}^{H}\mathbf{p}^{\mathrm{MRT}}_{k}|^{2}\sim \mathrm{Gamma}(1,1)$, $|\mathbf{e}^{H}_{j}\mathbf{p}^{\mathrm{MRT}}_{k}|^{2}\sim \mathrm{Gamma}(1,1)$, and $|\mathbf{e}^{H}_{k}\mathbf{p}^{\mathrm{MRT}}_{k}|^{2}\sim \mathrm{Gamma}(1,1)$. For the common stream precoder, it is assumed that $\mathbf{p}_{c}$ is independent of $\mathbf{h}_{k}$ and isotropically distributed on the unit sphere, and thus $|\mathbf{h}^{H}_{k}\mathbf{p}_{c}|^{2}\sim \mathrm{Gamma}(1,1)$.
Finally, we introduce a method to approximate the sum of Gamma-distributed r.v.s. Consider $X\hspace{-0.1cm}=\hspace{-0.1cm}\sum_{i}A_{i}$, where $A_{i}\hspace{-0.1cm} \sim\hspace{-0.1cm} \mathrm{Gamma}(D_{i}, \theta_{i})$. The second order approximation of $X$ using the moment matching method is $\widehat{X}\hspace{-0.1cm}\sim\hspace{-0.1cm}\mathrm{Gamma}(\widehat{D},\widehat{\theta})$, where $\widehat{D}\hspace{-0.1cm}=\hspace{-0.1cm}\frac{(\sum_{i}D_{i}\theta_{i})^{2}}{\sum_{i}D_{i}\theta_{i}^{2}}$, $\widehat{\theta}\hspace{-0.1cm}=\hspace{-0.1cm}\frac{\sum_{i}D_{i}\theta_{i}^{2}}{\sum_{i}D_{i}\theta_{i}}$ \cite{jaramilloramirez_2015}.
\vspace{-0.6cm}
\subsection{Rates with MRT Precoding}
\label{sec:ratesMRTprecoding}
\vspace{-0.1cm}
In this section, we derive approximations and bounds for the rates when MRT precoders are employed for the private streams under imperfect CSIT. 
Since the MRT precoder for user-$k$ is calculated at transmitter using imperfect channel knowledge, it is expressed as $\mathbf{p}^{\mathrm{MRT}}_{k}=\widehat{\mathbf{h}}_{k}/|\widehat{\mathbf{h}}_{k}|$, $\forall k \in \mathcal{K}$. We rewrite the rates in \eqref{eqn:rates_MRT} accordingly as \vspace{-0.1cm}
\begin{align}
	&R^{\mathrm{MRT}}_{c,i}(t)=
	\mathbb{E}\left\lbrace \log_{2}\hspace{-0.1cm}\left(1\hspace{-0.1cm}+\hspace{-0.1cm}P(1-t)Y^{\mathrm{MRT}}_{i} \right) \hspace{-0.1cm}\right\rbrace, \nonumber \\ 
	&R^{\mathrm{MRT}}_{k,i}(t)\hspace{-0.1cm}=\hspace{-0.1cm}\mathbb{E}\left\lbrace \log_{2}\hspace{-0.05cm}\left( 1\hspace{-0.1cm}+\hspace{-0.1cm} \frac{Pt}{N}X^{\mathrm{MRT}}_{k,i} \right) \right\rbrace,
	\label{eqn:rates_MRT_perfect_1}
\end{align}
where $X^{\mathrm{MRT}}_{k,i}\hspace{-0.1cm}=\hspace{-0.1cm}\frac{v_{k}|\mathbf{h}_{k}^{H}\mathbf{p}^{\mathrm{MRT}}_{k}|^{2}}{1+\frac{Pt}{K}v_{k}\sum_{\substack{j \in \mathcal{K} \\ j \neq k}}|\mathbf{h}_{k}^{H}\mathbf{p}^{\mathrm{MRT}}_{j}|^{2}}$, $Y^{\mathrm{MRT}}_{i}\hspace{-0.1cm}=\hspace{-0.1cm}\min_{k\in\mathcal{K}}Y^{\mathrm{MRT}}_{k,i}$, and $Y^{\mathrm{MRT}}_{k,i}\hspace{-0.1cm}=\hspace{-0.1cm}
	\frac{v_{k}|\mathbf{h}_{k}^{H}\mathbf{p}_{c}|^{2}}{1+\frac{Pt}{K}v_{k}\sum_{j \in \mathcal{K}}|\mathbf{h}_{k}^{H}\mathbf{p}^{\mathrm{MRT}}_{j}|^{2}}$.

The subscript $i$ in the terms defined above represent that the corresponding expressions are written for imperfect $\widehat{\mathbf{h}}_{k}$. Let us define the terms in \eqref{eqn:terms_i}, where $\gamma\approx0.577$ is the Euler-Mascheroni constant and $E_{v}(x)=\int_{1}^{\infty}\frac{e^{-tx}}{t^{v}}dt$, for $x > 0, \ v \in \mathbb{R}$ is the generalized exponential-integral function \cite{chiccoli_1990}. The following propositions give lower bounds for the rates in \eqref{eqn:rates_MRT_perfect_1} under the assumptions of small/moderate and large $Pt$. 

\begin{table*}[t!]
\begin{align}
	&\eta^{\mathrm{MRT}}_{i} \triangleq - \gamma - \log\left(\sum_{k=1}^{K}\frac{1}{v_{k}}\right)-e^{\frac{K\sum_{k=1}^{K}\frac{1}{v_{k}}}{P\theta^{\mathrm{MRT}}_{i}t}}\sum_{m=1}^{\lfloor D^{\mathrm{MRT}}_{i}K\rceil}E_{m}\left(\frac{K\sum_{k=1}^{K}\frac{1}{v_{k}}}{P\theta^{\mathrm{MRT}}_{i}t}\right), \ \ \theta^{\mathrm{MRT}}_{i}\triangleq \frac{N\epsilon^{4}+(1-\epsilon^{2})^{2}+K-1}{N\epsilon^{2}+(1-\epsilon^{2})+K-1},  \nonumber \\
	 &D^{\mathrm{MRT}}_{i}\triangleq\frac{\left(N\epsilon^{2}+(1-\epsilon^{2})+K-1\right)^{2}}{N\epsilon^{4}+(1-\epsilon^{2})^{2}+K-1}, \ \ \rho^{\mathrm{MRT}}_{i}\hspace{-0.1cm}\triangleq\hspace{-0.1cm}\frac{K}{\theta^{\mathrm{MRT}}_{i}\left(\lfloor D^{\mathrm{MRT}}_{i}K\rceil-1\right)}e^{-\gamma-\frac{1}{2\left(\lfloor D^{\mathrm{MRT}}_{i}K\rceil-1\right)}}, \ \ \tau^{\mathrm{MRT}}_{i} \triangleq \hspace{-0.9cm}\sum_{n=\frac{K\sum_{k=1}^{K}\frac{1}{v_{k}}}{P\theta^{\mathrm{MRT}}_{i}t}}^{\lfloor D^{\mathrm{MRT}}_{i}K\rceil+\frac{K\sum_{k=1}^{K}\frac{1}{v_{k}}}{P\theta^{\mathrm{MRT}}_{i}t}-1}\hspace{-1.3cm}1/n.
	\label{eqn:terms_i}
\end{align}
\vspace{-0.2cm}
\hrule
\end{table*} 
\vspace{-0.1cm}
\begin{proposition}   % Proposition 1
Let $\widetilde{R}^{\mathrm{MRT}}_{c,i}(t)$ be an approximation to $R^{\mathrm{MRT}}_{c,i}(t)$, {\sl i.e.}, $R^{\mathrm{MRT}}_{c,i}(t)\approx\widetilde{R}^{\mathrm{MRT}}_{c,i}(t)$. For $Pt \leq \frac{K\sum_{k=1}^{K}\frac{1}{v_{k}}}{P\theta^{\mathrm{MRT}}_{i}t}$, the following inequality holds.\vspace{-0.25cm}
\begin{align}
	\hspace{-0.3cm}\widetilde{R}^{\mathrm{MRT}}_{c,i}(t) &\hspace{-0.1cm}\geq \log_{2}(1+P(1-t)e^{\eta^{\mathrm{MRT}}_{i}}) \nonumber \\
	&\hspace{-0.1cm}\geq \log_{2}\hspace{-0.1cm}\left(\hspace{-0.1cm}1\hspace{-0.1cm}+\hspace{-0.1cm}P(1-t)e^{\left( -\gamma-\log\left(\sum_{k=1}^{K}\frac{1}{v_{k}}\right)-\tau^{\mathrm{MRT}}_{i}\right)}\right) \nonumber \\
	&\hspace{-0.1cm}\approx \log_{2}\hspace{-0.1cm}\left(\hspace{-0.1cm}1\hspace{-0.1cm}+\hspace{-0.1cm}\frac{Ke^{-\gamma}P(1-t)}{P\theta^{\mathrm{MRT}}_{i}\lfloor D^{\mathrm{MRT}}_{i}K\rceil t\hspace{-0.1cm}+\hspace{-0.1cm}K\sum_{k=1}^{K}\hspace{-0.1cm}\frac{1}{v_{k}}}\right)\hspace{-0.1cm}.
	\label{eqn:prop1}
\end{align}
\end{proposition}
\quad\textit{Proof:} See Appendix~A. \hspace{4.6cm} $\blacksquare$  

\begin{proposition}  %Proposition 2
Let $\widetilde{R}^{\mathrm{MRT}}_{c,i}(t)$ be an approximation to $R^{\mathrm{MRT}}_{c,i}(t)$, {\sl i.e.}, $R^{\mathrm{MRT}}_{c,i}(t)\approx\widetilde{R}^{\mathrm{MRT}}_{c,i}(t)$. For $Pt \rightarrow \infty$, the following inequality holds.
\begin{align}
	\widetilde{R}^{\mathrm{MRT}}_{c,i}(t) &\geq \log_{2}(1+P(1-t)e^{\eta^{\mathrm{MRT}}_{i}})\nonumber\\
	 &\approx \log_{2}\left(1-\rho^{\mathrm{MRT}}_{i}+\rho^{\mathrm{MRT}}_{i}/t\right).
	\label{eqn:prop2}
\end{align}
\end{proposition}
\quad\textit{Proof:} The proof follows from \cite[Lemma 4]{dizdar_2021} by taking into account the abovementioned properties of MRT precoders, and is omitted here for the sake of brevity. \hspace{2.8cm}  $\blacksquare$

\begin{proposition}	%Proposition 3
Let $\widetilde{R}^{\mathrm{MRT}}_{k,i}(t)$ be an approximation to $R^{\mathrm{MRT}}_{k,i}(t)$, {\sl i.e.}, $R^{\mathrm{MRT}}_{k,i}(t)\approx\widetilde{R}^{\mathrm{MRT}}_{k,i}(t)$. Then, the following inequality holds.
\begin{align}
	\widetilde{R}^{\mathrm{MRT}}_{k,i}(t) &\geq \log_{2}\hspace{-0.1cm}\left(\hspace{-0.1cm}1+\frac{Pt}{N}v_{k}e^{\left(\log(\theta^{\mathrm{MRT}}_{i})+\psi(D^{\mathrm{MRT}}_{i})\right)}\hspace{-0.1cm}\right) \nonumber \\
	&-\log_{2}\hspace{-0.1cm}\left(\hspace{-0.1cm}1+\frac{Pt}{K}v_{k}(K-1)\hspace{-0.1cm}\right).
	\label{eqn:prop3} 
\end{align}
\end{proposition}
\quad\textit{Proof:} The proof follows from \cite[Proposition 1]{dizdar_2021} by taking into account the abovementioned properties of MRT precoders, and is omitted here for the sake of brevity. \hspace{2.2cm}  $\blacksquare$

\subsection{Rates with ZF Precoding}
\label{sec:ratesZFprecoding}
\vspace{-0.1cm}
In this section, we derive approximations and bounds for the rates when ZF precoders are employed for the private streams under imperfect CSIT. 
The ZF precoder for user-$k$ is calculated at transmitter using imperfect channel knowledge, such that, it satisfies $|\widehat{\mathbf{h}}_{j}^{H}\mathbf{p}^{\mathrm{ZF}}_{k}|^{2}=0$, $\forall j,k \in \mathcal{G}_{1}$, $j \neq k$; and $|\mathbf{h}_{l}^{H}\mathbf{p}^{\mathrm{ZF}}_{k}|^{2}\sim \mathrm{Gamma}(1,1)$, $\forall k \in \mathcal{G}_{1}$, $\forall l \in \mathcal{G}_{2}$. Then, we re-write the rates in \eqref{eqn:rates_ZF} as \vspace{-0.2cm}
\begin{align}
	R^{\mathrm{ZF}}_{c,i}(t)&=\mathbb{E}\left\lbrace \log_{2}\hspace{-0.1cm}\left(1\hspace{-0.1cm}+\hspace{-0.1cm}P(1-t)Y^{\mathrm{ZF}}_{i} \right) \hspace{-0.1cm}\right\rbrace, \nonumber \\
	R^{\mathrm{ZF}}_{k,i}(t)&=\mathbb{E}\left\lbrace \log_{2}\hspace{-0.05cm}\left( 1\hspace{-0.1cm}+\hspace{-0.1cm} \frac{Pt}{N}X^{\mathrm{ZF}}_{k,i} \right) \right\rbrace,
	\label{eqn:rates_ZF_imperfect_1}
\end{align}
where $X^{\mathrm{ZF}}_{k,i}=\frac{v_{k}\left(\epsilon^{2}|\widehat{\mathbf{h}}_{k}^{H}\mathbf{p}^{\mathrm{ZF}}_{k}|^{2}+(1-\epsilon^{2})|\widehat{\mathbf{e}}_{k}^{H}\mathbf{p}^{\mathrm{ZF}}_{k}|^{2}\right)}{1+(1-\epsilon^{2})\frac{Pt}{N}v_{k}\sum_{\substack{j \in \mathcal{G}_{1} \\ j\neq k}}|\mathbf{e}_{k}^{H}\mathbf{p}^{\mathrm{ZF}}_{j}|^{2}}$, $Y^{\mathrm{ZF}}_{i}=\min_{k\in\mathcal{K}}Y^{\mathrm{ZF}}_{k,i}$, and \vspace{-0.4cm}
\begin{align} 
%&X^{\mathrm{ZF}}_{k,i}=\frac{v_{k}\left(\epsilon^{2}|\widehat{\mathbf{h}}_{k}^{H}\mathbf{p}^{\mathrm{ZF}}_{k}|^{2}+(1-\epsilon^{2})|\widehat{\mathbf{e}}_{k}^{H}\mathbf{p}^{\mathrm{ZF}}_{k}|^{2}\right)}{1+(1-\epsilon^{2})\frac{Pt}{N}v_{k}\sum_{\substack{j \in \mathcal{G}_{1} \\ j\neq k}}|\mathbf{e}_{k}^{H}\mathbf{p}^{\mathrm{ZF}}_{j}|^{2}}, \nonumber \\
%&Y^{\mathrm{ZF}}_{i}=\min_{k\in\mathcal{K}}Y^{\mathrm{ZF}}_{k,i}, \quad 
Y^{\mathrm{ZF}}_{k,i}\hspace{-0.1cm}=\hspace{-0.1cm}
\begin{cases}
	\hspace{-0.1cm}\frac{v_{k}|\mathbf{h}_{k}^{H}\mathbf{p}_{c}|^{2}}{1+\frac{Pt}{N}v_{k}\left(\epsilon^{2}|\widehat{\mathbf{h}}_{k}^{H}\mathbf{p}^{\mathrm{ZF}}_{k}|^{2}+(1-\epsilon^{2})\sum_{j \in \mathcal{G}_{1}}|\mathbf{e}_{k}^{H}\mathbf{p}^{\mathrm{ZF}}_{j}|^{2}\right)}, & \hspace{-0.4cm}\text{if $k \in \mathcal{G}_{1}$,} \\ 
	\hspace{1.45cm}\frac{v_{k}|\mathbf{h}_{k}^{H}\mathbf{p}_{c}|^{2}}{1+\frac{Pt}{N}v_{k}\sum_{j \in \mathcal{G}_{1}}|\mathbf{h}_{k}^{H}\mathbf{p}^{\mathrm{ZF}}_{j}|^{2}}\hspace{1.45cm}, & \hspace{-0.4cm}\text{if $k \in \mathcal{G}_{2}$.}
	\end{cases} 
	\label{eqn:rvs} 
\end{align}

Let us define the terms in \eqref{eqn:terms_zf}. The following propositions and lemmas give lower bounds for the rates in \eqref{eqn:rates_ZF_imperfect_1}. 
\begin{table*}[t!]
\begin{align}
	&\eta^{\mathrm{ZF}}_{i}\triangleq-\gamma-\log\left(\sum_{k=1}^{K}\frac{1}{v_{k}}\right)-e^{\frac{N}{Pt}\sum_{k=1}^{K}\frac{1}{v_{k}}}\sum_{m=1}^{\lfloor N(K-N+D^{\mathrm{ZF}}_{i})\rceil}E_{m}\left(\frac{N}{Pt}\sum_{k=1}^{K}\frac{1}{v_{k}}\right), \ \theta^{\mathrm{ZF}}_{i}\triangleq\frac{\epsilon^{4}(1+N)+(1-2\epsilon^{2})N}{\epsilon^{2}(1-N)+N}, \nonumber \\
	&D^{\mathrm{ZF}}_{i}\triangleq\frac{\left(\epsilon^{2}(1-N)+N\right)^{2}}{\epsilon^{4}(1+N)+(1-2\epsilon^{2})N}, \ \rho^{\mathrm{ZF}}_{i}\hspace{-0.12cm}\triangleq\hspace{-0.12cm}\frac{N}{\lfloor N(K-N+D^{\mathrm{ZF}}_{i})\rceil-1}e^{\hspace{-0.1cm}-\gamma-\frac{1}{2\left(\lfloor N(K-N+D^{\mathrm{ZF}}_{i})\rceil-1\right)}}, \ \ \ \ \tau^{\mathrm{ZF}}_{i}\triangleq\hspace{-1.3cm}\sum_{n=\frac{N}{Pt}\sum_{k=1}^{K}\frac{1}{v_{k}}}^{\lfloor N(K-N+D^{\mathrm{ZF}}_{i})\rceil+\frac{N}{Pt}\sum_{k=1}^{K}\frac{1}{v_{k}}-1}\hspace{-1.9cm}1/n.
	\label{eqn:terms_zf}
\end{align}
\vspace{-0.2cm}
\hrule
\vspace{-0.5cm}
\end{table*} 

\begin{proposition} %Proposition 4
Let $\widetilde{R}^{\mathrm{ZF}}_{c,i}(t)$ be an approximation to $R^{\mathrm{ZF}}_{c,i}(t)$, {\sl i.e.}, $R^{\mathrm{ZF}}_{c,i}(t)\approx\widetilde{R}^{\mathrm{ZF}}_{c,i}(t)$. Then, the following inequality holds.
\begin{align}
	\widetilde{R}^{\mathrm{ZF}}_{c,i}(t) \geq \log_{2}\left(1+P(1-t)e^{\eta^{\mathrm{ZF}}_{i}}\right). 
	\label{eqn:prop4} 
\end{align}
%\cite{milgram_1985,chiccoli_1990}.
\end{proposition}
\quad\textit{Proof:} See Appendix~\ref{appendix:prop4}. \hspace{4.6cm} $\blacksquare$

In the following lemmas, we provide simplified expressions for the bound in \eqref{eqn:prop4} under the assumptions of small/moderate and large $Pt$.

\begin{lemma}
For $Pt \leq N\sum_{k=1}^{K}\frac{1}{v_{k}}$, the following holds.
\begin{align}
	\widetilde{R}^{\mathrm{ZF}}_{c,i}(t) &\geq \log_{2}\hspace{-0.1cm}\left(1+P(1-t)e^{\eta^{\mathrm{ZF}}_{i}}\right)\hspace{-0.1cm} \nonumber\\
	&\geq \log_{2}\left(1+P(1-t)e^{\left( -\gamma-\log\left(\sum_{k=1}^{K}\frac{1}{v_{k}}\right)-\tau^{\mathrm{ZF}}_{i}\right)}\right) \nonumber \\
	&\approx \log_{2}\hspace{-0.1cm}\left(\hspace{-0.1cm}1+\frac{Ne^{-\gamma}P(1-t)}{P\lfloor N(N-K+D^{\mathrm{ZF}}_{i})\rceil t+N\sum_{k=1}^{K}\frac{1}{v_{k}}}\right).
	\label{eqn:lemma1}
\end{align}
\end{lemma}
\quad\textit{Proof:} The proof is similar to that of Proposition 1 and omitted here for the sake of brevity. \hspace{3.99cm} $\blacksquare$
%See Appendix~\ref{appendix:lemma1}. \hspace{11.6cm} $\blacksquare$}

\begin{lemma}
\hspace{-0.1cm} For $Pt\hspace{-0.1cm}\rightarrow \hspace{-0.1cm}\infty$, the following holds.
\begin{align}
	\widetilde{R}^{\mathrm{ZF}}_{c,i}(t) &\geq \log_{2}\left(1+P(1-t)e^{\eta^{\mathrm{ZF}}_{i}}\right) \nonumber \\
	&\approx \log_{2}\left(1-\rho^{\mathrm{ZF}}_{i}+\rho^{\mathrm{ZF}}_{i}/t\right),
	\label{eqn:lemma2}
\end{align}
\end{lemma}
\quad\textit{Proof:} The proof follows from \cite[Lemma 4]{dizdar_2021}. \hspace{1.4cm}  $\blacksquare$

\begin{proposition} %Proposition 5
Let $\widetilde{R}^{\mathrm{ZF}}_{k,i}(t)$ be an approximation to $R^{\mathrm{ZF}}_{k,i}(t)$. Then, the following inequality holds.\vspace{-0.25cm}
\begin{align}
	\widetilde{R}^{\mathrm{ZF}}_{k,i}(t) &\geq \log_{2}\hspace{-0.1cm}\left(\hspace{-0.1cm}1+\frac{Pt}{N}v_{k}e^{\left(\log(\theta^{\mathrm{ZF}}_{i})+\psi(D^{\mathrm{ZF}}_{i})\right)}\hspace{-0.1cm}\right) \nonumber \\
	&-\log_{2}\hspace{-0.1cm}\left(\hspace{-0.1cm}1+\frac{Pt}{N}v_{k}(N-1)(1-\epsilon^{2})\hspace{-0.1cm}\right).
	\label{eqn:prop5}
\end{align}
\end{proposition}
\quad\textit{Proof:} The proof follows from \cite[Proposition 1]{dizdar_2021}. \hspace{0.9cm}  $\blacksquare$

The derivations in this section will be used in the following sections to find the optimal precoder, power and rate allocation coefficients. We note that the results obtained in this section can be used for the perfect CSIT case by setting $\epsilon^{2}=1$. Accordingly, we define the following.
\begin{align}
&\rho^{\mathrm{MRT}}_{p}=\rho^{\mathrm{MRT}}_{i}\bigg\rvert_{\epsilon^{2}=1}\hspace{-0.6cm}, \ \ \rho^{\mathrm{ZF}}_{p}=\rho^{\mathrm{ZF}}_{i}\bigg\rvert_{\epsilon^{2}=1}\hspace{-0.6cm},  \ \ D^{\mathrm{MRT}}_{p}=N+K-1, \nonumber \\ 
&\theta^{\mathrm{MRT}}_{p}=1, \ \  D^{\mathrm{ZF}}_{p}=1, \ \ \theta^{\mathrm{ZF}}_{p}=1.
\label{eqn:perfect_parameters}
\end{align}

\section{Proposed Solution for Perfect CSIT}
\label{sec:solution_perfectcsit}
We start our investigation for a low-complexity solution for the problem in \eqref{eqn:problem2} by considering the perfect CSIT case. First, we derive closed-form solutions for optimal $\beta$ and $t$ ($\beta^{*}$ and $t^{*}$) under various assumptions using the lower bound expressions derived in Section~\ref{sec:boundRate}. Next, we employ the derived closed-form solutions to calculate ergodic rates using the corresponding expressions. Finally, we perform power and rate allocation according to the closed form solution that yields the largest rate calculated by the lower bounds.

\subsection{Solution Using ZF Precoding}
\label{sec:solution_perfectcsit_ZF}
Proposition 6 provides a condition which should be satisfied by $\beta^{*}$ and $t^{*}$. 
\begin{proposition}
The maximum of \eqref{eqn:ZFproblem_2} is attained by unique $\beta^{*}$ and $t^{*}$ satisfying 
\begin{align}
	\frac{1-N\beta^{*}}{K-N}R^{\mathrm{ZF}}_{c,p}(t^{*})=\beta^{*}R^{\mathrm{ZF}}_{c,p}(t^{*})+R^{\mathrm{ZF}}_{\tilde{k},p}(t^{*}). 
	\label{eqn:propx}
\end{align}
\end{proposition}
\quad\textit{Proof:} See Appendix~\ref{appendix:propx}. \hspace{4.6cm}$\blacksquare$

\begin{corollary}
	The solution $\beta^{*}$ satisfies $0\leq\beta^{*}\leq\frac{1}{K}$.
\end{corollary}
\quad\textit{Proof:} The proof is straightforward once \eqref{eqn:propx} is arranged as $\frac{1-K\beta^{*}}{K-N}R^{\mathrm{ZF}}_{c,p}(t^{*})=R^{\mathrm{ZF}}_{\tilde{k},p}(t^{*})\hspace{-0.1cm}\geq 0$. \hspace{3.8cm} $\blacksquare$

As Proposition~6 suggests, one can find the optimal solution for max-min fairness by substituting corresponding expressions for $R^{\mathrm{ZF}}_{c,p}(t)$ and $R^{\mathrm{ZF}}_{k^{\prime}}(t)$ into \eqref{eqn:propx} and solving for $\beta^{*}$ and $t^{*}$. 
We note that even if the system operates under the conditions where the noise is more dominant than the multi-user interference ({\sl i.e.}, noise-limited regime), allocating power to the common stream is essential to achieve a non-zero rate for each user, so that, $t<1$. Therefore, we need to find a solution for $\beta^{*}$ and $t^{*}$ at both moderate/low and high SNR regions. 
In the following, we use the bounds derived in Section~\ref{sec:boundRate} to find optimal solutions at each region. 

\subsubsection{High SNR}
\label{sec:perfectcsit_ZF_highsnr}
We consider the case where the power allocated to private streams is large, \emph{i.e.}, \mbox{$Pt\rightarrow\infty$}. Such a case is considered for scenarios where the transmit power is large enough that even when the it is distributed over the common and private streams, the received signal still has considerable amount of power for each stream. We start our analysis by finding the lower bound expression for $R^{\mathrm{ZF}}_{\tilde{k},p}(t)$. Using \eqref{eqn:prop5} with $\epsilon^{2}=1$ and $D^{\mathrm{ZF}}_{p}$ and $\theta^{\mathrm{ZF}}_{p}$ in \eqref{eqn:perfect_parameters}, it is straightforward to show that for given $v_{k}$, $\forall k \in \mathcal{G}_{1}$, $R^{\mathrm{ZF}}_{\tilde{k},p}(t)=\log_{2}(1+\sigma^{\mathrm{ZF}}_{\tilde{k},p}t)$, where $\sigma^{\mathrm{ZF}}_{\tilde{k},p}=v_{\tilde{k}}\frac{P}{N}e^{\psi(1)}$ and $\tilde{k}$ is redefined as \vspace{-0.3cm}
\begin{align}
	\tilde{k}=\argmin_{k \in \mathcal{G}_{1}}v_{k}.
	\label{eqn:tildek}
\end{align}
Using the above expressions with \eqref{eqn:lemma2} and $\rho^{\mathrm{ZF}}_{p}$ in \eqref{eqn:perfect_parameters}, we can rewrite \eqref{eqn:propx} as\vspace{-0.1cm}
\begin{align}
	&\frac{1-N\beta^{*}}{K-N}\log_{2}\left(1-\rho^{\mathrm{ZF}}_{p}+\rho^{\mathrm{ZF}}_{p}/t^{*}\right)\nonumber\\
	&=\beta^{*}\log_{2}\left(1-\rho^{\mathrm{ZF}}_{p}+\rho^{\mathrm{ZF}}_{p}/t^{*}\right)+\log_{2}(1+\sigma^{\mathrm{ZF}}_{\tilde{k},p}t^{*}).
	\label{eqn:perfectcsit_highsnr}
\end{align}

First, let us consider the scenario where $R^{\mathrm{ZF}}_{\tilde{k},p}(t^{*})$ is large enough, such that, $\log_{2}(1+\sigma^{\mathrm{ZF}}_{\tilde{k},p}t^{*})\approx\log_{2}(\sigma^{\mathrm{ZF}}_{\tilde{k},p}t^{*})$. We note that $\sigma^{\mathrm{ZF}}_{\tilde{k},p}>1$ in this scenario. Accordingly, the left-hand term in \eqref{eqn:perfectcsit_highsnr} is also expected to be large. Using this assumption and the fact that $1-\rho^{\mathrm{ZF}}_{p}\leq 1$, one can approximate $\log_{2}\left(1-\rho^{\mathrm{ZF}}_{p}+\rho^{\mathrm{ZF}}_{p}/t^{*}\right)\approx\log_{2}\left(\rho^{\mathrm{ZF}}_{p}/t^{*}\right)$. Then, 
\begin{align}
	\frac{1-N\beta^{*}}{K-N}\log_{2}\left(\rho^{\mathrm{ZF}}_{p}/t^{*}\right)&=\beta^{*}\log_{2}\left(\rho^{\mathrm{ZF}}_{p}/t^{*}\right)+\log_{2}\left(\sigma^{\mathrm{ZF}}_{\tilde{k},p}t^{*}\right), \nonumber 
	%\frac{1-K\beta^{*}}{K-N}\log_{2}\left(\frac{\rho^{\mathrm{ZF}}_{p}}{t^{*}}\right)&=\log_{2}(\sigma^{\mathrm{ZF}}_{\tilde{k},p}t^{*}), \nonumber \\
	%\frac{\rho^{\mathrm{ZF}}_{p}}{t^{*}}&=(\sigma^{\mathrm{ZF}}_{\tilde{k},p}t^{*})^{\frac{K-N}{1-K\beta^{*}}},\nonumber \\
	%t^{*}&=\left(\frac{(\rho^{\mathrm{ZF}}_{p})^{1-K\beta^{*}}}{(\sigma^{\mathrm{ZF}}_{\tilde{k},p})^{K-N}}\right)^{\frac{1}{1-K\beta^{*}+K-N}}.
	%\label{eqn:t1}
	\vspace{-0.4cm}
\end{align}
which can be re-arranged to find $t^{*}=\frac{(\rho^{\mathrm{ZF}}_{p})^{\frac{1-K\beta^{*}}{1-K\beta^{*}+K-N}}}{(\sigma^{\mathrm{ZF}}_{\tilde{k},p})^{\frac{K-N}{1-K\beta^{*}+K-N}}}\triangleq t^{(1)}_{p}(\beta^{*})$. 
We can find $\beta^{*}$ by maximizing either $\frac{1-N\beta}{K-N}R^{\mathrm{ZF}}_{c,p}(t^{(1)}_{p}(\beta))$ or $\beta R^{\mathrm{ZF}}_{c,p}(t^{(1)}_{p}(\beta))+R^{\mathrm{ZF}}_{\tilde{k},p}(t^{(1)}_{p}(\beta))$, since they are equal at $\beta^{*}$ as Proposition 6 suggests. We provide a useful result in Lemma 5 to obtain $\beta^{*}$ by maximizing $\frac{1-N\beta}{K-N}R^{\mathrm{ZF}}_{c,p}(t^{(1)}_{p}(\beta))$ with respect to $\beta$.

\begin{lemma}
The function $\frac{1-N\beta}{K-N}\log_{2}\left(1-\rho^{\mathrm{ZF}}_{p}+\rho^{\mathrm{ZF}}_{p}/t^{(1)}_{p}(\beta)\right)$ is a monotonic decreasing function of $\beta$ for $\beta\in\left[0, \frac{1}{K}\right]$ and $\sigma^{\mathrm{ZF}}_{\tilde{k},p}>1$. 
\end{lemma}
\quad\textit{Proof:} See Appendix~\ref{appendix:lemmax}. \hspace{4.6cm}$\blacksquare$

Lemma 3 shows that the maximum of $\frac{1-N\beta}{K-N}R^{\mathrm{ZF}}_{c,p}(t^{(1)}_{p}(\beta))$ is attained by choosing $\beta$ as small as possible. Thus, we conclude that $\beta^{*}=0$, implying the common stream should serve only the users in $G_{2}$ for optimal performance. Accordingly, we define the optimal rate and parameters for this scenario as in \eqref{eqn:1_perfect}. One can see that $ t^{(1)}_{p}\geq 0$ since $\sigma^{\mathrm{ZF}}_{\tilde{k},p}\geq 0$ and $\rho^{\mathrm{ZF}}_{p}\geq 0$.

\begin{remark}
The result $\beta^{*}=0$ is in line with the DoF analyses in \cite{joudeh_2017, clerckx_2021}, which state that the users in $G_{1}$ are served only by ZF precoded private streams for optimal DoF with perfect CSIT.
\end{remark}
\begin{remark}
RSMA allocates a fraction of total power to private streams in a way that their SINR values are not in the interference-limited regime (regime where the multi-user interference power, which depends on the transmit power, becomes dominant compared to the noise power with increasing transmit power) but the noise-limited regime \cite{clerckx_2023}. 
The result $\beta^{*}=0$ shows that for optimal performance, all transmit power needs to be allocated to the private streams of the users in $G_{1}$ at high SNR under perfect CSIT. Accordingly, we can conclude that the SINRs of the users in $G_{1}$ are not in the interference-limited regime but in the  the noise-limited regime even at high SNR under perfect CSIT. 
A direct implication of this conclusion is that the SINRs of the users in $G_{1}$ will still be in the noise-limited regime at moderate/low SNR (where the noise power is more dominant compared to the transmit power at high SNR case). Therefore, we set $\beta^{*}=0$ also for the other scenarios to be analysed in the rest of this section. 
\end{remark}

\begin{table*}
\begin{align}
		&t^{(1)}_{p}\triangleq\min\left\lbrace\left(\frac{\rho^{\mathrm{ZF}}_{p}}{(\sigma^{\mathrm{ZF}}_{\tilde{k},p})^{K-N}}\right)^{\frac{1}{1+K-N}}\hspace{-0.3cm},1\right\rbrace\hspace{-0.1cm}, \quad \beta^{(1)}_{p}= 0,\quad r^{(1)}_{mm,p}\triangleq\min\left\lbrace\hspace{-0.1cm}\frac{\log_{2}\hspace{-0.1cm}\left(\hspace{-0.08cm}1\hspace{-0.1cm}-\hspace{-0.1cm}\rho^{\mathrm{ZF}}_{p}\hspace{-0.1cm}+\hspace{-0.1cm}\rho^{\mathrm{ZF}}_{p}/t^{(1)}_{p}\hspace{-0.08cm}\right)}{K-N}, \log_{2}\left(\hspace{-0.08cm}1\hspace{-0.1cm}+\hspace{-0.1cm}\sigma^{\mathrm{ZF}}_{\tilde{k},p}t^{(1)}_{p}\hspace{-0.08cm}\right)\hspace{-0.1cm}\right\rbrace\hspace{-0.1cm}.		
		\label{eqn:1_perfect}
\end{align}
\vspace{-0.2cm}
\hrule
\vspace{-0.4cm}
\end{table*}

Next, let us consider the scenario where $R^{\mathrm{ZF}}_{\tilde{k},p}(t^{*})$ is small enough, such that, $\log_{2}\left(1+\sigma^{\mathrm{ZF}}_{\tilde{k},p}t^{*}\right)\approx\sigma^{\mathrm{ZF}}_{\tilde{k},p}t^{*}$. Again, we assume the left-hand term in \eqref{eqn:perfectcsit_highsnr} is large, such that, $\log_{2}\left(1-\rho^{\mathrm{ZF}}_{p}+\rho^{\mathrm{ZF}}_{p}/t^{*}\right)\approx\log_{2}\left(\rho^{\mathrm{ZF}}_{p}/t^{*}\right)$. Setting $\beta^{*}=0$, we rewrite \eqref{eqn:perfectcsit_highsnr} as \vspace{-0.2cm}
\begin{align}
	%\log_{2}\left(\frac{\rho^{\mathrm{ZF}}_{p}}{t^{*}}\right)&=(K-N)\sigma^{\mathrm{ZF}}_{\tilde{k},p}t^{*}, \nonumber \\
	\rho^{\mathrm{ZF}}_{p}&=t^{*}e^{\log(2)(K-N)\sigma^{\mathrm{ZF}}_{\tilde{k},p}t^{*}}. 
	\label{eqn:perfectcsit_highsnr_assump2}
\end{align}
We multiply both sides of \eqref{eqn:perfectcsit_highsnr_assump2} by $\delta_{p}\triangleq\log(2)(K-N)\sigma^{\mathrm{ZF}}_{\tilde{k},p}$ to obtain $\delta_{p}\rho^{\mathrm{ZF}}_{p}=\delta_{p}t^{*}e^{\delta_{p}t^{*}}$, which can be re-arranged as $\delta_{p}t^{*}=W_{k}(\delta_{p}\rho^{\mathrm{ZF}}_{p})$, 
where $W_{k}(x)$ is the $k$-th branch of Lambert-W function. As $\delta_{p}\rho^{\mathrm{ZF}}_{p}$ and $\delta_{p}t^{*}$ are real and $\delta_{p}\rho^{\mathrm{ZF}}_{p}>0$, it suffices to check only the principal branch $W_{0}(x)$: \vspace{-0.3cm}
\begin{align}
	\delta_{p}t^{*}=W_{0}(\delta_{p}\rho^{\mathrm{ZF}}_{p}).
	\label{eqn:perfectcsit_highsnr_assump2_lambert2}
\end{align}
We use the approximation \mbox{$W_{0}(x)\approx\log(x)-\log(\log(x))$} for  large $x$ \cite{corless_1996} in \eqref{eqn:perfectcsit_highsnr_assump2_lambert2} to obtain  \vspace{-0.1cm}
\begin{align}
	t^{*}=&\frac{\log\left(\delta_{p}\rho^{\mathrm{ZF}}_{p}\right)-\log\left(\log\left(\delta_{p}\rho^{\mathrm{ZF}}_{p}\right)\right)}{\delta_{p}}\triangleq t^{(2)}_{p}.
	\label{eqn:t2}
\end{align}
\begin{lemma}
	The power allocation coefficient $t^{(2)}_{p}$ satisfies $0\leq t^{(2)}_{p}\leq 1$ for $\delta_{p}\rho^{\mathrm{ZF}}_{p} \geq e$.
\end{lemma}
\quad\textit{Proof:} See Appendix~\ref{appendix:lemmay}. \hspace{4.6cm}$\blacksquare$

Accordingly, we define the optimal rate and parameters for this scenario as in \eqref{eqn:2_perfect}.
\begin{table*}
\begin{align} 
		&t^{(2)}_{p}\triangleq
		\begin{cases}
    		\frac{\log\left(\delta_{p}\rho^{\mathrm{ZF}}_{p}\right)-\log\left(\log\left(\delta_{p}\rho^{\mathrm{ZF}}_{p}\right)\right)}{\delta_{p}}, & \hspace{-0.2cm}\text{if $\delta_{p}\rho^{\mathrm{ZF}}_{p} \geq e$} \\
    		\hspace{1.65cm}1\hspace{1.65cm}, &\hspace{-0.2cm} \text{otherwise}.
    	\end{cases}\hspace{-0.1cm}, 
		\ \beta^{(2)}_{p}=0, \  r^{(2)}_{mm,p}\triangleq\min\left\lbrace\hspace{-0.1cm}\frac{\log_{2}\left(\hspace{-0.08cm}1\hspace{-0.1cm}-\hspace{-0.1cm}\rho^{\mathrm{ZF}}_{p}+\hspace{-0.1cm}\rho^{\mathrm{ZF}}_{p}/t^{(2)}_{p}\hspace{-0.08cm}\right)}{K-N}, \log_{2}\left(\hspace{-0.08cm}1\hspace{-0.1cm}+\hspace{-0.1cm}\sigma^{\mathrm{ZF}}_{\tilde{k},p}t^{(2)}_{p}\hspace{-0.08cm}\right)\hspace{-0.1cm} \right\rbrace\hspace{-0.1cm}.	
	\label{eqn:2_perfect}
\end{align}
\vspace{-0.1cm}
\hrule
\vspace{-0.4cm}
\end{table*}
\subsubsection{Moderate/Low SNR}
\label{sec:perfectcsit_ZF_lowsnr}
Next, we consider the case where the system operates at moderate/low SNR regime.   Accordingly, we can assume that the transmit power level allocated to private streams is moderate or small. Based on Remark 2, we set $\beta^{*}=0$ for the analysis in this section. We follow the procedure in Section~\ref{sec:perfectcsit_ZF_highsnr} and use \eqref{eqn:lemma1}, $D^{\mathrm{ZF}}_{p}$ in \eqref{eqn:perfect_parameters} to write \vspace{-0.2cm}
\begin{align}
	\frac{\log_{2}\left(1+\frac{e^{-\gamma}P(1-t^{*})}{P(K-N+1)t^{*}+\sum_{k=1}^{K}1/v_{k}}\right)}{K-N}=\log_{2}\left(1+\sigma^{\mathrm{ZF}}_{\tilde{k},p}t^{*}\right). \nonumber
\end{align}
Removing the logarithm from both sides yields $1+\frac{e^{-\gamma}P(1-t^{*})}{P(K-N+1)t^{*}+\sum_{k=1}^{K}1/v_{k}}=\left(1+\sigma^{\mathrm{ZF}}_{\tilde{k},p}t^{*}\right)^{\frac{1}{K-N}}$.
We benefit from the Binomial theorem $(1+x)^n=1+nx+\frac{n(n-1)}{2!}x^{2}+\ldots+x^{n}$. Assuming $\sigma^{\mathrm{ZF}}_{\tilde{k},p}t^{*} < 1$, we take the first two terms of Binomial expansion of $\left(1+\sigma^{\mathrm{ZF}}_{\tilde{k},p}t^{*}\right)^{\frac{1}{K-N}}$ and obtain \vspace{-0.2cm}
\begin{align}
	\frac{e^{-\gamma}P(1-t^{*})}{P(K-N+1)t^{*}+\sum_{k=1}^{K}1/v_{k}}=(K-N)\sigma^{\mathrm{ZF}}_{\tilde{k},p}t^{*}. 
	\label{eqn:t3_init}
\end{align}
From \eqref{eqn:t3_init}, we write the quadratic equation\vspace{-0.3cm}
\begin{align}
	&\underbrace{P(K-N+1)(K-N)\sigma^{\mathrm{ZF}}_{\tilde{k},p}}_{a_{3,p}}(t^{*})^{2} \nonumber \\
	&+\underbrace{((K-N)\sigma^{\mathrm{ZF}}_{\tilde{k},p}\sum_{k=1}^{K}1/v_{k}+e^{-\gamma}P)}_{b_{3,p}}t^{*}+\underbrace{(-e^{-\gamma}P)}_{c_{3,p}}=0. \nonumber
\end{align}
\vspace{-0.1cm}
Then, the roots are calculated as $s_{1/2}=\frac{-b_{3,p}\pm\sqrt{b_{3,p}^{2}-4a_{3,p}c_{3,p}}}{2a_{3,p}}$.	
Noting that $\sqrt{b_{3,p}^{2}-4a_{3,p}c_{3,p}}>0$ for $K\geq N$, we check the conditions $0 \leq t^{*} \leq 1$ to determine a single expression for $t^{*}$. The condition $t^{*} \geq 0$ requires the expression $s_{1}=\frac{-b_{3,p}+\sqrt{b_{3,p}^{2}-4a_{3,p}c_{3,p}}}{2a_{3,p}}$, since $a_{3,p},b_{3,p}>0$ and $c_{3,p}<0$. We can check whether the condition $ \frac{-b_{3,p}+\sqrt{b_{3,p}^{2}-4a_{3,p}c_{3,p}}}{2a_{3,p}} \leq 1$ is satisfied by rearranging it in the form  $\sqrt{b_{3,p}^{2}-4a_{3,p}c_{3,p}} \leq 2a_{3,p}+b_{3,p}$. Taking the squares of both sides and cancelling out the identical terms gives $
	e^{-\gamma}P \leq P(K-N+1)(K-N)\sigma^{\mathrm{ZF}}_{\tilde{k},p}+(K-N)\sigma^{\mathrm{ZF}}_{\tilde{k},p}\sum_{k=1}^{K}1/v_{k}+e^{-\gamma}P$,
which holds for all values of $N\geq 0$, $K\geq N$ and $P\geq 0$. Therefore, we can write the optimal rate and parameters for this scenario as in \eqref{eqn:3_perfect}.
\begin{table*}
\begin{align}
		t^{(3)}_{p}&\triangleq\frac{-b_{3,p}\hspace{-0.1cm}+\hspace{-0.1cm}\sqrt{b_{3,p}^{2}\hspace{-0.1cm}-\hspace{-0.1cm}4a_{3,p}c_{3,p}}}{2a_{3,p}}, \   \beta^{(3)}_{p}=0, \  r^{(3)}_{mm,p}\hspace{-0.05cm}\triangleq\hspace{-0.05cm}\min\hspace{-0.1cm}\left\lbrace\frac{\log_{2}\left(\hspace{-0.08cm}1\hspace{-0.1cm}+\hspace{-0.1cm}\frac{Ne^{-\gamma}P(1-t^{(3)}_{p})}{PN(K-N+1)t^{(3)}_{p}+N\sum_{k=1}^{K}1/v_{k}}\hspace{-0.08cm}\right)}{K-N}, \log_{2}\left(\hspace{-0.08cm}1\hspace{-0.1cm}+\hspace{-0.1cm}\sigma^{\mathrm{ZF}}_{\tilde{k},p}t^{(3)}_{p}\hspace{-0.08cm}\right)\hspace{-0.1cm}\right\rbrace\hspace{-0.1cm}.	
	\label{eqn:3_perfect}
\end{align}
\hrule
\vspace{-0.4cm}
\end{table*}
\vspace{-0.6cm}
\subsection{Solution Using MRT Precoding}
\label{sec:solution_perfectcsit_MRT}
In this section, we study the case where MRT precoding is used for the private streams. Different from the design in the previous section, each user is served by  at least one private stream in the considered design. Accordingly, allocating power to the common stream is not mandatory to achieve a non-zero rate for each user (which is not the case for the design with ZF precoding), so that, $t \leq 1$.
As done in the previous section, we solve the problem by using the lower bounds derived in Section~\ref{sec:boundRate} for the ergodic rates. However, the two conditions which will be studied in the following subsections are valid for different system settings. Thus, only one of them will be used for resource allocation (together with the allocations calculated in the previous subsection) and the choice of which version to use will be dependent on the system settings ({\sl e.g.}, $N$, $K$, $\epsilon$), as will be shown in Section~\ref{sec:numerical}.
\subsubsection{Case 1}
\label{sec:perfectcsit_MRT_highsnr}
We investigate the case for the scenario where optimal power splitting is performed when transmit power is large enough, so that, $Pt^{*}\rightarrow\infty$. Intuitively, this case captures the scenarios where the interference-limited regime is reached at high SNR, {\sl e.g.}, large $N$ and/or small $K-N$. 
Using \eqref{eqn:prop3} with $\epsilon^{2}=1$ and $D^{\mathrm{MRT}}_{i}$ and $\theta^{\mathrm{MRT}}_{i}$ in \eqref{eqn:perfect_parameters}, it is straightforward to show that for given $v_{k}$, $\forall k \in \mathcal{K}$, $R^{\mathrm{MRT}}_{\hat{k},p}(t)=\log_{2}(1+\alpha^{\mathrm{MRT}}_{p}t)-\log_{2}(1+\lambda^{\mathrm{MRT}}_{p}t)$, where $\alpha^{\mathrm{MRT}}_{p}=v_{\hat{k}}\frac{P}{K}e^{\psi(N+K-1)}$, $\lambda^{\mathrm{MRT}}_{p}=v_{\hat{k}}\frac{P(K-1)}{K}$ and $\hat{k}$ is redefined as  \vspace{-0.2cm}
\begin{align}
	\hat{k}=\argmin_{k \in \mathcal{K}}v_{k}. 
	\label{eqn:k_hat}
\end{align}
We use the expressions above with \eqref{eqn:prop2} and $\rho^{\mathrm{MRT}}_{p}$ in \eqref{eqn:perfect_parameters} to rewrite \eqref{eqn:MRTproblem} as\vspace{-0.2cm}
\begin{align}
	\max_{t\in (0,1]}\hspace{-0.09cm}\frac{1}{K}\log_{2}\hspace{-0.09cm}\left(\hspace{-0.1cm}1\hspace{-0.05cm}-\hspace{-0.05cm}\rho^{\mathrm{MRT}}_{p}\hspace{-0.1cm}+\hspace{-0.1cm}\frac{\rho^{\mathrm{MRT}}_{p}}{t}\hspace{-0.05cm}\right)\hspace{-0.1cm}+\hspace{-0.1cm}\log_{2}\hspace{-0.05cm}\left(\frac{1+\alpha^{\mathrm{MRT}}_{p}t}{1+\lambda^{\mathrm{MRT}}_{p}t}\hspace{-0.05cm}\right)\hspace{-0.05cm}.
	\label{eqn:MRTproblem_simple_2}
\end{align}
As in Section~\ref{sec:perfectcsit_ZF_highsnr}, we use the approximation $\log_{2}\left(1-\rho^{\mathrm{MRT}}_{p}+\rho^{\mathrm{MRT}}_{p}/t^{*}\right)\approx\log_{2}\left(\rho^{\mathrm{MRT}}_{p}/t^{*}\right)$ for simplicity. 
Taking the derivative of \eqref{eqn:MRTproblem_simple_2} with this approximation and equating to zero gives\vspace{-0.2cm}
\begin{align}
	&\underbrace{-\alpha^{\mathrm{MRT}}_{p}\lambda^{\mathrm{MRT}}_{p}}_{a_{1,p}}(t^{*})^{2}\nonumber\\
	&+\underbrace{\left[K(\alpha^{\mathrm{MRT}}_{p}-\lambda^{\mathrm{MRT}}_{p})-(\alpha^{\mathrm{MRT}}_{p}+\lambda^{\mathrm{MRT}}_{p})\right]}_{b_{1,p}}t^{*}-1=0. \nonumber
\end{align}
\vspace{-0.1cm}

Then, one can write the roots $s_{1/2}=\frac{-b_{1,p}\pm\sqrt{b_{1,p}^{2}+4a_{1,p}}}{2a_{1,p}}$. In order to have a real-valued $t^{*}\geq 0$, the conditions $\sqrt{b_{1,p}^{2}+4a_{1,p}}\geq 0$ and $b_{1,p}\geq 0$ should be satisfied. Accordingly, the optimal rate and parameters for this scenario are given in \eqref{eqn:45_perfect}.
\begin{table*}
\begin{align} 
		&t^{(4/5)}_{p}\triangleq
		\begin{cases}
    		\min\left\lbrace1,\frac{-b_{1,p}\pm\sqrt{b_{1,p}^{2}+4a_{1,p}}}{2a_{1,p}}\right\rbrace, & \hspace{-0.2cm}\text{if $\sqrt{b_{1,p}^{2}+4a_{1,p}} \geq 0$, $b_{1,p} \geq 0$,} \\
    		\hspace{1.85cm}1\hspace{1.85cm}, &\hspace{-0.2cm} \text{otherwise},
    	\end{cases}\hspace{-0.1cm},\ r^{(4/5)}_{mm,p}\triangleq\frac{1}{K}\log_{2}\hspace{-0.1cm}\left(\hspace{-0.08cm}1\hspace{-0.1cm}-\hspace{-0.1cm}\rho^{\mathrm{MRT}}_{p}\hspace{-0.1cm}+\hspace{-0.1cm}\frac{\rho^{\mathrm{MRT}}_{p}}{t^{(4/5)}_{p}}\hspace{-0.08cm}\right)\hspace{-0.1cm}+\hspace{-0.1cm}\log_{2}\left(\hspace{-0.08cm}\frac{1+\alpha^{\mathrm{MRT}}_{p}t^{(4/5)}_{p}}{1+\lambda^{\mathrm{MRT}}_{p}t^{(4/5)}_{p}}\hspace{-0.08cm}\right).	
	\label{eqn:45_perfect}
\end{align}
\vspace{-0.2cm}
\hrule
\vspace{-0.4cm}
\end{table*}
\vspace{-0.1cm}
\subsubsection{Case 2}
\label{sec:perfectcsit_MRT_lowsnr}
Next, we consider the alternative case for the system settings which require optimal power splitting to be performed at moderate/low SNR regions, so that, $Pt^{*}$ is moderate or small. Intuitively, this case captures the scenarios where the interference-limited regime is reached at moderate/low SNR, {\sl e.g.}, small $N$ or large $K-N$. We use \eqref{eqn:prop1} with $D^{\mathrm{MRT}}_{p}=1$ in \eqref{eqn:MRTproblem} to write
\begin{align} 
	\max_{t\in (0,1]}&\frac{1}{K}\log_{2}\hspace{-0.09cm}\left(1+\frac{Pe^{-\gamma}(1-t)}{(N+K-1)Pt+\sum_{k=1}^{K}\frac{1}{v_{k}}}\right)\nonumber \\
	&+\log_{2}\hspace{-0.05cm}\left(\frac{1+\alpha^{\mathrm{MRT}}_{p}t}{1+\lambda^{\mathrm{MRT}}_{p}t}\hspace{-0.05cm}\right)\hspace{-0.05cm}.
	\label{eqn:MRTproblem_simple_3}
\end{align}
Once again, taking the derivative and equating the result to zero yields
the optimal rate and parameters in \eqref{eqn:67_perfect} with the terms defined in \eqref{eqn:parameters}.
\begin{table*}
\begin{align} 
		&t^{(4/5)}_{p}\hspace{-0.08cm}\triangleq\hspace{-0.08cm}
		\begin{cases}
    		\hspace{-0.08cm}\min\left\lbrace1,\hspace{-0.08cm}\frac{-b_{2,p}\pm\sqrt{b_{2,p}^{2}-4a_{2,p}c_{2,p}}}{2a_{2,p}}\right\rbrace, & \hspace{-0.3cm}\text{if $\sqrt{b_{2,p}^{2}-4a_{2,p}c_{2,p}} \geq 0$, $\frac{-b_{2,p}\pm\sqrt{b_{2,p}^{2}-4a_{2,p}c_{2,p}}}{2a_{2,p}} \geq 0$} \\
    		\hspace{2.05cm}1\hspace{2.05cm}, &\hspace{-0.2cm} \text{otherwise}.
    	\end{cases}\nonumber \\
		&r^{(4/5)}_{mm,p}\hspace{-0.08cm}\triangleq\hspace{-0.08cm}\frac{1}{K}\log_{2}\hspace{-0.08cm}\left(\hspace{-0.08cm}1\hspace{-0.08cm}+\hspace{-0.08cm}\frac{Pe^{-\gamma}(1-t^{(4/5)}_{p})}{(N+K-1)Pt^{(4/5)}_{p}+\sum_{k=1}^{K}\frac{1}{v_{k}}}\hspace{-0.08cm}\right)\hspace{-0.08cm}+\hspace{-0.08cm}\log_{2}\hspace{-0.08cm}\left(\hspace{-0.08cm}\frac{1+\alpha^{\mathrm{MRT}}_{p}t^{(4/5)}_{p}}{1+\lambda^{\mathrm{MRT}}_{p}t^{(4/5)}_{p}}\hspace{-0.08cm}\right).	
	\label{eqn:67_perfect}
\end{align}
\vspace{-0.2cm}
\hrule
\vspace{-0.4cm}
\end{table*}
\begin{table*}[t!]
\begin{align}
a_{2,p}&\triangleq P^{2}(\alpha^{\mathrm{MRT}}_{p}-\lambda^{\mathrm{MRT}}_{p})(N+K-1)(N+K-1-e^{-\gamma})-\omega_{p}\alpha^{\mathrm{MRT}}_{p}\lambda^{\mathrm{MRT}}_{p}, \nonumber\\
b_{2,p}&\triangleq (\alpha^{\mathrm{MRT}}_{p}-\lambda^{\mathrm{MRT}}_{p})[P(N+K-1)(Pe^{-\gamma}+\sum_{k=1}^{K}\frac{1}{v_{k}})+P(N+K-1-e^{-\gamma})\sum_{k=1}^{K}\frac{1}{v_{k}}\biggr]-\omega_{p}(\alpha^{\mathrm{MRT}}_{p}+\lambda^{\mathrm{MRT}}_{p}), \nonumber\\
c_{2,p}&\triangleq (\alpha^{\mathrm{MRT}}_{p}-\lambda^{\mathrm{MRT}}_{p})\sum_{k=1}^{K}\frac{1}{v_{k}}(Pe^{-\gamma}+\sum_{k=1}^{K}\frac{1}{v_{k}})-\omega_{p},  \quad 
\omega_{p}\triangleq \frac{Pe^{-\gamma}(P(N+K-1)+\sum_{k=1}^{K}\frac{1}{v_{k}})}{K}.
\label{eqn:parameters}
\end{align}
\vspace{-0.2cm}
\hrule
\vspace{-0.5cm}
\end{table*}
\vspace{-0.4cm}
\subsection{Proposed Precoder and Resource Allocation}
\label{sec:perfectcsit_proposed}
\vspace{-0.1cm}
Finally, we describe the proposed power and precoder allocation for given parameters. The optimal power allocation $t_{p}$, rate allocation $\beta_{p}$, and precoder allocation for the private stream of user-$k$, $\mathbf{p}_{k,p}$, are performed as\vspace{-0.3cm}
\begin{align}
	&\hat{n}=\argmax_{n \in \{1,2,\ldots,5\}}r^{(n)}_{mm,p}, \  \mathbf{p}^{*}_{k,p}= 
	\begin{cases}
	 	\mathbf{p}^{\mathrm{ZF}}_{k}, & \hspace{-0.3cm}\text{if $\hat{n} \leq 3$ and $k \in \mathcal{G}_{1}$,} \\
	 	\hspace{0.25cm}\mathbf{0}\hspace{0.25cm}, & \hspace{-0.3cm}\text{if $\hat{n} \leq 3$ and $k \in \mathcal{G}_{2}$,} \\
	 	\mathbf{p}^{\mathrm{MRT}}_{k}\hspace{-0.15cm}, & \hspace{-0.3cm}\text{if $\hat{n} \geq 4$ and $k \in \mathcal{K}$,} \\
	\end{cases}\nonumber \\
		&t_{p}=t_{p}^{(\hat{n})},\ \beta_{p}=\begin{cases}
	 	0, &\hspace{-0.3cm} \text{if $\hat{n} \leq 3$,} \\
	 	\text{N/A}, &\hspace{-0.3cm} \text{otherwise} 
	\end{cases}\hspace{-0.1cm}. 
	\label{eqn:proposed_perfect}
\end{align}
For an enhanced performance, the common precoder, $\mathbf{p}_{c,p}$, can be calculated as the eigenvector corresponding to the largest eigenvalue of the combined channel matrix $\mathbf{H}=\left[\mathbf{h}_{1}, \mathbf{h}_{2}, \ldots, \mathbf{h}_{K}\right]$ instead of a random precoder.
\vspace{-0.3cm}
\section{Proposed Solution for Imperfect CSIT}
\label{sec:solution_imperfectcsit}
\vspace{-0.1cm}
In this section, we consider the scenario with imperfect CSIT. We follow the procedure in Section~\ref{sec:solution_perfectcsit} and find the optimal power, rate, and precoder allocation. 
\vspace{-0.3cm}
\subsection{Solution Using ZF Precoding}
\label{sec:solution_imperfectcsit_ZF}
\vspace{-0.1cm}
As done in Section~\ref{sec:solution_perfectcsit_ZF}, we define the user index groups $G_{1}$ and $G_{2}$, where the users with indexes in $G_{1}$ are served by private and common streams, while the users with indexes in $G_{2}$ are served only by the common stream. As shown by Proposition 6, the max-min problem for the considered scenario is written as \vspace{-0.1cm}
\begin{align}
	\frac{1-N\beta^{*}}{K-N}R^{\mathrm{ZF}}_{c,i}(t^{*})=\beta^{*}R^{\mathrm{ZF}}_{c,i}(t^{*})+R^{\mathrm{ZF}}_{\tilde{k},i}(t^{*}). 
	\label{eqn:problem_imperfect}
\end{align}
Similar to the cases in Section~\ref{sec:solution_perfectcsit_ZF} and \ref{sec:solution_perfectcsit_MRT}, it can be shown using \eqref{eqn:prop5} that $R^{\mathrm{ZF}}_{\tilde{k},i}(t)=\log_{2}(1+\sigma^{\mathrm{ZF}}_{\tilde{k},i}t)-\log_{2}(1+\nu^{\mathrm{ZF}}_{\tilde{k},i}t)$, where $\sigma^{\mathrm{ZF}}_{\tilde{k},i}=v_{\tilde{k}}\frac{P}{N}e^{(\log(\theta^{\mathrm{ZF}}_{i})+\psi(D^{\mathrm{ZF}}_{i}))}$, $\nu^{\mathrm{ZF}}_{\tilde{k},i}=v_{\tilde{k}}\frac{P(N-1)(1-\epsilon^{2})}{N}$, and, $\tilde{k}=\argmin_{k \in \mathcal{G}_{1}}v_{k}$ for given $v_{k}$.

\subsubsection{High SNR}
\label{sec:imperfectcsit_ZF_highsnr}
We start by investigating the case where the CSIT error is small enough, such that, $\nu^{\mathrm{ZF}}_{\tilde{k},i}t^{*} \ll 1$ and $(\sigma^{\mathrm{ZF}}_{\tilde{k},i}-\nu^{\mathrm{ZF}}_{\tilde{k},i})t^{*} \gg 1$. Using the approximation $\log(1+x)\approx \log(x)$ for large $x$, we write \vspace{-0.15cm}
\begin{align}
	\hspace{-0.3cm}\frac{1\hspace{-0.08cm}-\hspace{-0.08cm}N\beta^{*}}{K\hspace{-0.08cm}-\hspace{-0.08cm}N}\hspace{-0.08cm}\log_{2}\hspace{-0.08cm}\left(\rho^{\mathrm{ZF}}_{i}\hspace{-0.08cm}/t^{*}\hspace{-0.05cm}\right)\hspace{-0.09cm} &= \hspace{-0.05cm}\beta^{*}\hspace{-0.08cm}\log_{2}\hspace{-0.05cm}\left(\rho^{\mathrm{ZF}}_{i}\hspace{-0.08cm}/t^{*}\hspace{-0.05cm}\right)\hspace{-0.09cm} +\hspace{-0.05cm}\log_{2}\hspace{-0.03cm}(1\hspace{-0.06cm}+\hspace{-0.06cm}\sigma^{\mathrm{ZF}}_{\tilde{k},i}\hspace{-0.03cm}t^{*}\hspace{-0.03cm}).
	%\left(\hspace{-0.05cm}\frac{\rho^{\mathrm{ZF}}_{i}}{t^{*}}\hspace{-0.05cm}\right)^{\frac{1\hspace{-0.08cm}-\hspace{-0.08cm}K\beta^{*}}{K\hspace{-0.08cm}-\hspace{-0.08cm}N}} &= \sigma^{\mathrm{ZF}}_{\tilde{k},i}\hspace{-0.12cm}t^{*}, %\nonumber \\
	%t^{*}&=\frac{(\rho^{\mathrm{ZF}}_{i})^{\frac{1-K\beta^{*}}{1\hspace{-0.08cm}-\hspace{-0.08cm}K\beta^{*}+K-N}}}{\left(\sigma^{\mathrm{ZF}}_{\tilde{k},i}\right)^{\frac{K\hspace{-0.08cm}-\hspace{-0.08cm}N}{1-K\beta^{*}+K-N}}}\triangleq t^{(3)}_{i}(\beta^{*}).
	\label{eqn:imperfectcsit_highsnr_3}
\end{align} 
Removing logarithm from both sides of \eqref{eqn:imperfectcsit_highsnr_3} yields $t^{*}=(\rho^{\mathrm{ZF}}_{i})^{\frac{1-K\beta^{*}}{1-K\beta^{*}+K-N}}/(\sigma^{\mathrm{ZF}}_{\tilde{k},i})^{\frac{K\hspace{-0.08cm}-\hspace{-0.08cm}N}{1-K\beta^{*}+K-N}}\triangleq t^{(1)}_{i}(\beta^{*})$. One can immediately note that $t^{(1)}_{i}(\beta^{*}) \geq 0$.

In order to determine $\beta^{*}$, we maximize $\frac{1-N\beta}{K-N}R^{\mathrm{ZF}}_{c,i}(t^{(1)}_{i}(\beta))$. For simplicity, we use the approximation above for the common stream and write \vspace{-0.2cm}
\begin{align}
		%&\max_{\beta}\frac{1-N\beta}{K-N}\log_{2}\left(\frac{\rho^{\mathrm{ZF}}_{i}}{t^{(3)}_{i}(\beta)}\right)=
		&\max_{\beta}\frac{1-N\beta}{K-N}\log_{2}\left(\left(\sigma^{\mathrm{ZF}}_{\tilde{k},i}\rho^{\mathrm{ZF}}_{i}\right)^\frac{K-N}{1-K\beta+K-N}\right)\nonumber\\
		&=\max_{\beta}\frac{1-N\beta}{1-K\beta+K-N}\log_{2}\left(\sigma^{\mathrm{ZF}}_{\tilde{k},i}\rho^{\mathrm{ZF}}_{i}\right).
		\label{eqn:beta_3}
\end{align}
Taking the derivative of \eqref{eqn:beta_3} with respect to $\beta$ gives $\frac{(K-N)(1-N)}{(1-K\beta+K-N)^{2}}\log_{2}(\sigma^{\mathrm{ZF}}_{\tilde{k},i}\rho^{\mathrm{ZF}}_{i})$, which is negative for $\sigma^{\mathrm{ZF}}_{\tilde{k},i}\rho^{\mathrm{ZF}}_{i} > 1$ and positive for $\sigma^{\mathrm{ZF}}_{\tilde{k},i}\rho^{\mathrm{ZF}}_{i} < 1$. This implies that $\frac{1-N\beta}{K-N}R^{\mathrm{ZF}}_{c,i}(t^{(1)}_{i}(\beta))$ monotonically decreases with increasing $\beta$ for $\sigma^{\mathrm{ZF}}_{\tilde{k},i}\rho^{\mathrm{ZF}}_{i} > 1$ and monotonically increases for $\sigma^{\mathrm{ZF}}_{\tilde{k},i}\rho^{\mathrm{ZF}}_{i} < 1$.
Accordingly, we define the optimal rate and parameters for this scenario in \eqref{eqn:3_imperfect}. 
\begin{table*}
\begin{align} 
		&t^{(1)}_{i}\triangleq \min\left\lbrace\frac{(\rho^{\mathrm{ZF}}_{i})^{\frac{1-K\beta^{(1)}_{i}}{1\hspace{-0.08cm}-\hspace{-0.08cm}K\beta^{(1)}_{i}+K-N}}}{\left(\sigma^{\mathrm{ZF}}_{\tilde{k},i}\right)^{\frac{K\hspace{-0.08cm}-\hspace{-0.08cm}N}{1-K\beta^{(1)}_{i}+K-N}}}, 1\right\rbrace, \quad \beta^{(1)}_{i}\triangleq 
		\begin{cases}
    		0, & \hspace{-0.2cm}\text{if $\sigma^{\mathrm{ZF}}_{\tilde{k},i}\rho^{\mathrm{ZF}}_{i} > 1$} \\
    		\varepsilon\frac{1}{K}, &\hspace{-0.2cm} \text{otherwise}.
    	\end{cases}, \nonumber	\\  
		&r^{(1)}_{mm,p}\triangleq\min\left\lbrace\frac{1-N\beta^{(1)}_{i}}{K-N}\log_{2}\hspace{-0.1cm}\left(\hspace{-0.08cm}1\hspace{-0.1cm}-\hspace{-0.1cm}\rho^{\mathrm{ZF}}_{i}\hspace{-0.1cm}+\hspace{-0.1cm}\frac{\rho^{\mathrm{ZF}}_{i}}{t^{(1)}_{i}}\right), \  \beta^{(1)}_{i}\log_{2}\hspace{-0.1cm}\left(\hspace{-0.08cm}1\hspace{-0.1cm}-\hspace{-0.1cm}\rho^{\mathrm{ZF}}_{i}\hspace{-0.1cm}+\hspace{-0.1cm}\frac{\rho^{\mathrm{ZF}}_{i}}{t^{(1)}_{i}}\hspace{-0.08cm}\right)\hspace{-0.1cm}+\hspace{-0.1cm}\log_{2}\left(\hspace{-0.08cm}\frac{1\hspace{-0.1cm}+\hspace{-0.1cm}\sigma^{\mathrm{ZF}}_{\tilde{k},i}t^{(1)}_{i}}{1\hspace{-0.1cm}+\hspace{-0.1cm}\nu^{\mathrm{ZF}}_{\tilde{k},i}t^{(1)}_{i}}\hspace{-0.08cm}\right)\hspace{-0.1cm}\right\rbrace\hspace{-0.1cm}.			
		\label{eqn:3_imperfect}
\end{align}
\vspace{-0.2cm}
\hrule
\vspace{-0.4cm}
\end{table*}
We note that \eqref{eqn:3_imperfect} reduces to \eqref{eqn:1_perfect} for $\epsilon^{2}=1$ and $\varepsilon = 0$.

Secondly, let us consider the scenario where both the CSIT error and $v_{\tilde{k}}$ are small enough, such that, $\nu^{\mathrm{ZF}}_{\tilde{k},i}t^{*} \ll 1$ and $(\sigma^{\mathrm{ZF}}_{\tilde{k},i}-\nu^{\mathrm{ZF}}_{\tilde{k},i})t^{*} \ll 1$. Assuming a large rate for the common stream and using the approximation $\log(1+x)\approx x$ for small $x$, we re-write \eqref{eqn:problem_imperfect} as\vspace{-0.2cm}
  \begin{align}
	%&\frac{1\hspace{-0.08cm}-\hspace{-0.08cm}N\beta^{*}}{K\hspace{-0.08cm}-\hspace{-0.08cm}N}\hspace{-0.08cm}\log_{2}\hspace{-0.08cm}\left(\hspace{-0.05cm}\frac{\rho^{\mathrm{ZF}}_{i}}{t^{*}}\hspace{-0.05cm}\right)\hspace{-0.09cm} = \hspace{-0.05cm}\beta^{*}\hspace{-0.08cm}\log_{2}\hspace{-0.05cm}\left(\hspace{-0.08cm}\frac{\rho^{\mathrm{ZF}}_{i}}{t^{*}}\hspace{-0.05cm}\right)\hspace{-0.09cm} +\hspace{-0.05cm}\log_{2}\hspace{-0.05cm}\left(\hspace{-0.05cm}\hspace{-0.05cm}1\hspace{-0.09cm}+\hspace{-0.09cm}\frac{(\sigma^{\mathrm{ZF}}_{\tilde{k},i}\hspace{-0.12cm}-\hspace{-0.05cm}\nu^{\mathrm{ZF}}_{\tilde{k},i})t^{*}}{1+\nu^{\mathrm{ZF}}_{\tilde{k},i}t^{*}}\hspace{-0.05cm}\right)\hspace{-0.1cm}, \nonumber \\
	\hspace{-0.5cm}\frac{1\hspace{-0.08cm}-\hspace{-0.08cm}N\beta^{*}}{K\hspace{-0.08cm}-\hspace{-0.08cm}N}\hspace{-0.08cm}\log_{2}\hspace{-0.08cm}\left(\rho^{\mathrm{ZF}}_{i}/t^{*}\hspace{-0.05cm}\right)\hspace{-0.09cm} = \hspace{-0.08cm}\beta^{*}\log_{2}\hspace{-0.08cm}\left(\hspace{-0.05cm}\rho^{\mathrm{ZF}}_{i}\hspace{-0.08cm}/t^{*}\hspace{-0.05cm}\right)\hspace{-0.08cm}+\hspace{-0.03cm}\hspace{-0.05cm}(\sigma^{\mathrm{ZF}}_{\tilde{k},i}-\nu^{\mathrm{ZF}}_{\tilde{k},i})t^{*}\hspace{-0.1cm}, \hspace{-0.1cm}%\nonumber \\
	%&\hspace{2.2cm}\rho^{\mathrm{ZF}}_{i}=t^{*}e^{\frac{\log(2)(K-N)(\sigma^{\mathrm{ZF}}_{\tilde{k},i}-\nu^{\mathrm{ZF}}_{\tilde{k},i})}{1-K\beta^{*}}t^{*}}.
	\label{eqn:imperfectcsit_highsnr_2}
\end{align}
and re-arrange \eqref{eqn:imperfectcsit_highsnr_2} to obtain $\rho^{\mathrm{ZF}}_{i}=t^{*}e^{\frac{\log(2)(K-N)(\sigma^{\mathrm{ZF}}_{\tilde{k},i}-\nu^{\mathrm{ZF}}_{\tilde{k},i})}{1-K\beta^{*}}t^{*}}$. Let us define $\delta_{i}\triangleq\log(2)(K-N)(\sigma^{\mathrm{ZF}}_{\tilde{k},i}-\nu^{\mathrm{ZF}}_{\tilde{k},i})$. We multiply both sides of \eqref{eqn:imperfectcsit_highsnr_2} by $\frac{\delta_{i}}{1-K\beta^{*}}$ and use the Lambert-W function to obtain $t^{*}=\frac{1-K\beta^{*}}{\delta_{i}}W_{k}\left(\frac{\delta_{i}\rho^{\mathrm{ZF}}_{i}}{1-K\beta^{*}}\right)$,
Since $\frac{\delta_{i}\rho^{\mathrm{ZF}}_{i}}{1-K\beta^{*}}$ and $\frac{\delta_{i}}{1-K\beta^{*}}t^{*}$ are real and $\frac{\delta_{i}\rho^{\mathrm{ZF}}_{i}}{1-K\beta^{*}}>0$, it suffices to check only the principal branch $W_{0}(x)$. As done in Section~\ref{sec:perfectcsit_ZF_highsnr}, we use the approximation \mbox{$W_{0}(x)\approx\log(x)-\log(\log(x))$} to obtain  \vspace{-0.2cm}
\begin{align}
	t^{*}&=\frac{\log\left(\frac{\delta_{i}\rho^{\mathrm{ZF}}_{i}}{1-K\beta^{*}}\right)-\log\left(\log\left(\frac{\delta_{i}}{1-K\beta^{*}}\rho^{\mathrm{ZF}}_{i}\right)\right)}{\frac{\delta_{i}}{1-K\beta^{*}}} \triangleq t^{(2)}_{i}(\beta^{*}). \nonumber
	%\label{eqn:t2_imperfect}
\end{align} 
\vspace{-0.2cm}
\begin{lemma}
	The coefficient $t^{(2)}_{i}(\beta^{*})$ satisfies $0\leq t^{(2)}_{i}(\beta^{*})\leq 1$ for $\beta^{*} \in \left[0,\frac{1}{K}\right)$ and $\frac{\delta_{i}\rho^{\mathrm{ZF}}_{i}}{1-K\beta^{*}} \geq e$ .
\end{lemma}
\quad\textit{Proof:} The proof is similar to that of Lemma 4 and is omitted for the sake of brevity. \hspace{4.1cm}$\blacksquare$
%See Appendix~\ref{appendix:lemmaz}. \hspace{11.6cm}$\blacksquare$

We note that $\lim_{\beta^{*}\rightarrow 1/K}t^{(2)}_{i}(\beta^{*})=0$. Accordingly, $t^{(2)}_{i}(\beta^{*})$ and $\beta^{*}$ should satisfy $t^{(2)}_{i}(\beta^{*})>0$ and $\beta^{*}<\frac{1}{K}$, respectively, due to our initial assumption $Pt^{*}\rightarrow\infty$. On the other hand, it can be shown that $\lim_{\beta^{*}\rightarrow 1/K}\frac{1-N\beta^{*}}{K-N}\log(\rho^{\mathrm{ZF}}_{i}/t^{(2)}_{i}(\beta^{*}))\rightarrow\infty$.
This shows that one needs to choose $\beta^{*}$ as large as possible to maximize the minimum rate. Therefore, we define an arbitrary coefficient $\varepsilon<1$ to assign $\beta^{*}=\varepsilon\frac{1}{K}$. 
Accordingly, we define the optimal rate and parameters for this scenario as in \eqref{eqn:2_imperfect}. The parameter $\lambda$ in \eqref{eqn:2_imperfect} serves as a design parameter and checks the validity of the assumption for large $Pt^{(2)}_{i}(\beta^{*})$.
\begin{table*}
\begin{align} 
		&t^{(2)}_{i}\triangleq
		\begin{cases}
    		\frac{\log\left(\frac{\delta_{i}\rho^{\mathrm{ZF}}_{i}}{1-\varepsilon}\right)-\log\left(\log\left(\frac{\delta_{i}\rho^{\mathrm{ZF}}_{i}}{1-\varepsilon}\right)\right)}{\frac{\delta_{i}}{1-\varepsilon}}, & \hspace{-0.2cm}\text{if $\frac{\delta_{i}\rho^{\mathrm{ZF}}_{i}}{1-\varepsilon} \geq e$ and $Pt^{(2)}_{i}\left(\varepsilon\frac{1}{K}\right)\geq\lambda$} \\
    		\hspace{2.1cm}1\hspace{2.1cm}, &\hspace{-0.2cm} \text{otherwise}.
    	\end{cases},  \quad
    	\beta^{(2)}_{i}=\varepsilon\frac{1}{K}, \nonumber \\  
    	&r^{(2)}_{mm,p}\triangleq\min\left\lbrace\frac{1-N\beta^{(2)}_{i}}{K-N}\log_{2}\hspace{-0.1cm}\left(\hspace{-0.08cm}1\hspace{-0.1cm}-\hspace{-0.1cm}\rho^{\mathrm{ZF}}_{i}\hspace{-0.1cm}+\hspace{-0.1cm}\frac{\rho^{\mathrm{ZF}}_{i}}{t^{(2)}_{i}}\hspace{-0.08cm}\right), \  \beta^{(2)}_{i}\log_{2}\hspace{-0.1cm}\left(\hspace{-0.08cm}1\hspace{-0.1cm}-\hspace{-0.1cm}\rho^{\mathrm{ZF}}_{i}\hspace{-0.1cm}+\hspace{-0.1cm}\frac{\rho^{\mathrm{ZF}}_{i}}{t^{(2)}_{i}}\right)\hspace{-0.1cm}+\hspace{-0.1cm}\log_{2}\left(\hspace{-0.08cm}\frac{1\hspace{-0.1cm}+\hspace{-0.1cm}\sigma^{\mathrm{ZF}}_{i}t^{(2)}_{i}}{1\hspace{-0.1cm}+\hspace{-0.1cm}\nu^{\mathrm{ZF}}_{i}t^{(2)}_{i}}\hspace{-0.08cm}\right)\hspace{-0.1cm}\right\rbrace\hspace{-0.1cm}.	
	\label{eqn:2_imperfect}
\end{align}
\vspace{-0.2cm}
\hrule
\vspace{-0.6cm}
\end{table*}
We note that \eqref{eqn:2_imperfect} reduces to \eqref{eqn:2_perfect} for $\epsilon^{2}=1$ and $\varepsilon = 0$.

Finally, we consider the scenario where the CSIT error and $v_{\tilde{k}}$ is large enough, such that, $\sigma^{\mathrm{ZF}}_{\tilde{k},i}t^{*} \gg 1$ and $\nu^{\mathrm{ZF}}_{\tilde{k},i}t^{*} \gg 1$. We also assume that the rate of the common stream is large enough, such that, $\log_{2}\left(1-\rho^{\mathrm{ZF}}_{i}+\rho^{\mathrm{ZF}}_{i}/t^{*}\right)\approx\log_{2}\left(\rho^{\mathrm{ZF}}_{i}/t^{*}\right)$. From \eqref{eqn:problem_imperfect}, we write\vspace{-0.2cm}
\begin{align}
	\frac{1-N\beta^{*}}{K-N}&\log_{2}\left(\rho^{\mathrm{ZF}}_{i}/t^{*}\right) \nonumber \\[-6pt]
	&=\beta^{*}\log_{2}\left(\rho^{\mathrm{ZF}}_{i}/t^{*}\right)+\log_{2}\left(\sigma^{\mathrm{ZF}}_{\tilde{k},i}t^{*}/\nu^{\mathrm{ZF}}_{\tilde{k},i}t^{*}\right),  \nonumber 
	%\frac{1-K\beta^{*}}{K-N}\log_{2}\left(\frac{\rho^{\mathrm{ZF}}_{i}}{t^{*}}\right)&=\log_{2}\left(\frac{\sigma^{\mathrm{ZF}}_{i}t^{*}}{\nu^{\mathrm{ZF}}_{i}t^{*}}\right), \nonumber \\
	%\left(\frac{\rho^{\mathrm{ZF}}_{i}}{t^{*}}\right)^{\frac{1-K\beta^{*}}{K-N}}&=\frac{\sigma^{\mathrm{ZF}}_{i}}{\nu^{\mathrm{ZF}}_{i}}, \nonumber \\
	%t^{*}&=\frac{\rho^{\mathrm{ZF}}_{i}}{\left(\frac{\sigma^{\mathrm{ZF}}_{i}}{\nu^{\mathrm{ZF}}_{i}}\right)^\frac{K-N}{1-K\beta^{*}}}\triangleq t^{(1)}_{i}(\beta^{*}).
	%\label{eqn:imperfectcsit_highsnr_1}
\end{align}
\vspace{-0.15cm}
\hspace{-0.09cm}which yields $t^{*}=\rho^{\mathrm{ZF}}_{i}(\nu^{\mathrm{ZF}}_{\tilde{k},i}/\sigma^{\mathrm{ZF}}_{\tilde{k},i})^\frac{K-N}{1-K\beta^{*}}\triangleq t^{(3)}_{i}(\beta^{*})$. One can observe that $t^{(3)}_{i}(\beta^{*}) \geq 0$.

As done in Section~\ref{sec:perfectcsit_ZF_highsnr}, we maximize $\frac{1-N\beta}{K-N}R^{\mathrm{ZF}}_{c,i}\left(t^{(3)}_{i}(\beta)\right)$ to find $\beta^{*}$. For simplicity, we use the approximation above for the common stream and write \vspace{-0.2cm}
\begin{align}
		%&\max_{\beta}\frac{1-N\beta}{K-N}\log_{2}\left(\frac{\rho^{\mathrm{ZF}}_{i}}{t^{(3)}_{i}(\beta)}\right)=
		&\max_{\beta}\frac{1-N\beta}{K-N}\log_{2}\left(\left(\sigma^{\mathrm{ZF}}_{\tilde{k},i}/\nu^{\mathrm{ZF}}_{\tilde{k},i}\right)^\frac{K-N}{1-K\beta}\right) \nonumber \\
		=&\max_{\beta}\frac{1-N\beta}{1-K\beta}\log_{2}\left(\sigma^{\mathrm{ZF}}_{\tilde{k},i}/\nu^{\mathrm{ZF}}_{\tilde{k},i}\right). 
		\label{eqn:beta_1} 
\end{align}
\vspace{-0.05cm}
Taking derivative of \eqref{eqn:beta_1} with respect to $\beta$ gives $\frac{K-N}{(1-K\beta)^{2}}\log_{2}(\sigma^{\mathrm{ZF}}_{\tilde{k},i}/\nu^{\mathrm{ZF}}_{\tilde{k},i})$, which is positive for $\forall \beta \in \left[0,\frac{1}{K}\right]$. This means that $\frac{1-N\beta}{K-N}R^{\mathrm{ZF}}_{c,i}(t^{(3)}_{i}(\beta))$ is an increasing function of $\beta$ and $\beta^{*}$ should be chosen as large as possible in the region $\left[0,\frac{1}{K}\right]$. In order to find the maximum possible $\beta$, we refer to our initial assumption $\nu^{\mathrm{ZF}}_{\tilde{k},i}t^{*} \gg 1$, and write $\nu^{\mathrm{ZF}}_{\tilde{k},i}\rho^{\mathrm{ZF}}_{i}/(\sigma^{\mathrm{ZF}}_{\tilde{k},i}/\nu^{\mathrm{ZF}}_{\tilde{k},i})^\frac{K-N}{1-K\beta^{*}} \geq 1$, which yields $\beta \leq \frac{1}{K}-\frac{(K-N)\log\left(\sigma^{\mathrm{ZF}}_{\tilde{k},i}/\nu^{\mathrm{ZF}}_{\tilde{k},i}\right)}{K\log\left(\nu^{\mathrm{ZF}}_{\tilde{k},i}\rho^{\mathrm{ZF}}_{i}\right)}$.
We define optimal rate and parameters for this scenario in \eqref{eqn:1_imperfect}. 
\begin{table*}
\begin{align}
		&t^{(3)}_{i}\triangleq \min\left\lbrace\rho^{\mathrm{ZF}}_{i}\left(\frac{\nu^{\mathrm{ZF}}_{\tilde{k},i}}{\sigma^{\mathrm{ZF}}_{\tilde{k},i}}\right)^\frac{K-N}{1-K\beta^{(3)}_{i}}\hspace{-0.2cm}, 1\right\rbrace, \quad \beta^{(3)}_{i}\triangleq \max\left\lbrace\frac{1}{K}-\frac{(K-N)\log\left(\sigma^{\mathrm{ZF}}_{\tilde{k},i}/\nu^{\mathrm{ZF}}_{\tilde{k},i}\right)}{K\log\left(\nu^{\mathrm{ZF}}_{\tilde{k},i}\rho^{\mathrm{ZF}}_{i}\right)},0\right\rbrace, \nonumber	\\  
		&r^{(3)}_{mm,p}\triangleq\min\left\lbrace\frac{1-N\beta^{(3)}_{i}}{K-N}\log_{2}\hspace{-0.1cm}\left(\hspace{-0.08cm}1\hspace{-0.1cm}-\hspace{-0.1cm}\rho^{\mathrm{ZF}}_{i}\hspace{-0.1cm}+\hspace{-0.1cm}\frac{\rho^{\mathrm{ZF}}_{i}}{t^{(3)}_{i}}\right), \  \beta^{(3)}_{i}\log_{2}\hspace{-0.1cm}\left(\hspace{-0.08cm}1\hspace{-0.1cm}-\hspace{-0.1cm}\rho^{\mathrm{ZF}}_{i}\hspace{-0.1cm}+\hspace{-0.1cm}\frac{\rho^{\mathrm{ZF}}_{i}}{t^{(3)}_{i}}\right)+\log_{2}\hspace{-0.1cm}\left(\hspace{-0.08cm}\frac{1+\sigma^{\mathrm{ZF}}_{\tilde{k},i}t^{(3)}_{i}}{1+\nu^{\mathrm{ZF}}_{\tilde{k},i}t^{(3)}_{i}}\hspace{-0.08cm}\right)\hspace{-0.1cm}\right\rbrace\hspace{-0.1cm}.			
		\label{eqn:1_imperfect}
\end{align}
\vspace{-0.2cm}
\hrule
\vspace{-0.6cm}
\end{table*}

\subsubsection{Moderate/Low SNR}
\label{sec:imperfectcsit_ZF_lowsnr}
Next, we consider the case where the power allocated to private streams is moderate or small. We follow the procedure in Section~\ref{sec:perfectcsit_ZF_lowsnr} and use \eqref{eqn:lemma1} to write \vspace{-0.5cm}
\begin{align} 
	&\frac{1-K\beta^{*}}{K-N}\log_{2}\hspace{-0.1cm}\left(\hspace{-0.1cm}1+\frac{Ne^{-\gamma}P(1-t)}{P\lfloor N(N-K+D^{\mathrm{ZF}}_{i})\rceil t+N\sum_{k=1}^{K}\frac{1}{v_{k}}}\right)\nonumber \\
	&=\log_{2}\hspace{-0.1cm}\left(\hspace{-0.1cm}1+\frac{(\sigma^{\mathrm{ZF}}_{i}-\nu^{\mathrm{ZF}}_{i})t^{*}}{1+\nu^{\mathrm{ZF}}_{i}t^{*}}\right). \nonumber
	%\label{nonasymp_imperfect_1}
\end{align}
 \vspace{-0.1cm}
Applying the approximations $\log(1+x)\approx \log(x)$ for large $x$ and $\log(1+x)\approx x$ for the rates of the common stream and private stream, respectively, we obtain \vspace{-0.2cm}
\begin{align}
	&\log_{2}\hspace{-0.1cm}\left(\hspace{-0.2cm}\left(\hspace{-0.1cm}\frac{Ne^{-\gamma}P(1-t)}{P\lfloor N(N-K+D^{\mathrm{ZF}}_{i})\rceil t+N\sum_{k=1}^{K}\frac{1}{v_{k}}}\hspace{-0.1cm}\right)^{\hspace{-0.2cm}\frac{1-K\beta^{*}}{K-N}}\hspace{-0.1cm}\right)\hspace{-0.1cm}\nonumber\\
	&=\hspace{-0.1cm}\frac{(\sigma^{\mathrm{ZF}}_{\tilde{k},i}\hspace{-0.1cm}-\hspace{-0.05cm}\nu^{\mathrm{ZF}}_{\tilde{k},i})t^{*}}{\nu^{\mathrm{ZF}}_{\tilde{k},i}t^{*}+1}. 
	\label{eqn:mod_snr}
\end{align}
\vspace{-0.1cm}
Considering the moderate or small $Pt^{*}$ assumption, we further assume $P\lfloor N(N-K+D^{\mathrm{ZF}}_{i})\rceil t^{*} \ll N\sum_{k=1}^{K}1/v_{k}$ and $\nu^{\mathrm{ZF}}_{\tilde{k},i}t^{*}\ll 1$. Accordingly, we can re-arrange \eqref{eqn:mod_snr} to obtain $e^{-\frac{\delta_{i}t^{*}}{(1-K\beta^{*})}}(1-t^{*})=\frac{\sum_{k=1}^{K}\hspace{-0.1cm}\frac{1}{v_{k}}}{e^{-\gamma}P}$. Multiplying both sides by $\frac{\delta_{i}}{(1-K\beta^{*})}e^{\frac{\delta_{i}}{(1-K\beta^{*})}}$, we obtain
 \vspace{-0.3cm}
	\begin{align}
		%\frac{e^{-\gamma}P(1-t^{*})}{\hspace{-0.1cm}\sum_{k=1}^{K}\hspace{-0.1cm}\frac{1}{v_{k}}}&=e^{\frac{\delta_{i}t^{*}}{(1-K\beta^{*})}},  \\
		%e^{-\frac{\delta_{i}t^{*}}{(1-K\beta^{*})}}\frac{e^{-\gamma}P}{\hspace{-0.1cm}\sum_{k=1}^{K}\hspace{-0.1cm}\frac{1}{v_{k}}}(1-t^{*})&=1,  \\
		%&e^{-\frac{\delta_{i}t^{*}}{(1-K\beta^{*})}}(1-t^{*})=\frac{\sum_{k=1}^{K}\hspace{-0.1cm}\frac{1}{v_{k}}}{e^{-\gamma}P},  \\
		\frac{\delta_{i}(1-t^{*})}{(1-K\beta^{*})}e^{\frac{\delta_{i}(1-t^{*})}{(1-K\beta^{*})}}&=\frac{\delta_{i}\sum_{k=1}^{K}\hspace{-0.1cm}\frac{1}{v_{k}}}{(1-K\beta^{*})e^{-\gamma}P}e^{\frac{\delta_{i}}{(1-K\beta^{*})}},  %\\
		%\frac{\delta_{i}(1-t^{*})}{(1-K\beta^{*})}&=W_{k}\hspace{-0.1cm}\left(\hspace{-0.1cm}\frac{\delta_{i}\hspace{-0.1cm}\sum_{k=1}^{K}\hspace{-0.1cm}\frac{1}{v_{k}}}{(1-K\beta^{*})e^{-\gamma}P}e^{\frac{\delta_{i}}{(1-K\beta^{*})}}\hspace{-0.1cm}\right)\hspace{-0.1cm}, %\\
		%t^{*}\hspace{-0.05cm}&=\hspace{-0.05cm}1\hspace{-0.05cm}-\hspace{-0.05cm}\frac{W_{k}\hspace{-0.1cm}\left(\hspace{-0.1cm}\frac{\delta_{i}\sum_{k=1}^{K}\hspace{-0.1cm}\frac{1}{v_{k}}}{(1-K\beta^{*})e^{-\gamma}P}e^{\frac{\delta_{i}}{(1-K\beta^{*})}}\hspace{-0.1cm}\right)}{\frac{\delta_{i}}{(1-K\beta^{*})}}. 
		\label{eqn:t4_pre}
	\end{align}
\hspace{-0.02cm}which yields $\frac{\delta_{i}(1-t^{*})}{(1-K\beta^{*})}=W_{k}(\frac{\delta_{i}\hspace{-0.1cm}\sum_{k=1}^{K}\hspace{-0.1cm}\frac{1}{v_{k}}}{(1-K\beta^{*})e^{-\gamma}P}e^{\frac{\delta_{i}}{(1-K\beta^{*})}})$.
Checking the principal branch $W_{0}$ and benefiting from the approximation $W_{0}(x)\approx\log (x)-\log(\log(x))$, we can obtain
\begin{align}
	t^{*}&=1-\frac{\log\left(\hspace{-0.1cm}\frac{\delta_{i}\sum_{k=1}^{K}\hspace{-0.1cm}\frac{1}{v_{k}}}{(1-K\beta^{*})e^{-\gamma}P}e^{\frac{\delta_{i}}{(1-K\beta^{*})}}\hspace{-0.1cm}\right)}{\frac{\delta_{i}}{(1-K\beta^{*})}} \nonumber\\
	&+\frac{\log\left(\log\left(\hspace{-0.1cm}\frac{\delta_{i}\sum_{k=1}^{K}\hspace{-0.1cm}\frac{1}{v_{k}}}{(1-K\beta^{*})e^{-\gamma}P}e^{\frac{\delta_{i}}{(1-K\beta^{*})}}\hspace{-0.1cm}\right)\right)}{\frac{\delta_{i}}{(1-K\beta^{*})}}\triangleq t_{i}^{(4)}(\beta^{*}).\nonumber
\end{align}
\vspace{-0.2cm}
\begin{lemma}
	The coefficient $t^{(4)}_{i}(\beta^{*})$ satisfies $0\leq\hspace{-0.1cm}t^{(4)}_{i}(\beta^{*})\hspace{-0.1cm}\leq 1$ for $\beta^{*}\hspace{-0.1cm}\in\hspace{-0.1cm}\left[0,\frac{1}{K}\right]$ and $\frac{\delta_{i}\sum_{k=1}^{K}\hspace{-0.1cm}\frac{1}{v_{k}}}{(1-K\beta^{*})e^{-\gamma}P}\hspace{-0.1cm}\geq\hspace{-0.1cm}e$.
\end{lemma}
\quad\textit{Proof:}  The proof is similar to that of Lemma 4 with reverse conditions for $0 \leq t^{(2)}_{p}(\beta^{*})\leq 1$ and omitted here. \hspace{1.15cm}$\blacksquare$
\begin{table*}[t!]
	\begin{align}
		&t^{(4)}_{i}\triangleq
		\begin{cases}		
		1-\frac{\log\left(\hspace{-0.1cm}\frac{\delta_{i}\sum_{k=1}^{K}\hspace{-0.1cm}\frac{1}{v_{k}}}{e^{-\gamma}P}e^{\delta_{i}}\hspace{-0.1cm}\right)-\log\left(\log\left(\hspace{-0.1cm}\frac{\delta_{i}\sum_{k=1}^{K}\hspace{-0.1cm}\frac{1}{v_{k}}}{e^{-\gamma}P}e^{\delta_{i}}\right)\right)}{\delta_{i}}, & \hspace{-0.2cm}\text{if $\frac{\delta_{i}\sum_{k=1}^{K}\hspace{-0.1cm}\frac{1}{v_{k}}}{(1-K\beta^{*})e^{-\gamma}P} \geq e$ and $Pt^{(4)}_{i}\left(\varepsilon\frac{1}{K}\right)\geq\lambda$} \\
    		\hspace{3.2cm}1\hspace{3.2cm}, &\hspace{-0.2cm} \text{otherwise}.
    	\end{cases}, \quad
    	\beta^{(4)}_{i}=0, \nonumber\\
		&r^{(4)}_{mm,i}\hspace{-0.05cm}\triangleq\hspace{-0.05cm}\min\hspace{-0.1cm}\left\lbrace\frac{1}{K-N}\log_{2}\hspace{-0.1cm}\left(\hspace{-0.1cm}1\hspace{-0.05cm}+\hspace{-0.05cm}\frac{Ne^{-\gamma}P(1-t^{(4)}_{i})}{P\lfloor N(N-K+D^{\mathrm{ZF}}_{i})\rceil t^{(4)}_{i}\hspace{-0.05cm}+\hspace{-0.05cm}N\sum_{k=1}^{K}\frac{1}{v_{k}}}\right), \log_{2}\hspace{-0.1cm}\left(1+\sigma^{\mathrm{ZF}}_{\tilde{k},i}t^{(4)}_{i}\right) \right\rbrace\hspace{-0.05cm}.	
	\label{eqn:4_imperfect}
	\vspace{-0.5cm}
	\end{align} 
	\vspace{-0.2cm}
	\hrule
	\vspace{-0.5cm}
\end{table*}

Following the procedure in the previous cases, we determine $\beta^{*}$ by maximizing $\frac{1-N\beta}{K-N}R^{\mathrm{ZF}}_{c,i}(t^{(4)}_{i}(\beta^{*}))$. However, the expressions become difficult to track and finding $\beta^{*}$ becomes challenging in this case. Therefore, we set $\beta^{*}$ according to our observations. We start by noting that $\lim_{\beta^{*}\rightarrow 1/K}t^{(4)}_{i}(\beta^{*})=0$, which can be obtained by L'Hôpital's rule. This shows that $t^{(4)}_{i}(\beta^{*})$ decreases with increasing $\beta^{*}$. By taking into account the expression for $t^{(4)}_{i}(\beta^{*})$ and $R^{\mathrm{ZF}}_{c,i}\left(t\right)$, one can conclude that the term $R^{\mathrm{ZF}}_{c,i}(t)$ increases logarithmically with respect to $\beta^{*}$. On the other hand, the term $\frac{1-N\beta}{K-N}$ decreases linearly with increasing $\beta$. As the decrease rate is higher, we set $\beta^{*}=0$ to maximize $\frac{1-N\beta}{K-N}R^{\mathrm{ZF}}_{c,i}(t^{(4)}_{i}(\beta^{*}))$.
We can write the optimal rate and parameters for this scenario as in \eqref{eqn:4_imperfect}.
\vspace{-0.3cm}
\subsection{Solution Using MRT Precoding}
\label{sec:solution_imperfectcsit_MRT}
\vspace{-0.2cm}
We follow the same procedure as in Section~\ref{sec:solution_imperfectcsit_MRT} to obtain the rate and power allocation for MRT precoding. The explanations and derivations are identical to the those in Section~\ref{sec:solution_imperfectcsit_MRT}, with the terms in the expressions replaced by their counterparts for imperfect CSIT. Therefore, we skip the derivations and provide the resulting expressions directly for the sake of brevity.  
\subsubsection{Case 1}
\label{sec:imperfectcsit_MRT_highsnr}
This case captures the scenarios where the interference-limited regime is reached at high SNR, {\sl e.g.}, large $N$, small $K-N$, or large $\epsilon$. Under the assumption $Pt^{*}\rightarrow\infty$, the optimal rate and parameters for this scenario are written in \eqref{eqn:56_imperfect}, where $a_{1,i}=-\alpha^{\mathrm{MRT}}_{i}\lambda^{\mathrm{MRT}}_{i}$, $b_{1,i}=\left[K(\alpha^{\mathrm{MRT}}_{i}-\lambda^{\mathrm{MRT}}_{i})-(\alpha^{\mathrm{MRT}}_{i}+\lambda^{\mathrm{MRT}}_{i})\right]$, $\alpha^{\mathrm{MRT}}_{i}=v_{\hat{k}}\frac{P}{K}e^{\left(\psi(\theta^{\mathrm{MRT}}_{i})+\log(D^{\mathrm{MRT}}_{i})\right)}$, $\lambda^{\mathrm{MRT}}_{i}=v_{\hat{k}}\frac{P(K-1)}{K}$.
\begin{table*}[t!]
\begin{align} 
		&t^{(5/6)}_{i}\triangleq
		\begin{cases}
    		\min\left\lbrace 1,\frac{-b_{1,i}\pm\sqrt{b_{1,i}^{2}+4a_{1,i}}}{2a_{1,i}}\right\rbrace, & \hspace{-0.2cm}\text{if $\sqrt{b_{1,i}^{2}+4a_{1,i}} \geq 0$, $\frac{-b_{1,i}\pm\sqrt{b_{1,i}^{2}+4a_{1,i}}}{2a_{1,i}} \geq 0$} \\
    		\hspace{1.8cm}1\hspace{1.8cm}, &\hspace{-0.2cm} \text{otherwise}.
    	\end{cases}, \nonumber\\
		&r^{(5/6)}_{mm,p}\triangleq\frac{1}{K}\log_{2}\left(1-\rho^{\mathrm{MRT}}_{i}+\frac{\rho^{\mathrm{MRT}}_{i}}{t^{(5/6)}_{i}}\right)+\log_{2}\left(\frac{1+\alpha^{\mathrm{MRT}}_{i}t^{(5/6)}_{i}}{1+\lambda^{\mathrm{MRT}}_{i}t^{(5/6)}_{i}}\right).	
	\label{eqn:56_imperfect}
	\vspace{-0.5cm}
	\end{align}
\vspace{-0.2cm}
\hrule
\vspace{-0.6cm}
\end{table*}
\subsubsection{Case 2}
\label{sec:imperfectcsit_MRT_lowsnr}
Next, we consider the scenarios where the interference-limited regime is reached at moderate/low SNR, {\sl e.g.}, small $N$, large $K-N$, or small $\epsilon$. Following the derivations in Section~\ref{sec:imperfectcsit_MRT_lowsnr} with the expressions for imperfect CSIT, optimal rate and parameters for this scenario are written in \eqref{eqn:78_imperfect} with the parameters defined in \eqref{eqn:78_parameters}.
\begin{table*}[t!]
\begin{align} 
		&t^{(5/6)}_{i}\hspace{-0.08cm}\triangleq\hspace{-0.08cm}
		\begin{cases}
    		\min\left\lbrace1,\hspace{-0.08cm}\frac{-b_{2,i}\pm\sqrt{b_{2,i}^{2}-4a_{2,i}c_{2,i}}}{2a_{2,i}}\right\rbrace, & \text{if $\sqrt{b_{2,i}^{2}-4a_{2,i}c_{2,i}} \geq 0$, $\frac{-b_{2,i}\pm\sqrt{b_{2,i}^{2}-4a_{2,i}c_{2,i}}}{2a_{2,i}} \geq 0$} \\
    		\hspace{2.0cm}1\hspace{2.0cm}, &\hspace{-0.2cm} \text{otherwise}.
    	\end{cases},\nonumber \\ 
		&r^{(5/6)}_{mm,p}\hspace{-0.08cm}\triangleq\hspace{-0.08cm}\frac{1}{K}\log_{2}\hspace{-0.08cm}\left(\hspace{-0.08cm}1\hspace{-0.08cm}+\hspace{-0.08cm}\frac{PKe^{-\gamma}\left(1-t^{(5/6)}_{i}\right)}{P\theta^{\mathrm{MRT}}_{i}\lceil D^{\mathrm{MRT}}_{i}K\rfloor t^{(5/6)}_{i}+K\sum_{k=1}^{K}\frac{1}{v_{k}}}\hspace{-0.08cm}\right)\hspace{-0.08cm}+\hspace{-0.08cm}\log_{2}\hspace{-0.08cm}\left(\hspace{-0.08cm}\frac{1+\alpha^{\mathrm{MRT}}_{i}t^{(5/6)}_{i}}{1+\lambda^{\mathrm{MRT}}_{i}t^{(5/6)}_{p}}\hspace{-0.08cm}\right).	
	\label{eqn:78_imperfect}
	\vspace{-0.5cm}
\end{align}
\vspace{-0.2cm}
\hrule
\vspace{-0.5cm}
\end{table*}
\begin{table*}[t!]
\begin{align}
a_{2,i}&\triangleq (\alpha^{\mathrm{MRT}}_{i}-\lambda^{\mathrm{MRT}}_{i})\left[(P\theta^{\mathrm{MRT}}_{i}\lceil D^{\mathrm{MRT}}_{i}K\rfloor)^{2}-P^{2}\theta^{\mathrm{MRT}}_{i}\lceil D^{\mathrm{MRT}}_{i}K\rfloor Ke^{-\gamma}\right]-\omega_{i}\alpha^{\mathrm{MRT}}_{i}\lambda^{\mathrm{MRT}}_{i}, \nonumber\\
b_{2,i}&\triangleq (\alpha^{\mathrm{MRT}}_{i}-\lambda^{\mathrm{MRT}}_{i})PK  \biggl[\theta^{\mathrm{MRT}}_{i}\lceil D^{\mathrm{MRT}}_{i}K\rfloor e^{-\gamma} +2\theta^{\mathrm{MRT}}_{i}\lceil D^{\mathrm{MRT}}_{i}K\rfloor\sum_{k=1}^{K}\frac{1}{v_{k}}-Ke^{-\gamma}\sum_{k=1}^{K}\frac{1}{v_{k}}\biggr]-\omega_{i}(\alpha^{\mathrm{MRT}}_{i}+\lambda^{\mathrm{MRT}}_{i}), \nonumber\\
c_{2,i}&\triangleq (\alpha^{\mathrm{MRT}}_{i}-\lambda^{\mathrm{MRT}}_{i})K^{2}\sum_{k=1}^{K}\frac{1}{v_{k}}\left(Pe^{-\gamma}+\sum_{k=1}^{K}\frac{1}{v_{k}}\right)-\omega_{i},  \quad 
\omega_{i}\triangleq Pe^{-\gamma}\left(P\theta^{\mathrm{MRT}}_{i}\lceil D^{\mathrm{MRT}}_{i}K\rfloor+K\sum_{k=1}^{K}\frac{1}{v_{k}}\right).
\label{eqn:78_parameters}
\vspace{-0.5cm}
\end{align}
\vspace{-0.2cm}
\hrule
\vspace{-0.4cm}
\end{table*}

We note that \eqref{eqn:56_imperfect} and \eqref{eqn:78_imperfect} reduce to \eqref{eqn:45_perfect} and \eqref{eqn:67_perfect}, respectively, for $\epsilon^{2}=1$.
\vspace{-0.5cm}
\subsection{Proposed Precoder and Resource Allocation}
\label{sec:imperfectcsit_proposed}
\vspace{-0.1cm}
Finally, we describe the proposed power and precoder allocation for given parameters. The optimal power allocation $t_{i}$, rate allocation $\beta_{i}$, and precoder allocation for the private stream of user-$k$, $\mathbf{p}_{k,i}$, are performed as 
\begin{align}
	&\hat{n}=\argmax_{n \in \{1,2,\ldots,6\}}r^{(n)}_{mm,i}, \ \mathbf{p}^{*}_{k,i}= 
	\begin{cases}
	 	\mathbf{p}^{\mathrm{ZF}}_{k}\hspace{0.0cm}, & \hspace{-0.3cm}\text{if $\hat{n} \leq 4$ and $k \in \mathcal{G}_{1}$,} \\
	 	\hspace{0.45cm}\mathbf{0}\hspace{0.25cm}, & \hspace{-0.3cm}\text{if $\hat{n} \leq 4$ and $k \in \mathcal{G}_{2}$,} \\
	 	\mathbf{p}^{\mathrm{MRT}}_{k}\hspace{-0.2cm}, &\hspace{-0.3cm} \text{if $\hat{n} \geq 5$ and $k \in \mathcal{K}$,} \\
	\end{cases} \nonumber \\[-6pt]
	&t_{i}=t_{i}^{(\hat{n})}, \ \
	%&\beta_{i}=\begin{cases}
	% 	\max\left\lbrace\frac{1}{K}-\frac{(K-N)\log\left(\frac{\sigma^{\mathrm{ZF}}_{\tilde{k},i}}{\nu^{\mathrm{ZF}}_{\tilde{k},i}}\right)}{K\log\left(\nu^{\mathrm{ZF}}_{\tilde{k},i}\rho^{\mathrm{ZF}}_{i}\right)},0\right\rbrace\hspace{-0.15cm}, & \text{if $\hat{n} = 1$,} \\
	% 	\hspace{2.35cm}\varepsilon\frac{1}{K}\hspace{2.3cm}, & \text{if $\hat{n} = 2$ or ($\hat{n} = 3$ and $\sigma^{\mathrm{ZF}}_{\tilde{k},i}\rho^{\mathrm{ZF}}_{i} < 1$),} \\ 
	% 	\hspace{2.5cm}0\hspace{2.45cm}, & \text{if ($\hat{n} = 3$ and $\sigma^{\mathrm{ZF}}_{\tilde{k},i}\rho^{\mathrm{ZF}}_{i} > 1$) or $\hat{n} = 4$.} 
	%\end{cases}
	\beta_{i}=\begin{cases}
	 	\beta_{i}^{(\hat{n})}, &\hspace{-0.3cm} \text{if $\hat{n} \leq 4$,} \\
	 	\text{N/A}, &\hspace{-0.3cm} \text{otherwise.} 
	\end{cases}
	\label{eqn:proposed_imperfect}
\end{align}
As mentioned in the previous section, $\mathbf{p}_{c,i}$ can be calculated as the eigenvector corresponding to the largest eigenvalue of the combined channel matrix $\mathbf{H}=\left[\mathbf{h}_{1}, \mathbf{h}_{2}, \ldots, \mathbf{h}_{K}\right]$.
\vspace{-0.0cm}
\begin{figure}[t!]
	\begin{subfigure}{.5\textwidth}
		\centerline{\includegraphics[width=2.8in,height=2.8in,keepaspectratio]{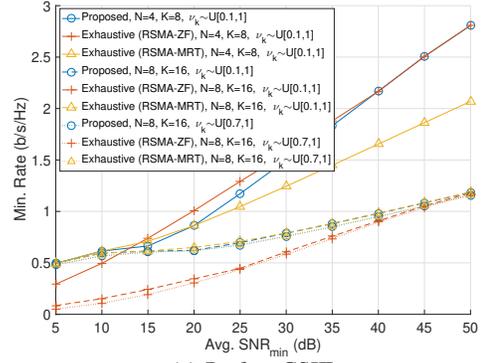}}
		\vspace{-0.2cm}
		\caption{Perfect CSIT.}
		\label{fig:PerfectCSIT_Exhaustive}
	\end{subfigure}
	\newline
	\begin{subfigure}{.5\textwidth}
		\centerline{\includegraphics[width=2.8in,height=2.8in,keepaspectratio]{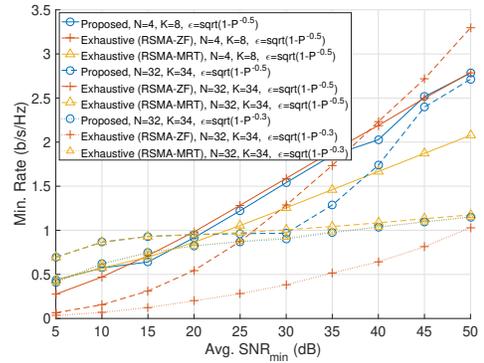}}
		\vspace{-0.2cm}
		\caption{Imperfect CSIT, $\nu_{k}\sim \mathrm{U}[0.1,1]$.}
		\label{fig:ImperfectCSIT_Exhaustive}
	\end{subfigure}
	\caption{Performance comparison with exhaustive search, $\lambda=2$.}
\end{figure}
\vspace{-0.2cm}
\section{Numerical Results}
\label{sec:numerical}
\vspace{-0.2cm}
In this section, we demonstrate the performance of the proposed scheme with numerical results. We perform Monte Carlo simulations over $100$ different channel realizations. We consider two types of CSIT error in our simulations. The first type is scaling error,  {\sl i.e.}, $\epsilon^{2}=1-P^{-\tau}$ \cite{mao_2018, mishra_2022_3}. This model represents the scenarios where the CSIT quality improves with increasing SNR, such as, CSIT quality depending on channel estimation errors. The coefficient $\tau$ determines the rate of change in CSIT quality with SNR. The second type is non-scaling error, {\sl i.e.}, $\epsilon^{2}$ is constant for varying SNR values. Such a model aims to model the CSIT error due to factors which prevent improvement in CSIT quality with increasing SNR, such as, mobility and network latency \cite{dizdar_2021, zhang_2009, papazafeiropoulos_2015, truong_2013}, and pilot contamination (which causes CSIT quality saturation with increasing uplink SNR) \cite{mishra_2022_5, papazafeiropoulos_2015}.    
\begin{figure*}[t!]
	\begin{subfigure}{.5\textwidth}
		\centerline{\includegraphics[width=2.8in,height=2.8in,keepaspectratio]{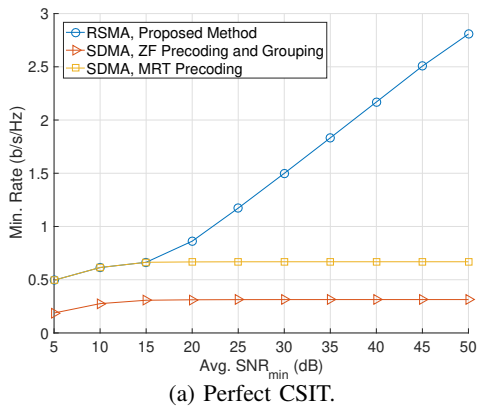}}
		\vspace{-0.2cm}
		\caption{Perfect CSIT.}
		\label{fig:N4K8_PerfectCSIT_Benchmarks}
	\end{subfigure}
	\begin{subfigure}{.5\textwidth}
		\centerline{\includegraphics[width=2.8in,height=2.8in,keepaspectratio]{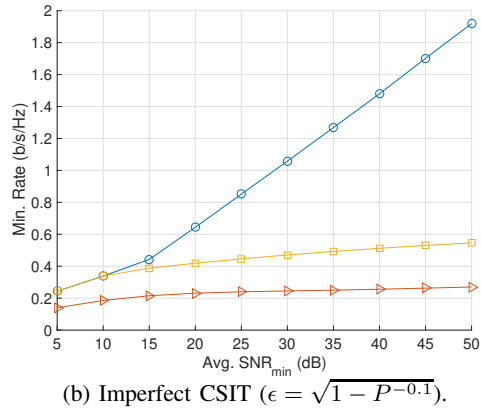}}
		\vspace{-0.2cm}
		\caption{Imperfect CSIT ($\epsilon=\sqrt{1-P^{-0.1}}$).}
		\label{fig:N4K8_ImperfectCSIT_ScalingAlpha01_Benchmarks}
	\end{subfigure}
	\newline
	\begin{subfigure}{.5\textwidth}
		\centerline{\includegraphics[width=2.8in,height=2.8in,keepaspectratio]{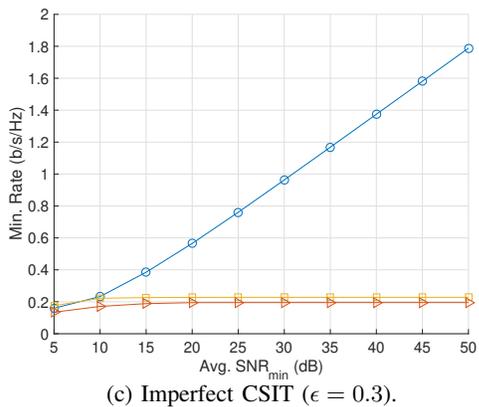}}
		\vspace{-0.2cm}
		\caption{Imperfect CSIT ($\epsilon=0.3$).}
		\label{fig:N4K8_ImperfectCSIT_Epssqare01_Benchmarks}
	\end{subfigure}
	\begin{subfigure}{.5\textwidth}
		\centerline{\includegraphics[width=2.8in,height=2.8in,keepaspectratio]{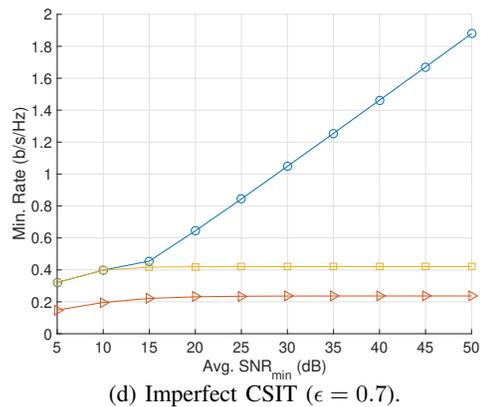}}
		\vspace{-0.2cm}
		\caption{Imperfect CSIT ($\epsilon=0.7$).}
		\label{fig:N4K8_ImperfectCSIT_Epssqare05_Benchmarks}
	\end{subfigure}
	\vspace{-0.5cm}
	\caption{Performance comparison with benchmark schemes, $N=4$, $K=8$, $\lambda=0.3$.}
	\vspace{-0.5cm}
\end{figure*}

\begin{figure*}[t!]
	\begin{subfigure}{.5\textwidth}
		\centerline{\includegraphics[width=2.8in,height=2.8in,keepaspectratio]{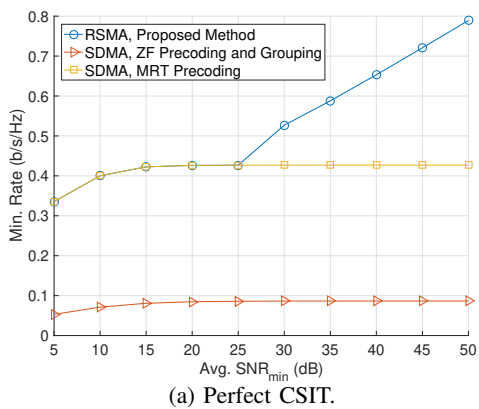}}
		\vspace{-0.2cm}
		\caption{Perfect CSIT.}
		\label{fig:N8K24_PerfectCSIT_Benchmarks}
	\end{subfigure}
	\begin{subfigure}{.5\textwidth}
		\centerline{\includegraphics[width=2.8in,height=2.8in,keepaspectratio]{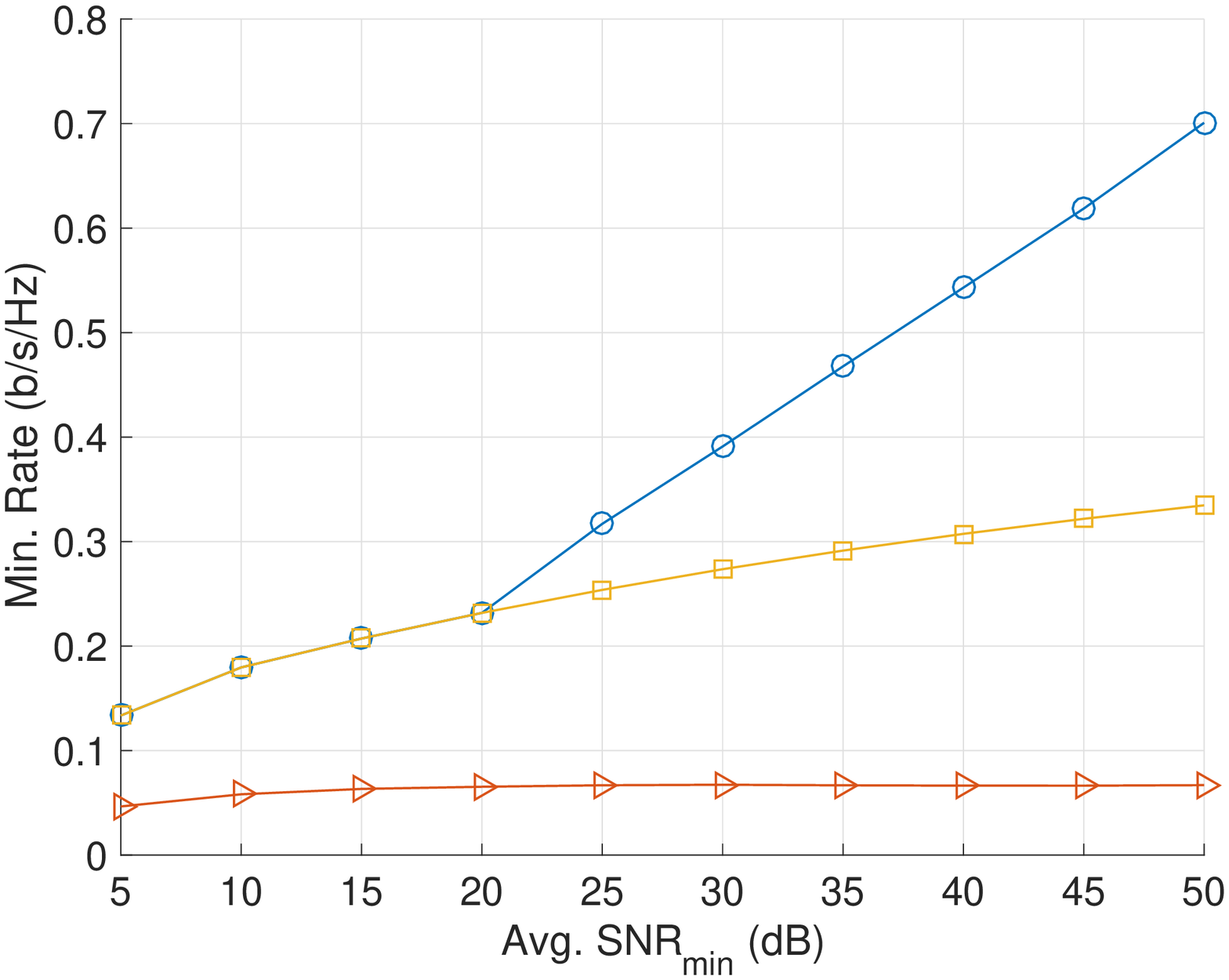}}
		\vspace{-0.2cm}
		\caption{Imperfect CSIT ($\epsilon=\sqrt{1-P^{-0.1}}$).}
		\label{fig:N8K24_ImperfectCSIT_ScalingAlpha01_Benchmarks}
	\end{subfigure}
	\newline
	\begin{subfigure}{.5\textwidth}
		\centerline{\includegraphics[width=2.8in,height=2.8in,keepaspectratio]{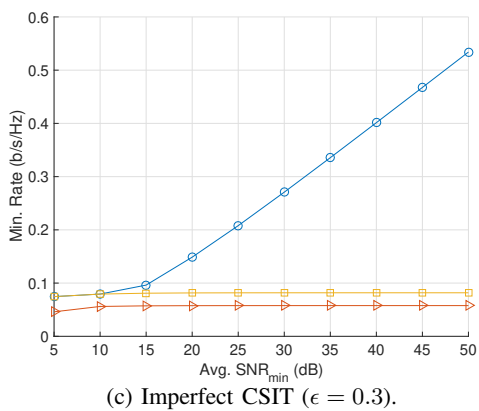}}
		\vspace{-0.2cm}
		\caption{Imperfect CSIT ($\epsilon=0.3$).}
		\label{fig:N8K24_ImperfectCSIT_Epssqare01_Benchmarks}
	\end{subfigure}
	\begin{subfigure}{.5\textwidth}
		\centerline{\includegraphics[width=2.8in,height=2.8in,keepaspectratio]{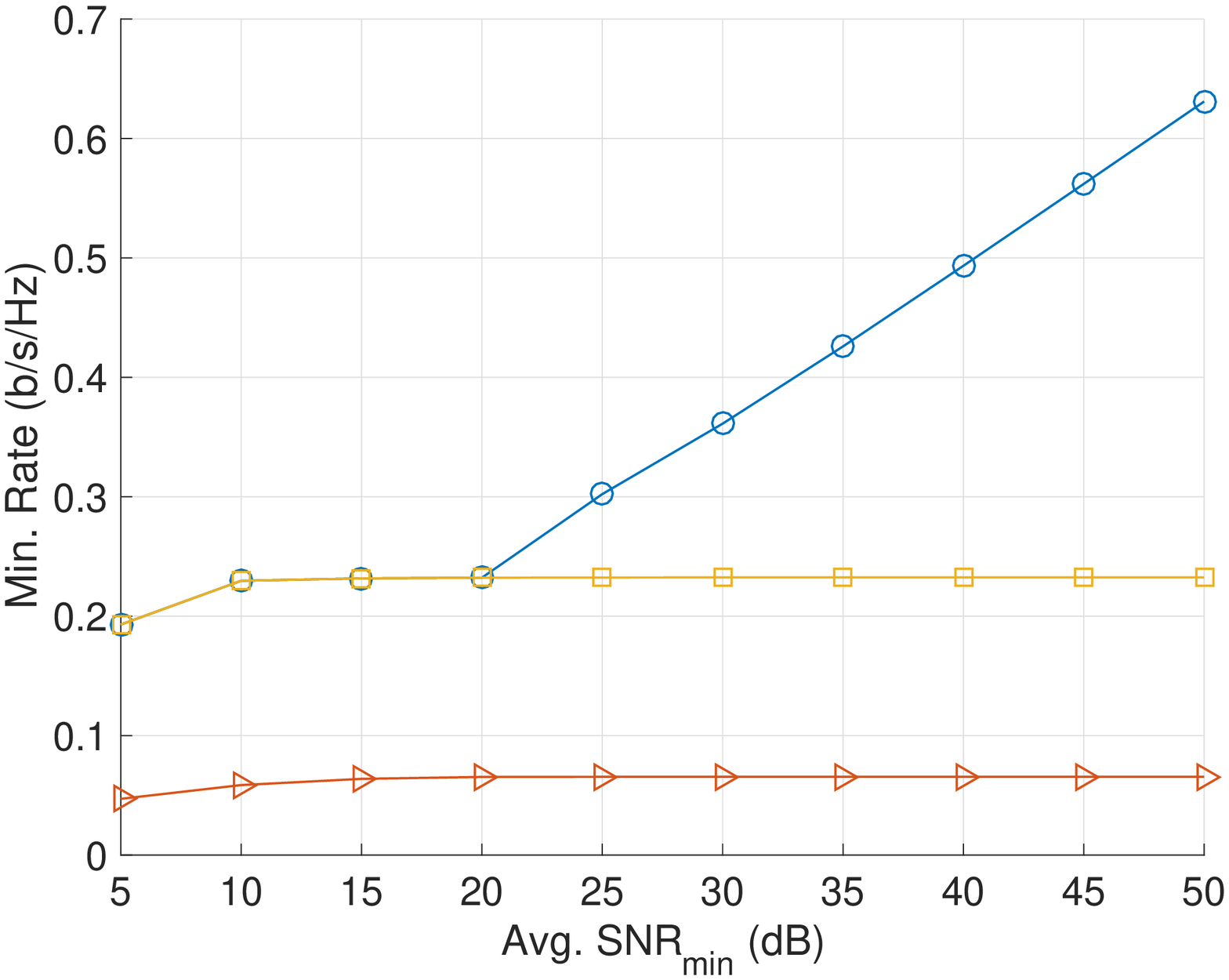}}
		\vspace{-0.2cm}
		\caption{Imperfect CSIT ($\epsilon=0.7$).}
		\label{fig:N8K24_ImperfectCSIT_Epssqare05_Benchmarks}
	\end{subfigure}
	\vspace{-0.5cm}
	\caption{Performance comparison with benchmark schemes, $N=8$, $K=24$, $\lambda=0.5/2$.}
	\vspace{-0.0cm}
\end{figure*}

\begin{figure*}[t!]
	\begin{subfigure}{.5\textwidth}
		\centerline{\includegraphics[width=2.8in,height=2.8in,keepaspectratio]{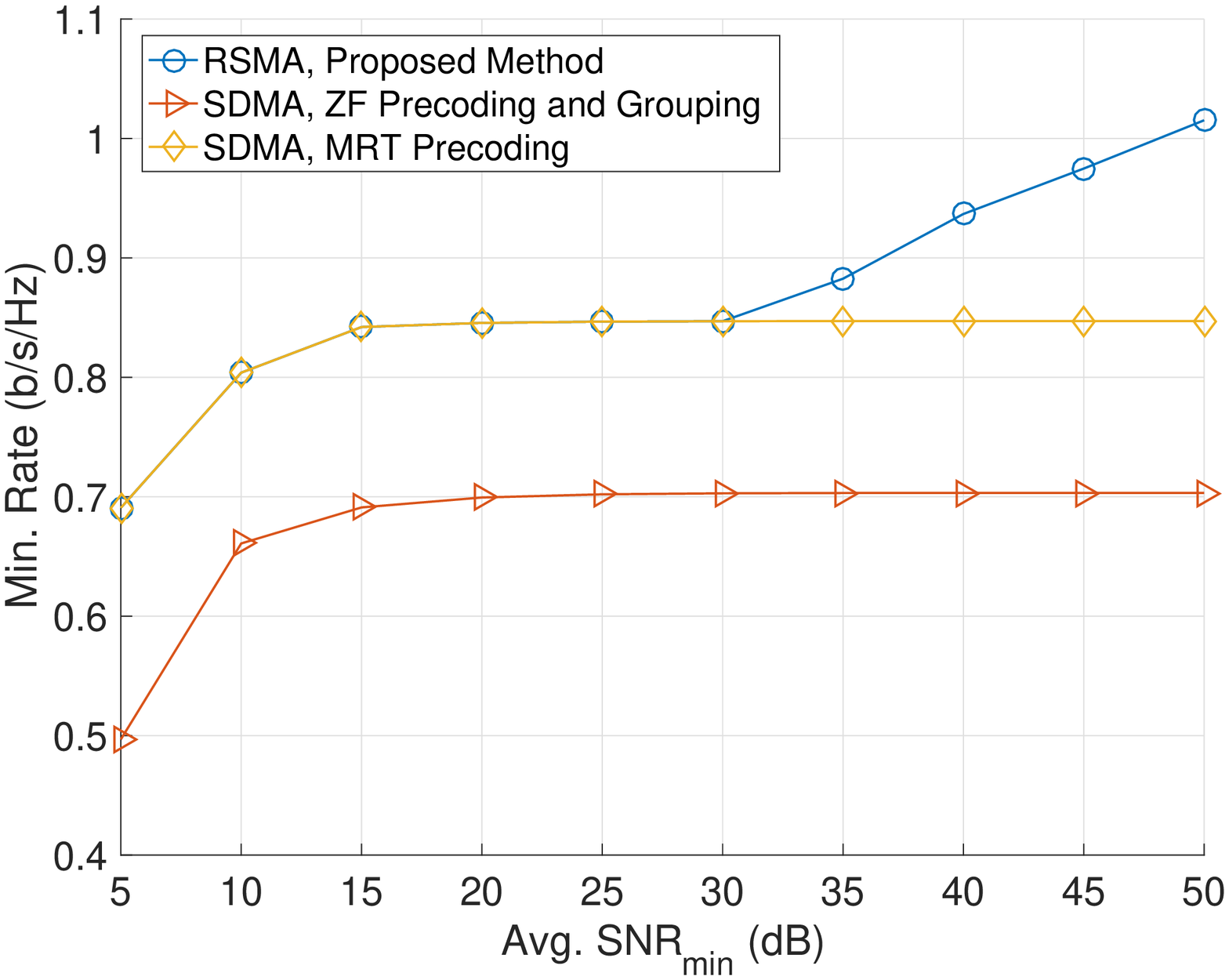}}
		\vspace{-0.2cm}
		\caption{Perfect CSIT.}
		\label{fig:N32K40_PerfectCSIT_Benchmarks}
	\end{subfigure}
	\begin{subfigure}{.5\textwidth}
		\centerline{\includegraphics[width=2.8in,height=2.8in,keepaspectratio]{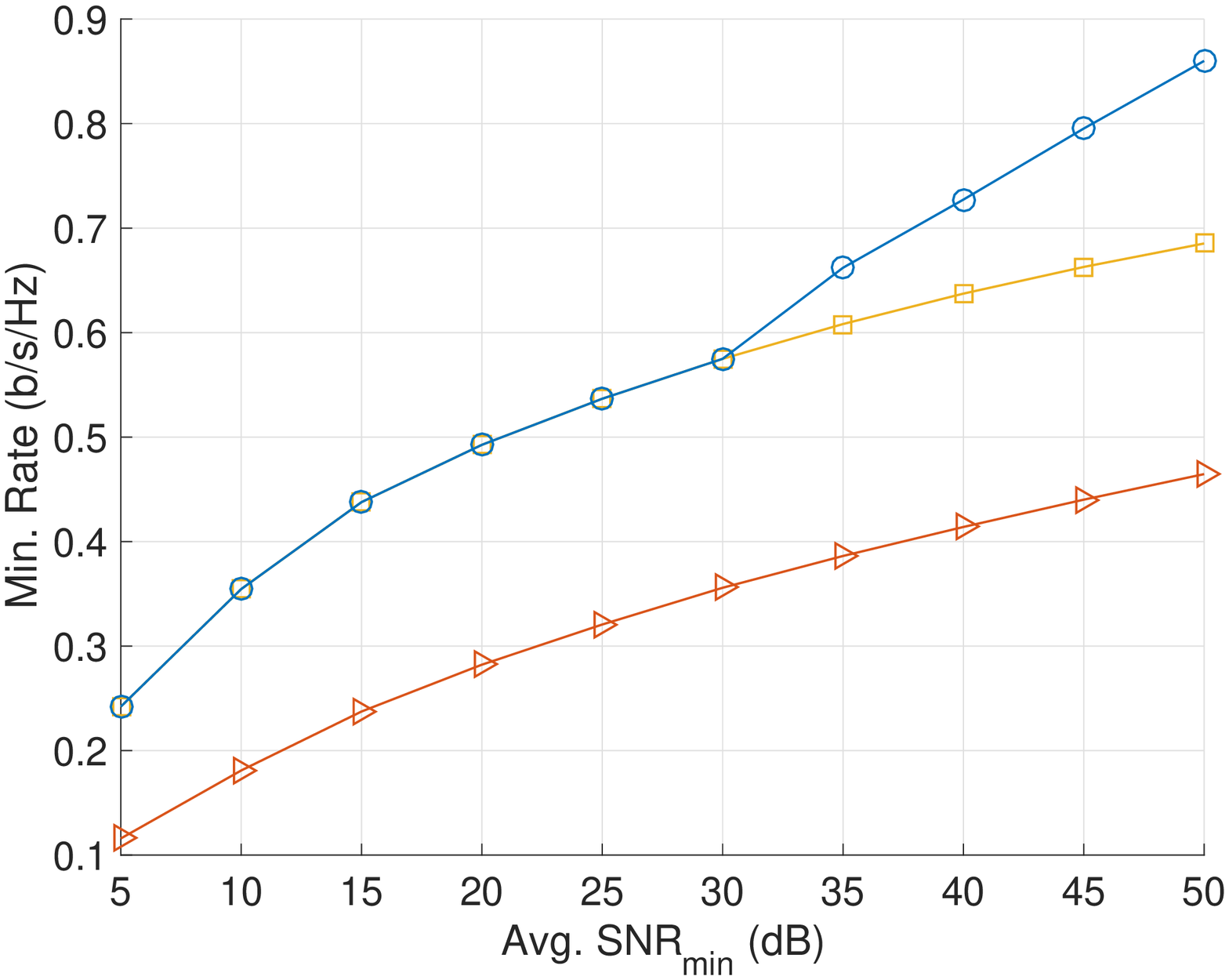}}
		\vspace{-0.2cm}
		\caption{Imperfect CSIT ($\epsilon=\sqrt{1-P^{-0.1}}$).}
		\label{fig:N32K40_ImperfectCSIT_ScalingAlpha01_Benchmarks}
	\end{subfigure}
	\newline
	\begin{subfigure}{.5\textwidth}
		\centerline{\includegraphics[width=2.8in,height=2.8in,keepaspectratio]{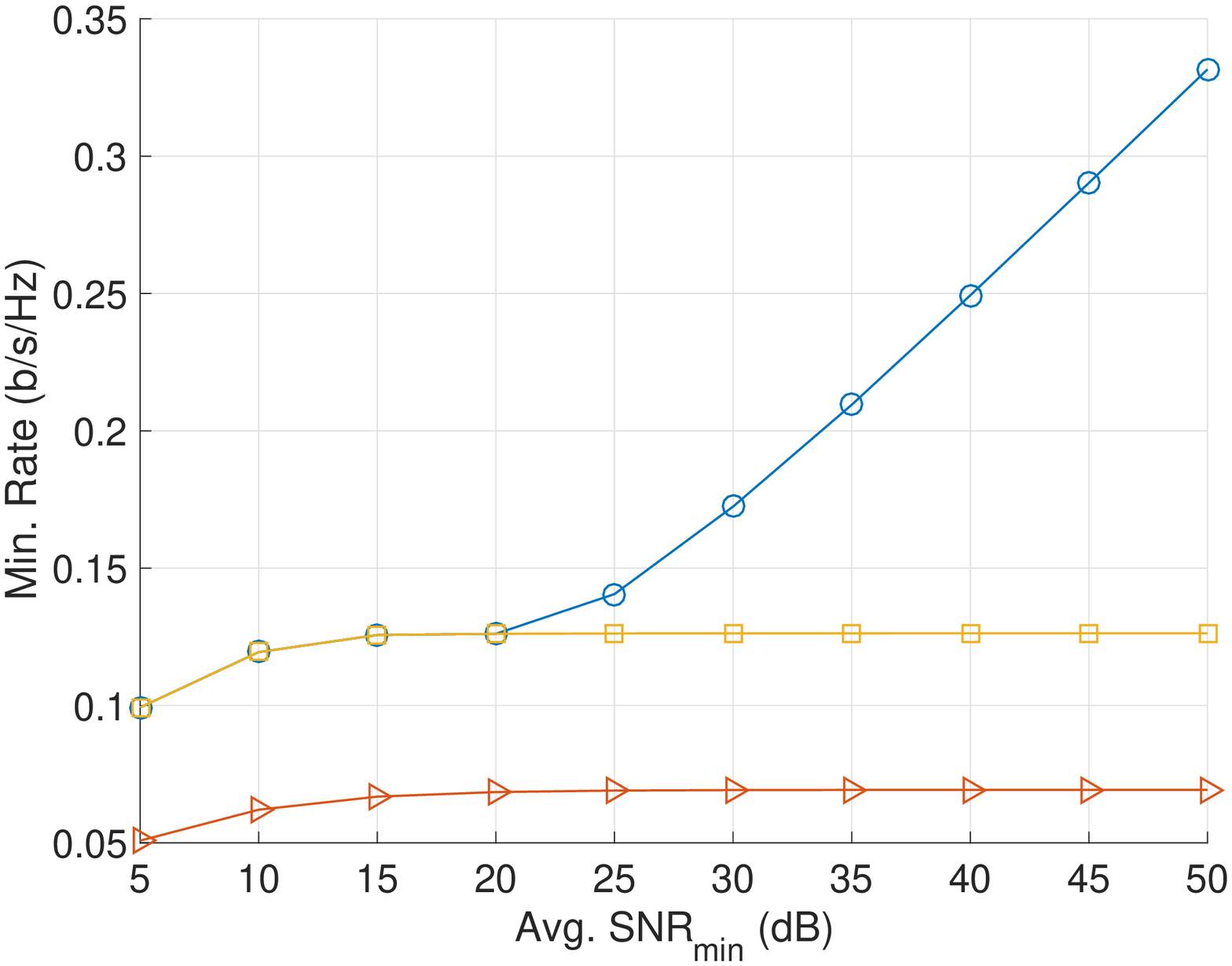}}
		\vspace{-0.2cm}
		\caption{Imperfect CSIT ($\epsilon=0.3$).}
		\label{fig:N32K40_ImperfectCSIT_Epssqare01_Benchmarks}
	\end{subfigure}
	\begin{subfigure}{.5\textwidth}
		\centerline{\includegraphics[width=2.8in,height=2.8in,keepaspectratio]{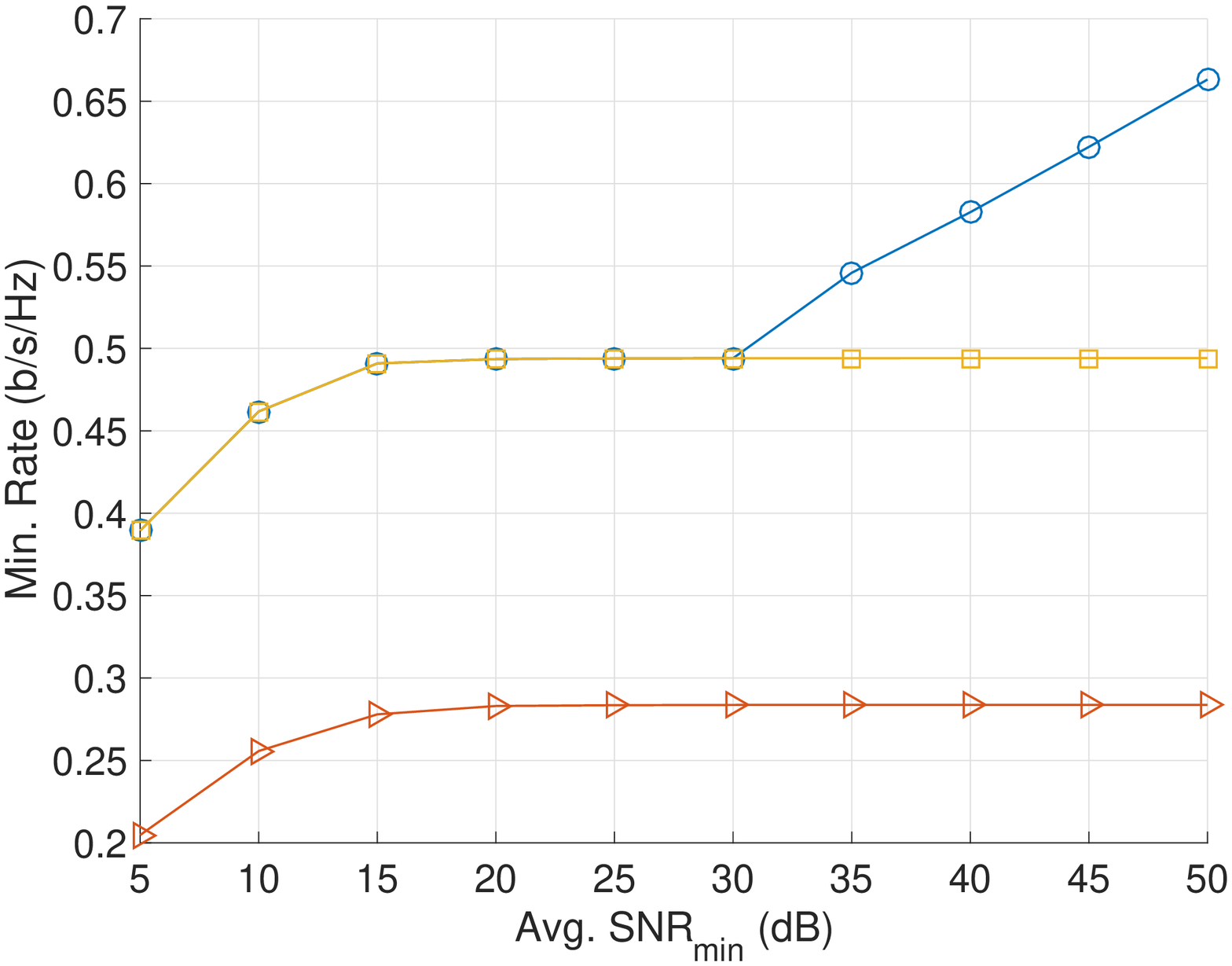}}
		\vspace{-0.2cm}
		\caption{Imperfect CSIT ($\epsilon=0.7$).}
		\label{fig:N32K40_ImperfectCSIT_Epssqare05_Benchmarks}
	\end{subfigure}
	\vspace{-0.5cm}
	\caption{Performance comparison with benchmark schemes, $N=32$, $K=40$, $\lambda=2$.}
	\vspace{-0.5cm}
\end{figure*}
First, we compare the performance of RSMA with the proposed scheme and RSMA using only ZF or MRT precoding with resource allocation by exhaustive search. The exhaustive search calculates the rate values $r^{(n)}_{mm,p}$ or $r^{(n)}_{mm,i}$ for all possible values of $t$ and $\beta$ with a certain resolution and choose the values that maximize the max-min rate. The search is performed over $t \in [10^{-6}, 1]$ and $\beta \in [0,1/K]$ with different resolutions in different intervals to reduce the complexity, so that $144$ values are considered for $t$ and $\lceil\frac{1}{0.001K}\rceil$ values are considered for $\beta$.

Fig.~\ref{fig:PerfectCSIT_Exhaustive} and \ref{fig:ImperfectCSIT_Exhaustive} show the performance comparison between the proposed method and exhaustive search with perfect and imperfect CSIT for various $N$, $K$, and $\epsilon$ values to investigate both small scale and massive MIMO. The x-axis of the figures represent the average SNR of the user with minimum large scale fading coefficient, {\sl i.e.}, $\min_{k \in \mathcal{K}}v_{k}P$. The values of $v_{k}$ are independently chosen by uniform distribution between $v_{min}$ and $1$, {\sl i.e.}, $v_{k}\sim \mathrm{U}[v_{min},1]$, with at least one user satisfying $v_{k}=v_{min}$ and one user satisfying $v_{k}=1$. Such a model aims to simulate randomly located users in the communication area with varying distances, with at least one user located at a maximum distance and one user close to the transmitter. For example, choosing $v_{min}=0.1$ simulates a scenario where the disparity between the maximum and minimum channel gains is $10$dB. The user channels $\mathbf{h}_{k}$ are independent r.v.s with Rayleigh distribution and unit variance.  We set $\varepsilon=0.98$. For simplicity, we determine $G_{1}$ for ZF precoding according to the user indexes, {\sl i.e.}, $\mathcal{G}_{1}={1, 2, \ldots, N}$. However, one should note that different grouping methods lead to different max-min rate performance. As one can observe from the figures, the proposed method performs very close to the optimal performance and can switch between ZF and MRT modes effectively. This verifies that the proposed method achieves close to optimal performance with low complexity. 

Next, we compare the performance of proposed design with benchmark schemes. First, we consider one-shot transmission schemes with no time-sharing as benchmarks. Such schemes are suitable for systems with tight scheduling and latency constraints \cite{yin_2021}. We use the following two schemes as benchmark:
\begin{enumerate}
	%\item \textit{SDMA with ZF precoding and distinct grouping:} Each user determines a distinct group of users randomly and calculates the ZF precoder from the channel estimates of the users in that group. 
	\item \textit{SDMA with ZF precoding and grouping:} Users are divided into $\lceil\frac{K}{N}\rceil$ groups with user index sets denoted by $\mathcal{K}_{i}$, $0 \leq i \leq \lceil\frac{K}{N}\rceil$, where each group consists of $K/\lceil\frac{K}{N}\rceil$ users. The users are grouped according to their user indexes, {\sl i.e.}, $\mathcal{K}_{1}=1,\ldots,K/\lceil\frac{K}{N}\rceil$, $\mathcal{K}_{2}=K/\lceil\frac{K}{N}\rceil+1,\ldots,2K/\lceil\frac{K}{N}\rceil$, and so on. Each user calculates the ZF precoder from the channel estimates of the other users in its own group. 
	\item \textit{SDMA with MRT precoding:} Users are served by MRT precoding.  
\end{enumerate}

Fig.~\ref{fig:N4K8_PerfectCSIT_Benchmarks}-\ref{fig:N32K40_ImperfectCSIT_Epssqare05_Benchmarks} show the performance comparison between the proposed method and the benchmark schemes with perfect and imperfect CSIT for various $N$, $K$, and $\epsilon$. For $N<64$ with MRT precoding, we employ the allocations in Case 2. For $N=64$, we employ the allocations in Case 1 for average $\mathrm{SNR}_{min}>25$dB. From the figures, one can observe the rate saturation for SDMA with respect to SNR due to multi-user interference. The saturation region cannot be observed for $\epsilon^{2}=1-P^{-0.1}$, however, it is evident that the saturation will occur at higher SNR since it occurs even for perfect CSIT. The rate saturation occurs since SDMA suffers from multiuser interference in both benchmark schemes in both perfect and imperfect CSIT cases. More specifically, SDMA with ZF precoding and grouping suffers from multiuser interference even in perfect CSIT case, since each user cancels the interference to other users in its own group but does not cancel the interference to the users in other groups. On the other hand, SDMA with MRT precoding does not attempt to cancel any interference at all, and thus, each user suffers from multiuser intererence from all other users. Additionally, multiuser interference becomes more severe under imperfect CSIT, where users with ZF precoding cannot cancel the interference fully even to the other users in their groups. Consequently, the achievable rate for SDMA is limited by the multiuser interference and saturates with increasing SNR.

On the other hand, RSMA achieves a non-saturating rate for every considered scenario. The gain achieved by RSMA varies depending on $N$ and $\epsilon$. As $N$ increases, the achieved gain becomes smaller as CSIT quality ($\epsilon$) increases. However, it is expected that improving CSIT quality becomes more challenging as $N$ increases (due to factors, such as, CSI feedback overhead and pilot contamination) in practice. Therefore, we can conclude that RSMA achieves a significant gain for any $N$, $K$, and $\epsilon$ value in practical overloaded MIMO systems.  

\begin{figure}[t!]
	\begin{subfigure}{.5\textwidth}
		\centerline{\includegraphics[width=2.8in,height=2.8in,keepaspectratio]{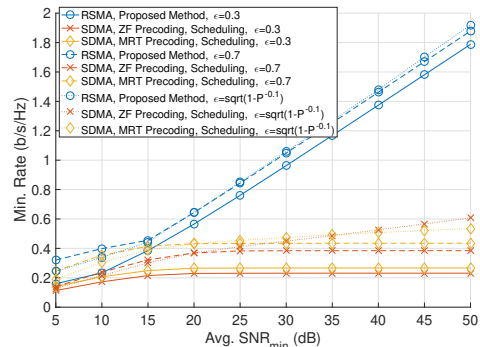}}
		\vspace{-0.2cm}
		\caption{$N=4$, $K=8$.}
		\label{fig:N4K8_ImperfectCSIT_AllEpssquare_SchedulingBenchmarks}
	\end{subfigure}
	\newline
	\begin{subfigure}{.5\textwidth}
		\centerline{\includegraphics[width=2.8in,height=2.8in,keepaspectratio]{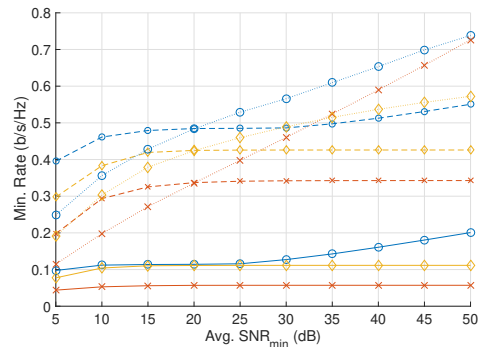}}
		\vspace{-0.2cm}
		\caption{$N=64$, $K=80$.}
		\label{fig:N64K80_ImperfectCSIT_AllEpssquare_SchedulingBenchmarks}
	\end{subfigure}
	\vspace{-0.5cm}
	\caption{Performance comparison with benchmark schemes with scheduling, $\lambda=0.3$.}
	\vspace{-0.5cm}
\end{figure}

Finally, we compare the performance of RSMA in overloaded networks with SDMA when users are grouped into $\lceil\frac{K}{N}\rceil$ groups and each group is scheduled in a different time slot. Consequently, the SDMA system becomes underloaded in each time slot. The user group in a particular time slot is served by ZF or MRT precoders.  We refer to this scheme as SDMA with scheduling. Fig.~\ref{fig:N4K8_ImperfectCSIT_AllEpssquare_SchedulingBenchmarks} and \ref{fig:N64K80_ImperfectCSIT_AllEpssquare_SchedulingBenchmarks} demonstrate the performance comparison of overloaded RSMA and SDMA with scheduling for imperfect CSIT with $N=4$, $K=8$ and $N=64$, $K=80$, respectively. The users are scheduled according to their user indexes in order to limit the computational complexity of SDMA with scheduling. The figures clearly show that the performance of RSMA even surpass that of SDMA with user  scheduling under imperfect CSIT in overloaded networks.  
 
\vspace{-0.5cm}
\section{Conclusion}
\label{sec:conclusion}
\vspace{-0.2cm}
In this work, we study RSMA in overloaded MIMO networks with perfect and imperfect CSIT for downlink multiple-access. We formulate a max-min fairness problem to find optimal precoder and resource allocation scheme. We provide a low-complexity solution for the formulated problem by using ZF and MRT precoders and deriving closed-form solutions for power and rate allocation between the common and private streams. By comparing its achievable rate with the rate obtained by exhaustive search, we demonstrate that the proposed low-complexity design achieves near-optimal performance. Numerical results show that the proposed low-complexity RSMA design achieves significantly higher rate than SDMA with ZF and MRT precoding when transmission is performed in a single time slot under perfect and imperfect CSIT. Furthermore, the proposed design achieves a significantly higher rate than SDMA under imperfect CSIT even when users are grouped and scheduled in different time slots. Future work includes the design of a low-complexity precoder and resource allocation scheme for RSMA in overloaded networks with mmWave and Rician fading channels, and considering the impact of imperfect SIC in the system design. 

\appendices
\vspace{-0.3cm}
\section{Proof of Proposition 1} 
\label{appendix:prop1}
We start the proof by defining a lower bound for $R^{\mathrm{MRT}}_{c,i}(t)$. Let us define an approximation $\widetilde{R}^{\mathrm{MRT}}_{c,i}(t)$ for $R^{\mathrm{MRT}}_{c,i}(t)$. Then, it can be shown that the following inequality holds.
\begin{align}
	R^{\mathrm{MRT}}_{c,i}(t)\approx\widetilde{R}^{\mathrm{MRT}}_{c,i}(t) \geq \log_{2}\left(1+P(1-t)e^{\eta^{\mathrm{MRT}}_{i}}\right),
	\label{eqn:prop11}
\end{align}
where $\eta^{\mathrm{MRT}}_{i}$ is defined as in \eqref{eqn:terms_i}.
The inequality \eqref{eqn:prop3} can be proved by using \cite[Proposition 2]{dizdar_2021} and taking into account the properties of MRT precoders stated in Section~\ref{sec:boundRate}, and is omitted here for the sake of brevity. Then, we use the upper bound $e^{x}E_{v}\left(x\right) \leq \frac{1}{v+x-1}$ \cite{chiccoli_1990} to obtain
\vspace{-0.1cm}
\begin{align}
	&e^{\frac{K\sum_{k=1}^{K}\frac{1}{v_{k}}}{P\theta^{\mathrm{MRT}}_{i}t}}\sum_{m=1}^{\lfloor D^{\mathrm{MRT}}_{i}K\rceil}E_{m}\left(\frac{K\sum_{k=1}^{K}\frac{1}{v_{k}}}{P\theta^{\mathrm{MRT}}_{i}t}\right) \nonumber \\
	&\leq \sum_{m=1}^{\lfloor D^{\mathrm{MRT}}_{i}K\rceil} \frac{1}{m+\frac{K\sum_{k=1}^{K}\frac{1}{v_{k}}}{P\theta^{\mathrm{MRT}}_{i}t}-1}=\hspace{-1.1cm}\sum_{n=\frac{K\sum_{k=1}^{K}\frac{1}{v_{k}}}{P\theta^{\mathrm{MRT}}_{i}t}}^{\lfloor D^{\mathrm{MRT}}_{i}K\rceil+\frac{K\sum_{k=1}^{K}\frac{1}{v_{k}}}{P\theta^{\mathrm{MRT}}_{i}t}-1}\hspace{-1.3cm}1/n \triangleq \tau^{\mathrm{MRT}}_{i}.\vspace{-0.1cm} \nonumber
\end{align}
Substituting $\tau^{\mathrm{MRT}}_{i}$ in \eqref{eqn:prop3}, we obtain
\vspace{-0.1cm}
\begin{align}
&\widetilde{R}^{\mathrm{MRT}}_{c,i}(t) \geq \log_{2}\left(1+P(1-t)e^{\eta^{\mathrm{MRT}}_{i}}\right) \nonumber \\[-6pt] 
&\geq \log_{2}\left(1+P(1-t)e^{\left( -\gamma-\log\left(\sum_{k=1}^{K}\frac{1}{v_{k}}\right)-\tau^{\mathrm{MRT}}_{i}\right)}\right).
\label{eqn:prop1_proof}
\end{align} 
\vspace{-0.1cm}
Using the approximation $\sum_{m=1}^{M}1/m\approx \log(M)+\gamma$, we approximate $\tau^{\mathrm{MRT}}_{i}$ by 
\begin{align}
	%\tau_{upper} \hspace{0.1cm}\approx \log\left(\lfloor N(K-N+D^{\mathrm{ZF}}_{i})\rceil+\frac{N}{Pt}\sum_{k=1}^{K}\frac{1}{v_{k}}\right)-\log\left(\frac{N}{Pt}\sum_{k=1}^{K}\frac{1}{v_{k}}\right), \\
	\hspace{-0.1cm}\tau^{\mathrm{MRT}}_{i} \hspace{-0.1cm}\approx\hspace{-0.1cm} \log\hspace{-0.1cm}\left(\hspace{-0.1cm}\lfloor D^{\mathrm{MRT}}_{i}K\rceil\hspace{-0.1cm}+\hspace{-0.1cm}\frac{K\hspace{-0.1cm}\sum_{k=1}^{K}\hspace{-0.1cm}\frac{1}{v_{k}}}{P\theta^{\mathrm{MRT}}_{i}t}\hspace{-0.1cm}\right)\hspace{-0.1cm}-\hspace{-0.1cm}\log\hspace{-0.1cm}\left(\hspace{-0.1cm}\frac{K\hspace{-0.1cm}\sum_{k=1}^{K}\hspace{-0.1cm}\frac{1}{v_{k}}}{P\theta^{\mathrm{MRT}}_{i}t}\hspace{-0.1cm}\right)\hspace{-0.1cm},
	\label{eqn:expint_approx1}
\end{align}
for $\lfloor D^{\mathrm{MRT}}_{i}K\rceil+\frac{K\sum_{k=1}^{K}\frac{1}{v_{k}}}{P\theta^{\mathrm{MRT}}_{i}t} \gg\hspace{-0.12cm} 1$ and $\frac{K\sum_{k=1}^{K}\frac{1}{v_{k}}}{P\theta^{\mathrm{MRT}}_{i}t} \geq\hspace{-0.1cm} 1$. Substituting \eqref{eqn:expint_approx1} into \eqref{eqn:prop1_proof} completes the proof. 

\section{Proof of Proposition 4}
\label{appendix:prop4}
\vspace{-0.2cm}
We start by deriving the CDF of $Y^{\mathrm{ZF}}_{i}$. Using \cite[Lemma 2]{dizdar_2021} and \eqref{eqn:rvs}, $Y^{\mathrm{ZF}}_{i}$ can be approximated by a r.v. $\widetilde{Y}^{\mathrm{ZF}}_{i}$ with CDF\vspace{-0.2cm}
\begin{align}
F_{\widetilde{Y}^{\mathrm{ZF}}_{i}}(y)\hspace{-0.1cm}=\hspace{-0.1cm}
	1\hspace{-0.1cm}-\hspace{-0.1cm}\frac{e^{-y\sum_{k=1}^{N}\frac{1}{v_{k}}}}{\left(y\frac{P\theta_{i}^{\mathrm{ZF}}t}{N}+1 \right)^{ND_{i}^{\mathrm{ZF}}}}\frac{e^{-y\sum_{k=N+1}^{K}\frac{1}{v_{k}}}}{\left(y\frac{Pt}{N}+1 \right)^{N(K-N)}}, 
\end{align}
where $D^{\mathrm{ZF}}_{i}\triangleq\frac{\left(\epsilon^{2}(1-N)+N\right)^{2}}{\epsilon^{4}(1+N)+(1-2\epsilon^{2})N}$ and $\theta^{\mathrm{ZF}}_{i}\triangleq\frac{\epsilon^{4}(1+N)+(1-2\epsilon^{2})N}{\epsilon^{2}(1-N)+N}$.
It is straightforward to show $\theta^{\mathrm{ZF}}_{i} \leq 1$.
Next, let us define a new r.v. $\widehat{Y}^{\mathrm{ZF}}_{i}$ with the CDF\vspace{-0.2cm}
\begin{align}
F_{\widehat{Y}^{\mathrm{ZF}}_{i}}(y)=
	1-\frac{e^{-y\sum_{k=1}^{K}\frac{1}{v_{k}}}}{\left(y\frac{Pt}{N}+1 \right)^{N(K-N+D_{i}^{\mathrm{ZF}})}}.
\end{align}
From the fact that $\theta^{\mathrm{ZF}}_{i} \leq 1$, the CDFs of the r.v.s $\widetilde{Y}^{\mathrm{ZF}}_{i}$ and $\widehat{Y}^{\mathrm{ZF}}_{i}$ satisfy $F_{\widehat{Y}^{\mathrm{ZF}}_{i}}(y)\geq F_{\widetilde{Y}^{\mathrm{ZF}}_{i}}(y)$ in the support set $[0,\infty )$. Consequently, we can write $\mathbb{E}\{\widehat{Y}^{\mathrm{ZF}}_{i}\} \leq \mathbb{E}\{\widetilde{Y}^{\mathrm{ZF}}_{i}\}$ and $\mathbb{E}\{\log(\widehat{Y}^{\mathrm{ZF}}_{i})\} \leq \mathbb{E}\{\log(\widetilde{Y}^{\mathrm{ZF}}_{i})\}$ \cite[Lemma 4]{hao_2015}.

Let us define the rate $\widetilde{R}^{\mathrm{ZF}}_{c,i}(t)=\log_{2}\left( 1+P(1-t) \widetilde{Y}^{\mathrm{ZF}}_{i}\right)$. Then, we can write the following: \vspace{-0.2cm}
\begin{align}
	R^{\mathrm{ZF}}_{c,i}(t)\hspace{-0.1cm}\approx\hspace{-0.1cm}\widetilde{R}^{\mathrm{ZF}}_{c,i}(t) \geq \log_{2}\left(1+P(1-t)e^{\mathbb{E}\left\lbrace\log\left(\widetilde{Y}^{\mathrm{ZF}}_{i}\right)\right\rbrace}\right).
	\label{eqn:prop1_proof1}
\end{align}
The inequality in \eqref{eqn:prop1_proof1} follows from Jensen's inequality and the fact that $\log_{2}(1+ae^{x})$ is a convex function of $x$ for any $a>0$. Recalling that $\mathbb{E}\{\log(\widehat{Y}^{\mathrm{ZF}}_{i})\} \leq \mathbb{E}\{\log(\widetilde{Y}^{\mathrm{ZF}}_{i})\}$, we can write \vspace{-0.2cm}
\begin{align}
	\widetilde{R}^{\mathrm{ZF}}_{c,i}(t) &\geq \log_{2}\left(1+P(1-t)e^{\mathbb{E}\left\lbrace\log\left(\widetilde{Y}^{\mathrm{ZF}}_{i}\right)\right\rbrace}\right)\nonumber\\
	&\geq \log_{2}\left(1+P(1-t)e^{\mathbb{E}\left\lbrace\log\left(\widehat{Y}^{\mathrm{ZF}}_{i}\right)\right\rbrace}\right).
	\label{eqn:prop1_proof2}
\end{align}
Next, we benefit from \cite[Lemma 3]{dizdar_2021} to obtain \vspace{-0.2cm}
\begin{align}
&\mathbb{E}\left\lbrace\log\left(\widehat{Y}^{\mathrm{ZF}}_{i}\right)\right\rbrace=-\gamma-\log\left(\sum_{k=1}^{K}\frac{1}{v_{k}}\right)\nonumber \\
&-e^{\frac{N}{Pt}\sum_{k=1}^{K}\frac{1}{v_{k}}}\hspace{-0.2cm}\sum_{m=1}^{\lfloor N(K-N+D^{\mathrm{ZF}}_{i})\rceil}\hspace{-0.6cm}E_{m}\left(\frac{N}{Pt}\sum_{k=1}^{K}\frac{1}{v_{k}}\right)\triangleq\eta^{\mathrm{ZF}}_{i}. \label{eqn:prop1_proof3}
\end{align}
Finally, combining \eqref{eqn:prop1_proof1}, \eqref{eqn:prop1_proof2}, and \eqref{eqn:prop1_proof3} completes the proof.

\section{Proof of Proposition 6}
\label{appendix:propx}
We prove Proposition 6 by contradiction. Assume that $\beta^{*}$ and $t^{*}$ are the points that attain the maximum for problem \eqref{eqn:ZFproblem_2}, and thus, are unique \cite[Theorem 1.2.2]{boudec_2021}. 
Let us assume that $\beta^{*}$ and $t^{*}$ satisfy $\frac{1-N\beta^{*}}{K-N}R^{\mathrm{ZF}}_{c,p}(t^{*})<\beta^{*}R^{\mathrm{ZF}}_{c,p}(t^{*})+R^{\mathrm{ZF}}_{\tilde{k},p}(t^{*})$ for $0<\beta^{*}<1/N$ and $0\leq t^{*}<1$. Then, one can find $\beta^{\prime}=\beta^{*}-\epsilon$ for arbitrarily small $\epsilon>0$, such that, $\frac{1-N\beta^{\prime}}{K-N}R^{\mathrm{ZF}}_{c,p}(t^{*})>\frac{1-N\beta^{*}}{K-N}R^{\mathrm{ZF}}_{c,p}(t^{*})$ and $\frac{1-N\beta^{\prime}}{K-N}R^{\mathrm{ZF}}_{c,p}(t^{*})<\beta^{\prime}R^{\mathrm{ZF}}_{c,p}(t^{*})+R^{\mathrm{ZF}}_{\tilde{k},p}(t^{*})$, which contradicts the initial assumption.
Next, we assume that maximum is attained by $\beta^{*}=0$ and $t^{*}$ satisfying $\frac{1}{K-N}R^{\mathrm{ZF}}_{c,p}(t^{*})<R^{\mathrm{ZF}}_{\tilde{k},p}(t^{*})$ for $0<t^{*}<1$. We note that $t$ is strictly positive in this case to obtain a non-zero maximum value under this assumption. By observing the first derivatives, it can be shown that $R^{\mathrm{ZF}}_{c,p}(t)$ is monotonically decreasing and $R^{\mathrm{ZF}}_{k^{\prime}}(t)$ is monotonically increasing with $t$. Consequently, we can find $t^{\prime}=t^{*}-\epsilon$ for arbitrarily small $\epsilon>0$, such that, $\frac{1}{K-N}R^{\mathrm{ZF}}_{c,p}(t^{\prime})>\frac{1}{K-N}R^{\mathrm{ZF}}_{c,p}(t^{*})$ and $\frac{1}{K-N}R^{\mathrm{ZF}}_{c,p}(t^{\prime})<R^{\mathrm{ZF}}_{\tilde{k},p}(t^{\prime})$, which contradicts the initial assumption. 

In order to complete the proof, one can repeat the procedure above in a similar fashion for the scenarios where maximum is attained by $\beta^{*}$ and $t^{*}$ satisfying $\frac{1-N\beta^{*}}{K-N}R^{\mathrm{ZF}}_{c,p}(t^{*})>\beta^{*}R^{\mathrm{ZF}}_{c,p}(t^{*})+R^{\mathrm{ZF}}_{\tilde{k},p}(t^{*})$, which is omitted here for brevity.   
\vspace{-0.4cm}
\section{Proof of Lemma 3}
\label{appendix:lemmax} 
\vspace{-0.2cm}
We prove Lemma 3 by showing that the derivative of $\frac{1-N\beta}{K-N}R^{\mathrm{ZF}}_{c,p}(t^{(1)}_{p}(\beta))$ with respect to $\beta$, which is given in \eqref{eqn:lemmax_derivative}, is strictly negative. 
\begin{table*}
	\begin{align}
		\frac{-N}{K-N}\frac{1}{\log(2)}\log\hspace{-0.05cm}\left(\hspace{-0.05cm}1\hspace{-0.05cm}-\hspace{-0.05cm}\rho_{p}\hspace{-0.05cm}+\hspace{-0.05cm}(\rho_{p}\sigma^{\mathrm{ZF}}_{\tilde{k},p})^{\frac{K-N}{1-K\beta+K-N}}\hspace{-0.05cm}\right)\hspace{-0.05cm}+\hspace{-0.05cm}\frac{1-N\beta}{K-N}\frac{1}{\log(2)}\frac{(\rho_{p}\sigma^{\mathrm{ZF}}_{\tilde{k},p})^{\frac{K-N}{1-K\beta+K-N}}\log(\rho_{p}\sigma^{\mathrm{ZF}}_{\tilde{k},p})\frac{N(K-N)}{(1-K\beta+K-N)^{2}}}{1-\rho_{p}+\left(\rho_{p}\sigma^{\mathrm{ZF}}_{\tilde{k},p}\right)^{\frac{K-N}{1-K\beta+K-N}}}.
		\label{eqn:lemmax_derivative} 
	\end{align}
	\vspace{-0.2cm}
	\hrule
	\vspace{-0.5cm}
\end{table*}
One can see that the first term in \eqref{eqn:lemmax_derivative} is strictly negative for $(\rho_{p}\sigma^{\mathrm{ZF}}_{\tilde{k},p})^{\frac{K-N}{1-K\beta+K-N}}>\rho_{p}$, which is satisfied for $\sigma^{\mathrm{ZF}}_{\tilde{k},p}>1$.
The second term is non-positive for $\sigma^{\mathrm{ZF}}_{\tilde{k},p}\rho_{p}\leq1$, which results in the overall derivative to be negative. For the case $\sigma^{\mathrm{ZF}}_{\tilde{k},p}\rho_{p}>1$, we need to show that $N\log\left(1-\rho_{p}+\left(\rho_{p}\sigma^{\mathrm{ZF}}_{\tilde{k},p}\right)^{\frac{K-N}{1-K\beta+K-N}}\right) > \frac{(1-N\beta)\left(\rho_{p}\sigma^{\mathrm{ZF}}_{\tilde{k},p}\right)^{\frac{K-N}{1-K\beta+K-N}}\log\left(\rho_{p}\sigma^{\mathrm{ZF}}_{\tilde{k},p}\right)\frac{(K-N)}{(1-K\beta+K-N)^{2}}}{1-\rho_{p}+\left(\rho_{p}\sigma^{\mathrm{ZF}}_{\tilde{k},p}\right)^{\frac{K-N}{1-K\beta+K-N}}}$. For this purpose, we rearrange the inequality as in \eqref{eqn:lemmax_derivative_rearranged}.
\begin{table*}
	\begin{align}
		\left(\hspace{-0.05cm}1\hspace{-0.05cm}-\hspace{-0.05cm}\rho_{p}\hspace{-0.05cm}+\hspace{-0.05cm}\left(\rho_{p}\sigma^{\mathrm{ZF}}_{\tilde{k},p}\right)^{\frac{K-N}{1-K\beta+K-N}}\hspace{-0.05cm}\right)\hspace{-0.05cm}\log\hspace{-0.05cm}\left(\hspace{-0.05cm}1\hspace{-0.05cm}-\hspace{-0.05cm}\rho_{p}\hspace{-0.05cm}+\hspace{-0.05cm}\left(\rho_{p}\sigma^{\mathrm{ZF}}_{\tilde{k},p}\right)^{\frac{K-N}{1-K\beta+K-N}}\hspace{-0.05cm}\right)\hspace{-0.05cm}>\hspace{-0.05cm}\left(\rho_{p}\sigma^{\mathrm{ZF}}_{\tilde{k},p}\right)^{\frac{K-N}{1-K\beta+K-N}}\hspace{-0.05cm}\log\hspace{-0.05cm}\left(\hspace{-0.05cm}\rho_{p}\sigma^{\mathrm{ZF}}_{\tilde{k},p}\right)\hspace{-0.05cm}\frac{(1\hspace{-0.05cm}-\hspace{-0.05cm}N\beta)(K\hspace{-0.05cm}-\hspace{-0.05cm}N)}{(1\hspace{-0.05cm}-\hspace{-0.05cm}K\beta\hspace{-0.05cm}+\hspace{-0.05cm}K\hspace{-0.05cm}-\hspace{-0.05cm}N)^{2}}.
		\label{eqn:lemmax_derivative_rearranged} 	
	\end{align}
	\vspace{-0.2cm}
	\hrule
	\vspace{-0.5cm}
\end{table*}
Since $1-\rho_{p} \geq 0$, the first term in \eqref{eqn:lemmax_derivative_rearranged} is lower bounded by $(\rho_{p}\sigma^{\mathrm{ZF}}_{\tilde{k},p})^{\frac{K-N}{1-K\beta+K-N}}\frac{(K-N)\log\left(\rho_{p}\sigma^{\mathrm{ZF}}_{\tilde{k},p}\right)}{1-K\beta+K-N}$.
Applying the lower bound for the first term in \eqref{eqn:lemmax_derivative_rearranged} and cancelling out the identical terms on both sides, we can obtain $\frac{(1-N\beta)}{(1-K\beta+K-N)}<1$,
which holds since $K-N>1$ and $\beta \leq \frac{1}{K}$. Therefore, we can conclude that $\frac{\partial}{\partial\beta}\frac{1-N\beta}{K-N}R^{\mathrm{ZF}}_{c,p}(t^{(1)}_{p}(\beta)) <0$ and $\frac{1-N\beta}{K-N}R^{\mathrm{ZF}}_{c,p}(t^{(1)}_{p}(\beta))$ is monotonically decreasing with $\beta$. 
\vspace{-0.4cm}
\section{Proof of Lemma 4}
\label{appendix:lemmay}
In order to have $t^{(2)}_{p} \geq 0$, its denominator should satisfy $\log\left(\delta_{p}\rho^{\mathrm{ZF}}_{p}\right)-\log\left(\log\left(\delta_{p}\rho^{\mathrm{ZF}}_{p}\right)\right) \geq 0$, which holds since $\log\left(\delta_{p}\rho^{\mathrm{ZF}}_{p}\right) \geq \log\left(\log\left(\delta_{p}\rho^{\mathrm{ZF}}_{p}\right)\right)$ for $\delta_{p}\rho^{\mathrm{ZF}}_{p} \hspace{-0.08cm}\geq \hspace{-0.08cm}e$. 
To show that $t^{(2)}_{p}\leq 1$, we write
\begin{align}
		t^{(2)}_{p} = \frac{\log\left(\frac{\delta_{p}\rho^{\mathrm{ZF}}_{p}}{\log\left(\delta_{p}\rho^{\mathrm{ZF}}_{p}\right)}\right)}{\delta_{p}} < \frac{\log\left(\frac{\delta_{p}\rho^{\mathrm{ZF}}_{p}}{\log\left(\delta_{p}\rho^{\mathrm{ZF}}_{p}\right)}\right)}{\delta_{p}\rho^{\mathrm{ZF}}_{p}} &< \frac{\log\left(\delta_{p}\rho^{\mathrm{ZF}}_{p}\right)}{\delta_{p}\rho^{\mathrm{ZF}}_{p}}\nonumber\\
		&< 1. \label{eqn:t2_proof_2}
\end{align}
The first inequality follows from $\rho^{\mathrm{ZF}}_{p}\hspace{-0.2cm}<\hspace{-0.1cm}1$ and the second inequality follows from $\log(\delta_{p}\rho^{\mathrm{ZF}}_{p})\hspace{-0.1cm}\geq\hspace{-0.1cm}1$. 

\ifCLASSOPTIONcaptionsoff
  \newpage
\fi

\end{document}